\documentclass[aps,prx,twocolumn,balance,superscriptaddress,floats,longbibliography]{revtex4-2}
\pdfoutput=1

\usepackage[a4paper, total={7.2in, 10in}]{geometry}
\usepackage{latexsym}
\usepackage{dcolumn}
\usepackage{amssymb,amsmath}
\usepackage{epsf}
\usepackage{float}
\usepackage{hyperref}
\usepackage{mathrsfs}
\usepackage{booktabs}
\usepackage{tabularx}
\usepackage{makecell}

\usepackage[toc,page,titletoc]{appendix}
\usepackage[capitalise,nameinlink]{cleveref}

\usepackage[pdftex]{graphicx}
\usepackage{epstopdf}
\usepackage{upgreek}
\usepackage{xcolor}
\usepackage{soul}
\usepackage{colortbl}

\definecolor{lgray}{gray}{0.9}
\definecolor{mgray}{gray}{0.78}

\newcommand{\ra}[1]{\renewcommand{\arraystretch}{#1}}

\newcolumntype{A}{>{\hsize=0.5\hsize\centering\arraybackslash}X}
\newcolumntype{B}{>{\hsize=0.75\hsize\centering\arraybackslash}X}

\newcolumntype{C}{>{\hsize=1.3\hsize\centering\arraybackslash}X}
\newcolumntype{E}{>{\hsize=1.15\hsize\centering\arraybackslash}X}
\newcolumntype{F}{>{\hsize=0.73\hsize\centering\arraybackslash}X}
\newcolumntype{G}{>{\hsize=0.82\hsize\centering\arraybackslash}X}
\newcolumntype{H}{>{\hsize=1.35\hsize\centering\arraybackslash}X}

\newcolumntype{J}{>{\hsize=1.6\hsize\raggedright\arraybackslash}X}
\newcolumntype{K}{>{\hsize=0.25\hsize\centering\arraybackslash}X}
\newcolumntype{L}{>{\hsize=0.8\hsize\centering\arraybackslash}X}

\begin{document}


\title{Modulation of Brillouin optomechanical interactions via acoustoelectric phonon-electron coupling}

\author{Nils T. Otterstrom}
\thanks{These authors contributed equally to this work}
\affiliation{Microsystems Engineering, Science, and Applications, Sandia National Laboratories, Albuquerque, New Mexico, USA}

\author{Matthew J. Storey}
\thanks{These authors contributed equally to this work}
\affiliation{Microsystems Engineering, Science, and Applications, Sandia National Laboratories, Albuquerque, New Mexico, USA}

\author{Ryan O. Behunin}
\thanks{These authors contributed equally to this work}
\affiliation{Department of Applied Physics and Materials Science, Northern Arizona University, Flagstaff, Arizona 86011, USA}
\affiliation{Center for Materials Interfaces in Research and Applications, Northern Arizona University, Flagstaff, Arizona 86011, USA}

\author{Lisa Hackett}
\affiliation{Microsystems Engineering, Science, and Applications, Sandia National Laboratories, Albuquerque, New Mexico, USA}

\author{Peter T. Rakich}
\affiliation{Department of Applied Physics, Yale University, New Haven, Connecticut 06520, USA}

\author{Matt Eichenfield}
\email{meichen@sandia.gov}
\affiliation{Microsystems Engineering, Science, and Applications, Sandia National Laboratories, Albuquerque, New Mexico, USA}


\date{\today}

\begin{abstract}

Optomechanical Brillouin nonlinearities---arising from the coupling between traveling photons and phonons---have become the basis for a range of powerful optical signal processing and sensing technologies. The dynamics of such interactions are largely set and limited by the host material's elastic, optical, and photo-elastic properties, which are generally considered intrinsic and static. Here we show for the first time that it is feasible to dynamically reconfigure the Brillouin nonlinear susceptibility in transparent semiconductors through acoustoelectric phonon-electron coupling.  Acoustoelectric interactions permit a wide range of tunability of the phonon dissipation rate and velocity, perhaps the most influential parameters in the Brillouin nonlinear susceptibility. We develop a Hamiltonian-based analysis that yields self-consistent dynamical equations and noise coupling, allowing us to explore the physics of such acoustoelectrically enhanced Brillouin (AEB) interactions and show that they give rise to a dramatic enhancement of the performance of Brillouin-based photonic technologies. Moreover, we show that these AEB effects can drive systems into new regimes of fully-coherent scattering that resemble the dynamics of optical parametric processes, dramatically different than the incoherent traditional Brillouin limit. We propose and computationally explore a particular semiconductor heterostructure in which the acoustoelectric interaction arises from a piezoelectric phonon-electron coupling. We find that this system provides the necessary piezoelectric and carrier response ($k^2\approx 6 \%$), favorable semiconductor materials properties, and large optomechanical confinement and coupling ($|g_0|\approx8000$ (rad/s)$\sqrt{\text{m}}$) sufficient to demonstrate these new AEB enhanced optomechanical interactions.

\end{abstract}
\maketitle

\section{Introduction}

\begin{figure*}[ht]
\centering
\includegraphics[width=1\linewidth]{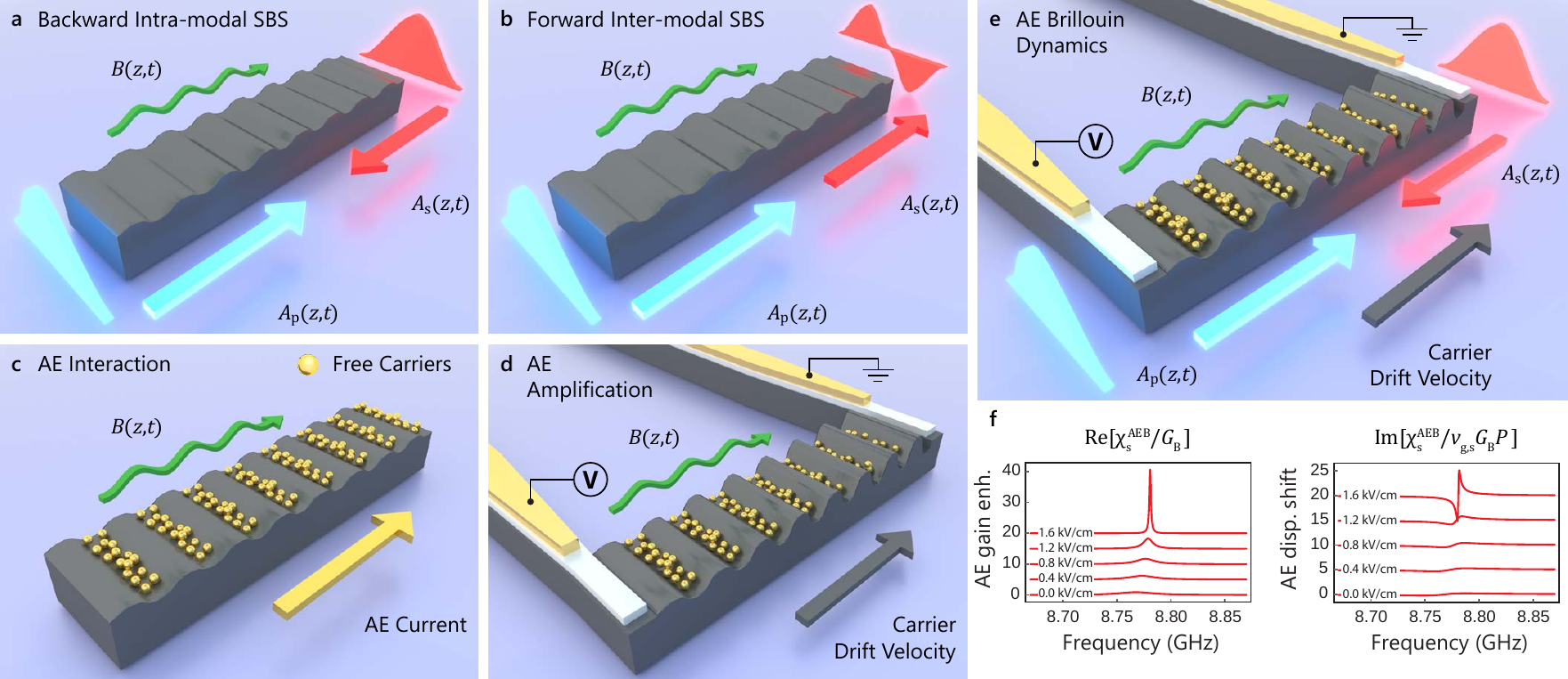}
\caption{Essential physical system for acoustoelectrically (AE) enhanced stimulated Brillouin scattering (SBS). $B(z,t)$, $A_{\rm s}(z,t)$ and $A_{\rm p}(z,t)$ represent the phonon, Stokes, and pump fields, respectively. (a) Backward intra-modal and (b) forward inter-modal SBS.  (c) Elastic wave creating an acoustoelectric current through the acoustoelectric effect.  (d) AE gain applied with an external DC field.  (e) Composite system permitting AE-enhanced stimulated Brillouin scattering. (f) Bulk model prediction of the relative change of (left) the SBS gain spectrum with AE amplification as compared to the peak ABS gain without AE amplification, and (right) the SBS dispersion compared to peak SBS gain rate ($v_{\rm g,s} G_{\rm B} P_{\rm p}$) without AE effects. For clarity, the curves with drift fields of (1.6,1.2,0.8,0.4,0.0) kV/cm are respectively displaced vertically by (20,15,10,5,0).}
\label{fig:physsyst2}
\end{figure*}

Optomechanical Brillouin interactions are unique amongst nonlinear optical processes in that they are enabled by phonons and consequently depend sensitively on the elastic properties of a material \cite{brillouin1922,gross1930change,chiao1964stimulated}.  In optical waveguides, the supported phonon modes determine the types of allowed stimulated Brillouin scattering (SBS) processes, including the traditional intra-modal backward (Fig.~\ref{fig:physsyst2}a), intra-modal forward, or inter-modal forward SBS (Fig.~\ref{fig:physsyst2}b), as well as the characteristic SBS frequencies, nonlinear coupling strengths, and bandwidths.  Leveraging these distinctive properties, SBS has become the basis for a range of flexible and powerful optical signal processing technologies such as amplifiers \cite{ippen1972stimulated,choudhary2016advanced,otterstrom2019resonantly}, lasers \cite{takuma1964stimulated,hill1976cw,li2012characterization,morrison2017compact,otterstrom2018silicon,gundavarapu2019sub,gundavarapu2019sub}, filters \cite{tanemura2002narrowband,shin2015control,marpaung2015low,kittlaus2018rf,gertler2020tunable}, nonreciprocal devices \cite{huang2011complete,kang2011reconfigurable,poulton2012design,kittlaus2018non}, and optical delay lines \cite{okawachi2005tunable,merklein2017chip,jaksch2017brillouin}. Despite the wide application space for SBS devices, however, the dynamics and resulting utility of these interactions have historically been limited by the host material's intrinsic elastic and optical properties \cite{kharel2016noise}. In particular, two fundamental material properties that determine the Brillouin susceptibility are the phononic dissipation rate and velocity. Although these properties and the resulting Brillouin susceptibility can be significantly enhanced by patterning materials into waveguides \cite{rakich2012giant}, the underlying material properties are generally considered intrinsic and static, and in a given material they ultimately limit the strength of these scattering processes and the performance of devices utilizing them.

In this context, the ability to electrically control phononic loss, or even induce phononic gain, could enable dynamic reconfigurability of the Brillouin susceptibility, suppressing or enhancing the nonlinear response by orders of magnitude on demand. Coulomb drag interactions between electrons and phonons in semiconductors provide a powerful mechanism to deterministically modify the phonon velocity and dissipation \cite{Parmenter}. This Coulomb drag effect is achieved in practice by applying external, quasi-static electric fields to the semiconductor charge carriers to produce a drift current; the drifting carriers then interact with and become spatially polarized by the generated charge distribution of the phonons via the electric fields they induce, as depicted in Fig. \ref{fig:physsyst2}c-d. The resulting coupling between the two moving charge distributions ultimately allows non-reciprocal amplification, attenuation, and velocity modification of the phonons. Although the acoustoelectric effect can be produced by different mechanisms in many materials \cite{weinreich1959acoustoelectric,kalameitsev2019valley}, the strongest interactions were predicted to exist \cite{Parmenter} and then observed in piezoelectric semiconductors \cite{wang1962strong}. Going beyond the effects intrinsic to single materials, it was proposed that heterostructures that combine the strongest piezoelectric materials with high-mobility semiconductors could produce exceptionally large acoustoelectric effects \cite{gulyaev1965amplification}, and this ultimately allowed demonstrations of devices such as radio frequency acoustic amplifiers \cite{collins1968amplification,lakin1969100,coldren1971amp}, convolvers \cite{cafarella1976acoustoelectric, leonberger1978gap}, and correlators \cite{bers1974surface,ingebrigtsen1976schottky, ingebrigtsen1975coherent}. Recent advances in semiconductor epitaxy, heterogeneous integration, and nanofabrication have led to a new class of acoustoelectric heterostructures, enabling ultra-compact acoustic amplifiers \cite{hackett2019amp,hackett2021amp,Malocha_AEGain_2020,Ghosh_FDSOI_2019,Mansoorzare_LNOSI_2020}, circulators \cite{hackett2021amp,Ghosh_Circulator_2020}, and switches \cite{Storey_Switch_2021}, with performance that greatly supercedes that which was previously possible.

Here, for the first time, we describe how Brillouin interactions can be modified by the acoustoelectric effect using externally applied electric fields, enabling new degrees of control over the nonlinear optical susceptibility. We show that straight-forward modifications of our recently developed heterostructure platform for acoustoelectric radio-frequency signal processing devices \cite{hackett2019amp,hackett2021amp} provide a powerful platform for demonstrating and using these effects, simultaneously allowing large optomechanical confinement and coupling ($|g_0|\approx8000$ (rad/s)$\sqrt{\text{m}}$) and acoustoelectric coupling ($k^2\approx 6\%$). We derive a general Hamiltonian framework that describes the coupled acoustoelectric and optomechanical interactions and show that acoustoelectric modifications to both the real and imaginary parts of the Brillouin susceptibility lead to novel effects and dynamics that would not be expected with intrinsic material properties. Using the proposed heterostructure as an example system for concrete physical predictions, we show how modification of the phonon dissipation rates through application of quasistatic electric fields in the semiconductor can drastically improve the performance of archetypal Brillouin photonic devices such as Brillouin amplifiers, lasers, nonreciprocal devices, and delay lines. Moreover, we show that these acoustoelectrically enhanced Brillouin interactions allow Brillouin scattering processes to go from a regime where the phonon coherence lengths ($\sim 100$ $\upmu$m) are significantly less than those of the photons ($\sim$ cm)---behaving effectively as incoherent scatterers---to one in which the phonon coherence lengths can achieve parity with and even exceed those of the photons, enabling fully coherent scattering processes that resemble the dynamics of optical parametric amplification \cite{yariv1965quantum}.  We then discuss future prospects for achieving these effects in other materials platforms, novel devices, and applications, as well as their application to cavity optomechanical systems.

\section{Proposed Candidate Physical System} \label{sec:system}

To enable reconfigurable Brillouin optomechanics through acoustoelectric control of phonon dissipation, we seek a traveling-wave optomechanical physical system that (1) guides and tightly confines optical and elastic waves, (2) supports non-zero Brillouin coupling, and (3) exhibits an acoustoelectric coupling between phonons and electrons---as depicted in Fig. \ref{fig:physsyst2}e. This combination of system properties allows us to modulate the Brillouin susceptibility using an applied electric field.

\begin{figure*}[ht]
\centering
\includegraphics[width=1\linewidth]{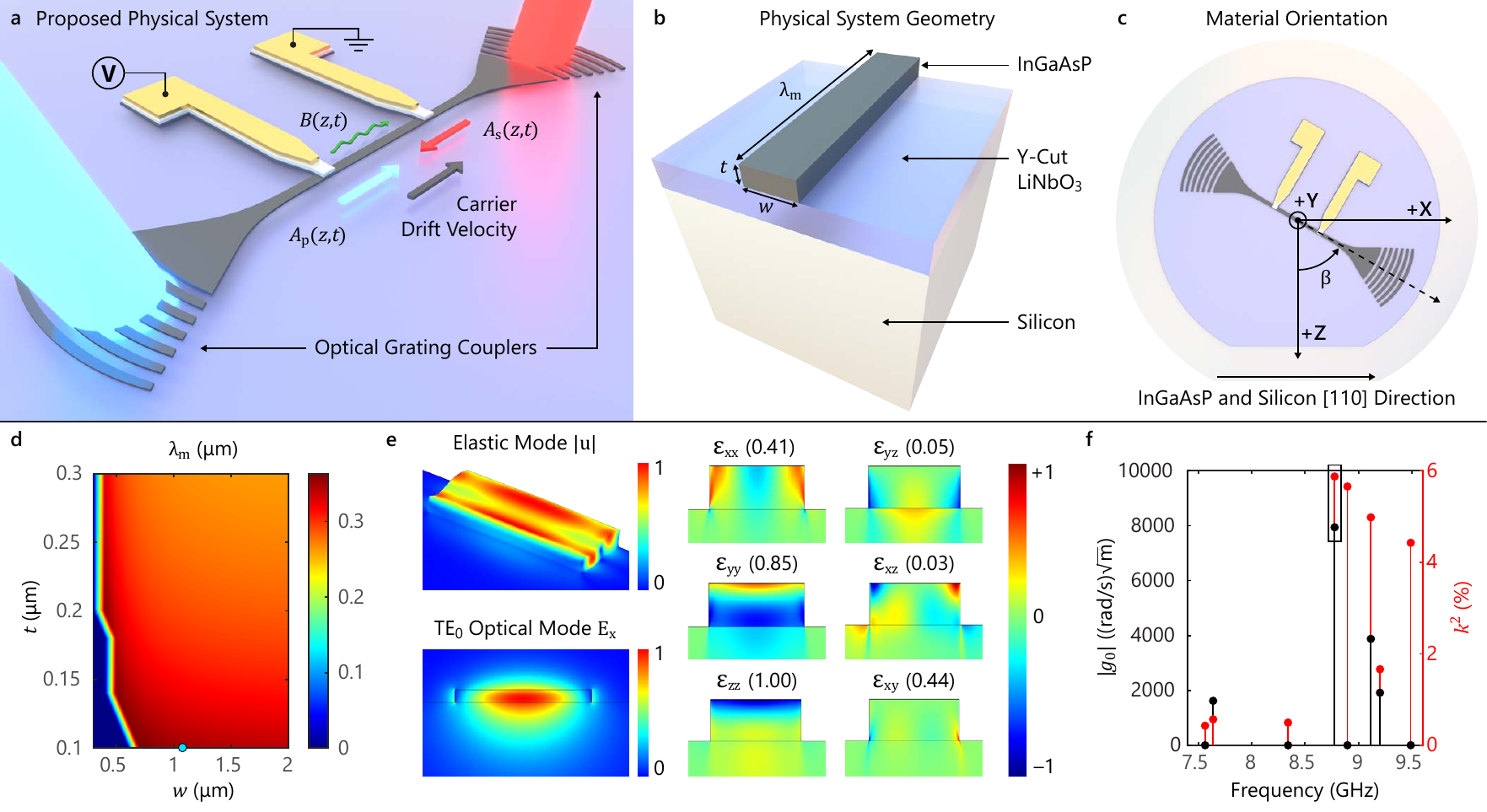}
\caption{(a) Illustration of the InGaAsP on Y-cut lithium niobate on silicon material system for acoustoelectrically enhanced Brillouin devices (not to scale). (b) The key geometric design parameters are shown as well as (c) the defined material orientation parameters (not to scale). (d) Phase-matched phonon wavelength as a function of waveguide geometry parameters for backward intra-modal Brillouin scattering (blue circle highlights the chosen waveguide dimensions). (e) Simulated elastic and optical mode shapes with profiles of each strain component. For each normalized strain profile, the relative magnitude between the strain components is shown in parenthesis. (f) The calculated optomechanical coupling coefficient (black) and piezoelectric coupling coefficient (red) for each phase-matched elastic mode simulated with the chosen waveguide dimensions. The elastic mode with the highest optomechanical and piezoelectric coupling is highlighted (black box).}
\label{fig:InGaAsPsystem}
\end{figure*}

Building on our recent work in acoustoelectric heterostructures \cite{hackett2019amp, hackett2021amp}, we propose the following physically realizable system to explore and harness these new acoustoelectric Brillouin dynamics. The system seen in Fig.~\ref{fig:InGaAsPsystem}a-c consists of a hybrid photonic-phononic waveguide patterned out of epitaxially grown In\textsubscript{0.712}Ga\textsubscript{0.288}As\textsubscript{0.625}P\textsubscript{0.375}, which is bonded to a Y-cut lithium niobate thin film on a silicon substrate (InGaAsP-LN-Si).  Here, the InGaAsP thin film is used as the optical guiding layer in addition to the semiconductor layer that provides the necessary free carriers for the acoustoelectric effect. Because a small carrier density ($\sim 10^{16}$ $\rm cm^{-3}$) is sufficient to produce a strong acoustoelectric gain, these free carriers have negligible to minimal impact on the linear optical losses \cite{hackett2021amp,hava1993theoretical}. For the elastic modes and quaternary composition considered in this work, the longitudinal and shear acoustic phase velocity in InGaAsP is smaller than in lithium niobate, allowing for a high degree of acoustic confinement. The higher thermal conductivity of the silicon substrate improves the heat dissipation of the system, which reduces the thermal effects of the Joule heating in the InGaAsP and allows for continuous operation of the acoustoelectrically enhanced Brillouin devices (see Supplementary Section~\ref{sec:CW}). The optomechanical coupling, $g_0$, is a function of the overlap between the co-localized optical and elastic modes (see Supplementary Section \ref{sec:SIdesign}), while the acoustoelectric gain is governed by the interaction of the electric potential of the elastic wave in the piezoelectric and the free carriers in the semiconductor. Given these considerations, the InGaAsP-LN-Si heterostructure provides the essential acoustoelectric and optomechanical ingredients for acoustoelectrically reconfigurable Brillouin optomechanics.

To determine the accessible experimental performance of such a waveguide structure, we carry out finite element simulations of the optical modes, the electric and strain fields of the elastic modes, along with $g_0$ and the relevant piezoelectric coupling $k^2$ (see Supplementary Section \ref{sec:SIdesign}), for a variety of Brillouin optomechanical interactions and device geometries. Figure~\ref{fig:InGaAsPsystem}b-c highlights key geometric parameters to consider when designing an acoustoelectrically enhanced Brillouin device (see Supplementary Section~\ref{sec:SIdesign} for more details). The width ($w$) and thickness ($t$) of the InGaAsP waveguide not only determines the confinement and effective index of the optical modes, but it also determines the lateral and vertical confinement of the elastic mode in the InGaAsP-LN-Si system. The phonon wavelength ($\lambda$\textsubscript{m}) sets the operating phonon frequency and elastic mode shape given a particular waveguide cross section. Finally, since this physical system includes anisotropic materials, the propagation angle ($\beta$) in the device plane and the acoustic polarization play a critical role in determining the acoustoelectric and optomechanical coupling.

With these considerations, we simulate the relevant optical modes, determine the phonon wavelength through the phase-matching condition, simulate the elastic mode with the appropriate periodic boundary condition, and then calculate the optomechanical overlap integrals and piezoelectric coupling ($k^2$) (see Fig.~\ref{fig:InGaAsPsystem}d-f and Supplementary Section~\ref{sec:SIdesign} for more details). Figure~\ref{fig:InGaAsPsystem}e highlights a particular mode triplet (required by the 3-wave Brillouin process) that demonstrates excellent Brillouin optomechanical and acoustoelectric properties, with a distributed optomechanical coupling rate of $|g_0|=7943$ (rad/s)$\rm \sqrt{m}$ and electromechanical coupling $k^2=5.87$\% (for details on these calculations, see Supplementary Section~\ref{sec:SIdesign}). The elastic mode that mediates this backward Brillouin interaction at telecom wavelengths is a Rayleigh-like waveguide mode with a center frequency of 8.78 GHz and an acoustic velocity of 3072 $\rm m/s$.

\begin{figure}[t]
\centering
\includegraphics[width=1\linewidth]{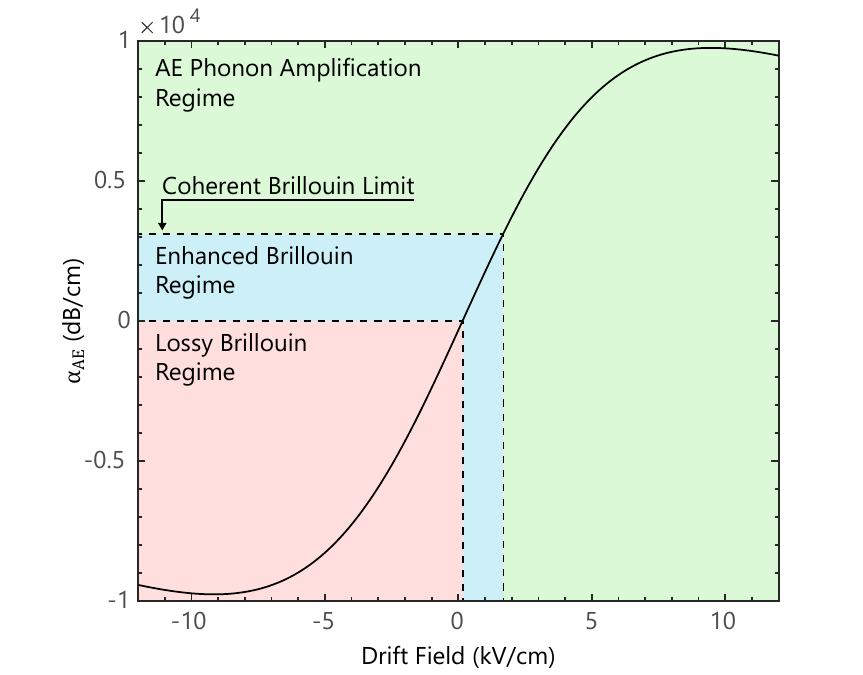}
\caption{Acoustoelectric phonon amplification as a function of drift field, highlighting the accessible regimes of dynamics.}
\label{fig:AEregions}
\end{figure}

Equipped with the optomechanical and acoustoelectric couplings, we now have the necessary parameters to examine the accessible dynamics for this particular mode triplet (see Eq. \ref{eq:eom}). For acoustoelectric gain, a key parameter is the electric field at which the free carrier drift velocity, $v_\text{d}$, equals the phonon phase velocity, $v_\text{m}$, which we call the equal-velocity point ($v_\text{d}=v_\text{m}$). At electric fields lower (higher) than this point, the acoustoelectric material causes phononic loss (gain). Using the normal mode theory developed by Kino and Reeder \cite{kino_normal_1971}, the acoustoelectric phonon gain as a function of applied drift field near this equal-velocity point is plotted in Fig.~\ref{fig:AEregions}. Here we assume an electron mobility of 2000~cm\textsuperscript{2}/Vs and an intrinsic phonon spatial decay rate of 3100 dB/cm, corresponding to an acoustic quality factor of 250. This loss rate defines the distinct, acoustoelectric-enabled regimes of operation that will be explored in Section~\ref{sec:enveloptheory}.  With these conditions, the drift field needed to reach the equal velocity point is 150 V/cm. Increasing the drift field further, we enter the acoustoelectrically enhanced Brillouin (AEB) regime in which the phonon loss is reduced by the net acoustoelectric gain.  As the strength of the drift field increases, we approach the acoustoelectric coherent Brillouin limit, defined as the point at which the acoustoelectric gain matches the intrinsic phonon losses; as we will see in Section~\ref{sec:parametric}, this yields parametric-like three-wave dynamics. If, by contrast, we apply a field below the equal-velocity point or with a negative polarity, additional phonon loss is introduced (lossy Brillouin regime).  In this case, the Brillouin coupling can either be dramatically enhanced or reduced, allowing for dynamical switching of the optomechanical response.

\section{Acoustoelectric Brillouin Dynamics}\label{sec:enveloptheory}

We present a general analysis of the effects of acoustoelectric phonon gain on the nonlinear susceptibility and spatio-temporal dynamics of optomechanical Brillouin interactions using the InGaAsP-LN-Si physical system as a concrete example for physical predictions (see Section~\ref{sec:system}).
In this system, the Hamiltonian governing a stimulated (Stokes) optomechanical Brillouin response can be expressed as \cite{kharel2016noise},

\begin{equation}
\begin{aligned}
H_{\rm B}^{\rm int}=\hbar \int dz \big(g_0 A^\dagger_{\rm p}(z,t)A_{\rm s}(z,t)B(z,t)\big)e^{i (q_{\rm m}-\Delta k_{\rm s})z}+h.c.
\end{aligned}
\label{eq:eom}
\end{equation}
where $A_{\rm p}(z,t)$, $A_{\rm s}(z,t)$, and $B(z,t)$ are the envelope operators for the pump, Stokes, and phonon fields, respectively, $g_0$ is the distributed optomechanical coupling of the waveguide system, and $q_{\rm m}-\Delta k_{\rm s}$ represents the phase-mismatch between the wavevectors of the phonon ($q_{\rm m}$) and the optical beat pattern ($\Delta k_{\rm s}=k_{\rm p}-k_{\rm s}$) \cite{kharel2016noise}.  

When phase-matching is satisfied (i.e., $q_{\rm m}-\Delta k_s=0$), the Heisenberg equations of motion yield
\begin{equation}
\begin{aligned}
\frac{\partial \bar{B}}{\partial t}&=-i (\Omega_{\rm m}-\Omega) \bar{B}-\frac{\Gamma}{2}\bar{B} + v_{\rm g,b} \frac{\partial \bar{B}}{\partial z}-i g_{0}^* \bar{A}_{\rm s}^{\dagger} \bar{A}_{\rm p}+\eta\\
\frac{\partial \bar{A}_{\rm p}}{\partial t}&=-\frac{\gamma_{\rm p}}{2} \bar{A}_{\rm p} + v_{\rm g,p} \frac{\partial \bar{A}_{\rm p}}{\partial z}-i g_{0} \bar{A}_{\rm s} \bar{B}+\xi_{\rm p}\\
\frac{\partial \bar{A}_{\rm s}}{\partial t}&=-\frac{\gamma_{\rm s}}{2} \bar{A}_{\rm s} - v_{\rm g,s} \frac{\partial \bar{A}_{\rm s}}{\partial z}-i g^*_0 \bar{A}_{\rm p} \bar{B}^\dagger +\xi_{\rm s}
\end{aligned}
\label{eq:eom-rot}
\end{equation}
where $\bar{B}(z,t)=B(z,t)\exp(i \Omega t)$, $\bar{A}_{\rm p}(z,t)=A_{\rm p}(z,t)\exp(i \omega_{\rm p} t)$, and $\bar{A_{\rm s}}(z,t)=A_{\rm p}(z,t)\exp(i \omega_{\rm s} t)$ are the slowly varying phonon, pump, and Stokes envelopes in the rotating frame, with the condition $\Omega=\omega_{\rm p}-\omega_{\rm s}$;  $\Omega_{\rm m}$ is the natural mechanical frequency of the phonon mode, while $v_{\rm g,b}$, $v_{\rm g,p}$, and $v_{\rm g,s}$ represent the group velocities for the phonon, pump, and Stokes fields.  We also note that Eq. \ref{eq:eom} includes the effects of dissipation---denoted by $\Gamma$, $\gamma_{\rm p}$, and $\gamma_{\rm s}$ for the phonon, pump, and Stokes fields, respectively---and therefore requires thermal and vacuum noise terms $\eta$, $\xi_{\rm p}$, and $\xi_{\rm s}$, according to the fluctuation-dissipation theorem.  

We next examine the impact of the acoustoelectric phonon gain or loss on these dynamics.  
For this purpose we develop a Hamiltonian formulation for the coupled dynamics of the electric potential, drift current, and elastic waves. This treatment provides (1) a general description for the acoustoelectric coupling for waveguides of arbitrary cross-sectional geometry, (2) yields the acoustoelectric gain, loss and dispersion in terms of coupling parameters (see $\kappa_{\omega,\ell}$ below) quantified by (overlap) integrals of products of the piezoelectric coupling and elastic and electric potential modes over the waveguide cross section, and (3) naturally lends itself to noise analysis leveraging the fluctuation-dissipation theorem \cite{callen1951irreversibility}. These acoustoelectric dynamics can be captured by the Hamiltonian given by 
\begin{equation}
\begin{aligned}
H_{\rm AE} = \hbar \sum_\ell \int d\omega & \int dz \bigg[ 
\Phi_{\omega \ell}^\dag(z)  \hat{\omega} \Phi_{\omega \ell }(z)
\\
 & +  (\kappa_{\omega, \ell}\Phi_{\omega j}(z) B^\dag(z) + H.c.)  \bigg],
\end{aligned}
\label{eq:HAE}
\end{equation}
where $\Phi_{\omega \ell}$ and $B$ are the envelope operators for the electric potential and phonon fields, respectively, $j$ labels bulk and surface modes of the potential and $\kappa_{\omega, \ell}$, quantifies the acoustoelectric coupling defined in Supplementary Section~\ref{sec:AEdynamics}. Within this coupled-envelope framework, the oscillation frequency of the mode $\hat{\omega} = \omega+v_{\rm d} q_{\rm m} -i {\bf v}_{\rm d} \cdot \nabla$ is operator valued, capturing the effects of temporal oscillations, current drift (given by ${\bf v}_{\rm d}=\mu {\bf E}_{\rm DC}$, where ${\bf E}_{\rm DC}$ is the applied drift field) and slowly-varying spatial dynamics of the potential amplitude.  Here, $\omega$ denotes the normal mode frequency of the potential in the absence of a drift current of speed $v_{\rm d}$ and $q_{\rm m}$ is the carrier wavevector of the phonon and potential within the slowly-varying envelope approximation. 

When an applied electric field causes free carriers to drift at a velocity greater than the acoustic wave, the acoustoelectric effect produces a spatial phonon gain given by $\alpha_{\rm AE} = - \sum_\ell 2\pi |\kappa_{\omega-v_{\rm d}q_{\rm m},\ell}|^2/v_{\rm g,b}$. Noting that $|\kappa_{\omega-v_{\rm d}q_{\rm m},\ell}|^2 \propto (v_{\rm m} - v_{\rm d})$ where $v_{\rm m}$ is the phonon phase velocity, we recover the essential features of acoustoelectric gain. Namely, acoustoelectric amplification (excess attenuation) occurs when the drift velocity exceeds (falls below) the phonon phase velocity. In the case of a Brillouin-active optomechanical system, the effects of acoustoelectric gain and the accompanying velocity shift as illustrated by Fig. \ref{fig:physsyst2}f can be described by a modified complex dissipation $\Gamma \rightarrow \tilde{\Gamma}$ within the coupled envelope equations (Eq.~\ref{eq:eom-rot}), where $\tilde{\Gamma}=2i\Delta \Omega_{\rm AE} + \Gamma-G_{\rm AE}$.  Here $\Delta \Omega_{\rm AE}$ denotes the acoustoelectric shift in the resonance frequency and $G_{\rm AE}$ quantifies the phonon gain as a time rate such that $G_{\rm AE}=v_{\rm g,b} \alpha_{\rm AE}$ (see Supplementary information for more details).

To elucidate the impact of acoustoelectric gain on the nonlinear Brillouin susceptibility, we will semi-classically treat this phenomena in the undepleted pump limit, for the moment ignore the noise terms (for noise analysis see Supplementary Section \ref{sec:noise}), and move to the Fourier domain ($f[\omega]=\int_{-\infty}^{\infty}f(t)\exp(i\omega t) dt$), which yields

\begin{equation}
\begin{aligned}
\left(-i \omega +\frac{\gamma_{\rm s}}{2}\right) \bar{A}_{\rm s}[z,\omega]+v_{\rm g,s} \frac{\partial \bar{A}_{\rm s}[z,\omega]}{\partial z}= -i g_0^* \bar{A}_{\rm p} \bar{B}^\dagger[z,\omega]\\
i(\Omega-\Omega_{\rm m}-\Delta \Omega_{\rm AE}-\omega)\bar{B}[z,\omega]+\frac{\Gamma-G_{\rm AE}}{2}\bar{B}[z,\omega]+\\v_{\rm g,b} \frac{\partial \bar{B}[z,\omega]}{\partial z}=-ig_0^*\bar{A}_{\rm p}\bar{A}^\dagger_{\rm s}[z,\omega].
\end{aligned}
\end{equation}
These spatial equations in the frequency domain are the basis for a range of new acoustoelectric Brillouin device physics that we explore below.

\subsection{Acoustoelectrically enhanced Brillouin (AEB) limit} \label{sec:tradlimit}

We now focus on the particular case in which the spatial coherence length of the phonon field, despite gain from the acoustoelectric effect, is significantly shorter than that of the optical fields \cite{boyd2020nonlinear}.  We will explore the case of near equal coherence lengths in Section \ref{sec:parametric}.  In this acoustoelectrically enhanced Brillouin (AEB) limit, the spatial dynamics of the phonon field can be adiabatically eliminated (i.e., $ v_{\rm g,b} \frac{\partial \bar{B}[z,\omega]}{\partial z}\approx 0$) \cite{boyd2020nonlinear}, yielding 

\begin{equation}
\begin{aligned}
\bar{B}[z,\omega]=-i g_0^* \chi^{\rm AE}_{\rm B}[\omega] \bar{A}_{\rm p} \bar{A}_{\rm s}^\dagger[z,\omega].
\end{aligned}
\label{eq:ad}
\end{equation}

\noindent
Where the acoustoelectrically modified phonon susceptibility is given by $\chi^{\rm AE}_{\rm B}[\omega]=(i(\Omega-\Omega_{\rm m}-\Delta \Omega_{\rm AE}-\omega)+(\Gamma-G_{\rm AE})/2)^{-1}$.	Through the stimulated Brillouin process, this modification translates into an acoustoelectric Brillouin nonlinear optical susceptibility $\chi_{\rm s}^{\rm AEB}[\omega]$ in the dynamics of the Stokes wave such that

\begin{equation}
\begin{aligned}
-i \omega \bar{A}_{\rm s}[\omega,z]&+\frac{\alpha v_{\rm g,s}}{2}\bar{A}_{\rm s}[\omega,z]\\&+v_{\rm g,s} \frac{\partial \bar{A}_{\rm s}[z,\omega]}{\partial z}=\chi_{\rm s}^{\rm AEB}[\omega] \bar{A}_{\rm s}[\omega,z],
\end{aligned}
\label{eq:sol1}
\end{equation}

\noindent
where $\alpha$ is the spatial optical decay rate and $\chi_{\rm s}^{\rm AEB}[\omega]=|g_0|^2 |\bar{A}_{\rm p}|^2 \chi_{\rm B}^{\rm AE}[\omega]$.

Thus, through this acoustoelectrically enhanced stimulated Brillouin process, a Stokes field with initial conditions $\bar{A}_{\rm s}[0,\omega]$ is amplified as 

\begin{equation}
\begin{aligned}
\bar{A}_{\rm s}[z,\omega]=\bar{A}_{\rm s}[0,\omega]\exp \bigg[\bigg(\frac{-i\omega}{v_{\rm g,s}}-\frac{\alpha}{2}+\frac{ G_{\rm B} P \Gamma \chi_{\rm B}^{\rm AE*}[\omega]}{4}\bigg)z\bigg],
\end{aligned}
\label{eq:sol2}
\end{equation}

\noindent
where the Brillouin gain coefficient power product $G_{\rm B} P$ is given by $G_{\rm B} P=4 |g_0|^2 |\bar{A}_{\rm p}|^2/(\Gamma v_{\rm s})$.
Inspecting the resonance condition (i.e., $\omega=0,\Omega=\Omega_{\rm m}+\Delta \Omega_{\rm AE}$), reveals a gain (loss) factor that is exponentially increased (decreased) by the acoustoelectric modified phonon dissipation, specifically

\begin{equation}
\begin{aligned}
\frac{|\bar{A}_{\rm s}[z,0]|^2}{|\bar{A}_{\rm s}[0,0]|^2}=\exp\bigg[\bigg(G_{\rm B} P \frac{\Gamma}{\Gamma-G_{\rm AE}}-\alpha \bigg)z\bigg],
\end{aligned}
\label{eq:sol3}
\end{equation}

\noindent
revealing that the ability to modify the nonlinear susceptibility through acoustoelectric coupling enables \textit{in situ} reconfigurability of the optical SBS gain. 

In addition to these gain dynamics, we explore thermal-mechanical Brillouin noise under the influence of acoustoelectric gain and find that the SBS noise factor from these thermal fluctuations scales as

\begin{equation}
\begin{aligned}
F\approx 1+n_{\rm th} \left( \frac{\Gamma}{\Gamma-G_{\rm AE}} \right),
\end{aligned}
\label{eq:facresult}
\end{equation}
where $n_{\rm th}$ is the thermal occupation of the phonon mode given by the Bose-Einstein distribution (for a detailed derivation, see Supplementary Section \ref{sec:noise}). When the intrinsic acoustoelectric noise is small relative to thermomechanical noise (see Supplementary Section \ref{sec:noise} for more details), these results suggest that near quantum limited amplification may be possible in the limit of low temperatures (i.e., when $k_{\rm B} T\ll \hbar \Omega_{\rm m}$).

\begin{figure*}[ht]
\centering
\includegraphics[width=\linewidth]{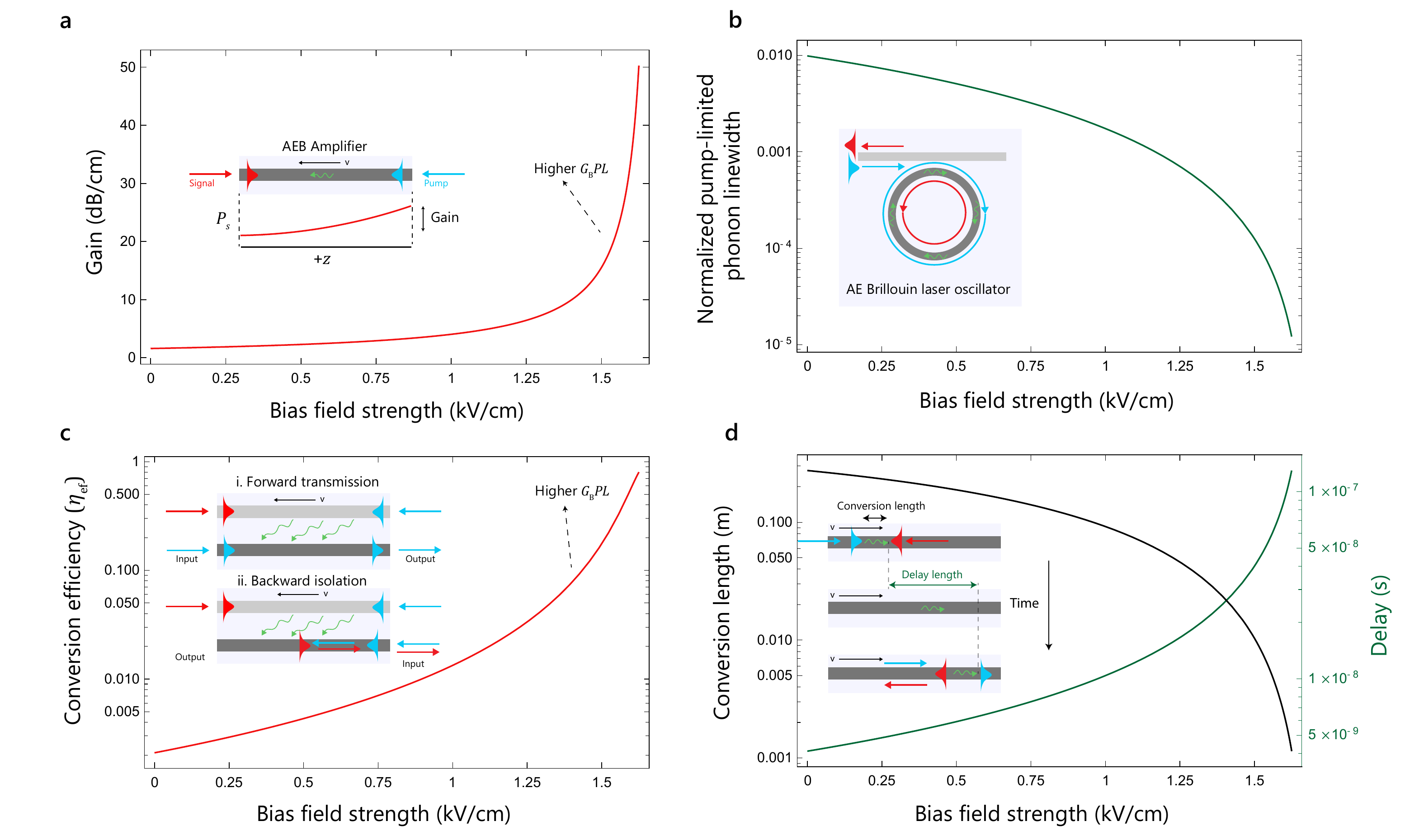}
\caption{Performance enhancement from acoustoelectric coupling for archetypal Brillouin devices in the acoustoelectrically enhanced Brillouin (AEB) limit using the selected mode triplet of the InGaAsP-LN-Si heterostructure waveguide as an example (see Fig. \ref{fig:InGaAsPsystem}). In addition to the simulated optomechanical and electromechanical coupling and acoustic quality factor of 250 (see Supplementary Section \ref{sec:SIdesign} for more details), for (a) and (c) we assume a pump power of 50 mW and a device length of 1 cm.   (a) Gain improvement in a backward SBS amplifier as a function of bias field strength.   (b) Potential improvement in pump-limited phonon linewidth for an acoustoelectrically enhanced Brillouin optomechanical oscillator operating in the phonon linewidth narrowing regime. Here the phonon linewidth is not limited by Schawlow-Townes narrowing. The phonon linewidth is normalized by the pump linewidth. (c) Nonreciprocal scattering efficiency for Brillouin-based AE-enhanced isolator as a function of bias field strength (neglecting optical loss). The particular implementation of this enhanced nonreciprocal scattering process is shown in insets i-ii, and requires optomechanical waveguides that are coupled phononically.  Counter-propagating pump and signal waves in the drive waveguide (light gray) set up an acousto-optic grating in the modulator waveguide (dark gray) that can be enhanced by the acoustoelectric effect. (i) When light at the pump frequency is injected into the modulator waveguide in the forward direction, it is not phase-matched to and passes unaffected by the acousto-optic grating. (ii) By contrast, in the backward direction, the light can experience complete acousto-optic scattering providing effective optical isolation.    (d) Phonon delay enhancement and conversion length reduction due to the acoustoelectric effect.  As diagrammed in the inset, pump (blue) and write (red) fields set up an acoustoelectrically enhanced acoustic-grating, with a conversion efficiency of 1 (i.e., one phonon from one photon) over the so-called conversion length (black).  At the same time, the acoustoelectric effect increases the effective delay (green) for the phononic memory.  After the delay, the write information can be recovered by the pump pulse, which upon acousto-optic scattering, creates a back-propagating read pulse (red).}
\label{fig:device}
\end{figure*}

\subsection{Performance of acoustoelectric Brillouin-based devices}\label{sec:bdevices}

Acoustoelectric phonon gain directly modifies the nonlinear optical susceptibility through an enhancement in the effective Brillouin gain coefficient, which is inversely proportional to the phonon dissipation rate.  These enhanced dynamics have important consequences for Brillouin-based devices such as amplifiers, lasers, nonreciprocal devices, optomechanical delay among others---especially in chip-scale systems where the accessible levels of Brillouin gain have been historically limited.

\subsubsection{Brillouin amplifiers}

The performance of Brillouin-based amplifiers---used in a range of signal processing \cite{eggleton2019brillouin}, filtering \cite{tanemura2002narrowband,choudhary2016advanced}, and optical nonreciprocity applications \cite{kang2011reconfigurable,otterstrom2019resonantly}---can be radically improved as acoustoelectric phonon gain dramatically reduces the effective phonon lifetime.	The effective exponential gain enhancement is highlighted by the power dynamics of a small-signal Stokes wave, given by

\begin{equation}
\begin{aligned}
P_{\rm s}[z]= P_{\rm s}[0]\exp\bigg[\big( G_{\rm B} P \frac{\Gamma}{\Gamma-G_{\rm AE}}-\alpha \big)z\bigg].
\end{aligned}
\label{eq:sol4}
\end{equation}

Figure~\ref{fig:device}a plots AEB-enhanced SBS gain as a function of applied drift field for the InGaAsP-LN-Si physical system under consideration. Without acoustoelectric gain, this Brillouin-optomechanical system might struggle to yield even net amplification (Brillouin gain compensating for propagation loss), but with it, 50 dB or more of dynamically reconfigurable amplification may be possible on-chip.  Moreover, this degree of AEB amplification is accompanied by an increasingly narrow gain bandwidth (from $\sim 35$~MHz to $\sim 300$~kHz), enabling selective narrowband operations in RF-photonic filtering applications.

\subsubsection{Brillouin-based laser oscillators}

The ability to dynamically reconfigure phonon dissipation rates and Brillouin gain by orders of magnitude with DC electric fields also can be leveraged in Brillouin laser oscillator applications.
For one, acoustoelectric phonon gain directly reduces the pump power required to reach self-oscillation.   Specifically, the threshold in this case is given by

\begin{equation}
\begin{aligned}
P^{\rm AEB}_{\rm th}=\frac{\Gamma- G_{\rm AE}}{\Gamma}P_{\rm th}.
\end{aligned}
\label{eq:sol5}
\end{equation}

\noindent
Thus, provided the acoustoelectric interaction can produce net phonon amplification, acoustoelectric gain enables near arbitrary control of the self-oscillation threshold. For example, a drift field of 1.62~kV/cm applied to the InGaAsP would reduce the threshold condition by approximately $30\times$.  

Second, acoustoelectric gain allows one to precisely shape the linewidth narrowing dynamics of Brillouin lasers.  The well-known ability of a Brillouin laser to produce ultra-low noise self-oscillation (even in cases where the pump is spectrally much broader) is contingent on a sufficient asymmetry between the optical and acoustic temporal dissipation rates.  Specifically, systems in which the optical dissipation rate exceeds that of the acoustic (and vice versa) can produce acoustic (optical) linewidth narrowing. Using external DC fields to change the phonon dissipation rate through the acoustoelectric effect permits flexible \textit{in} \textit{situ} control of the linewidth narrowing dynamics.  

\begin{figure*}[ht]
\centering
\includegraphics[width=\linewidth]{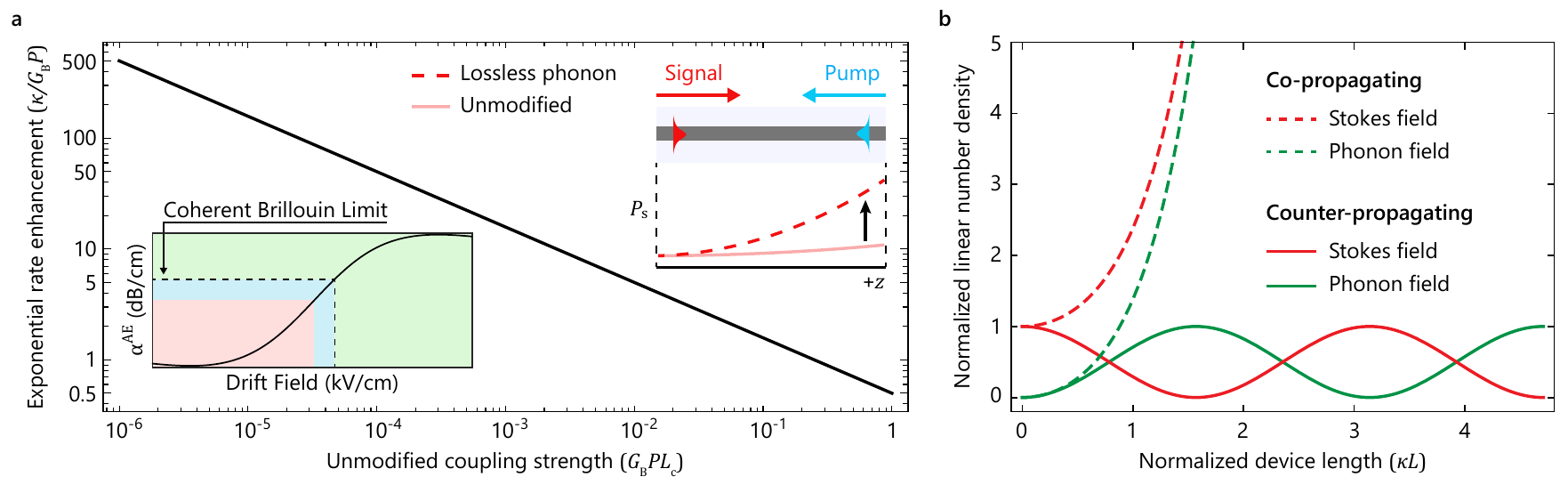}
\caption{(a) Exponential growth rate enhancement ($\kappa /(G_{\rm B} P)$) as a function of unmodified coupling strength ($G_{\rm B} P L_{\rm c}$ at the coherent Brillouin limit, where $L_{\rm c}$ is the intrinsic phonon coherence length $L_{\rm c}=v_{\rm g,b}/\Gamma$ ). The potential improvement by operating in this lossless regime is accentuated in the case of low intrinsic Brillouin coupling. The insets illustrate the acoustoelectric phonon amplification as a function of drift field, highlighting the coherent Brillouin limit (left) and the exponential growth rate enhancement by entering the coherent Brillouin regime (right). (b) Stokes and phonon number densities as a function of device length, normalized independently for illustration. Disparate parametric-like dynamics emerge in the case of co- and counter-propagating Stokes and phonon fields, as described by Eq. \ref{eq:par4},\ref{eq:par6}.}
\label{fig:parametric}
\end{figure*}

For oscillator applications, it may be desirable to achieve high degrees of spectral compression of the microwave-frequency phonon field.	In the case of a linewidth limited by pump noise, this ‘noise eating’ process can be quantified by the ratio of phonon linewidth ($\Delta \nu_{\rm B}$) to the input pump linewidth ($\Delta \nu_{\rm p}$), given by 

\begin{equation}
\begin{aligned}
\frac{\Delta \nu_{\rm B}}{\Delta \nu_{\rm p}}=\frac{1}{(1+\frac{\gamma}{\Gamma-G_{\rm AE}})^2},
\end{aligned}
\label{eq:sol6}
\end{equation}
in analogy with the optical linewidth narrowing predicted by \cite{debut2000linewidth}. Figure \ref{fig:device}b plots the pump-limited phonon linewidth as a function of acoustoelectric gain, revealing phonon linewidth narrowing beyond what is possible without the acoustoelectric effect. Looking forward, this material stack may be compatible with quantum well systems that yield optical gain in addition to acoustoelectric phonon gain.  In principle, this could allow complete control over the photonic and phononic dissipation rates, enabling dynamics ranging from the significant optical linewidth narrowing to the acoustic linewidth narrowing regime described above.

\subsubsection{Brillouin-based acousto-optic isolators}

Acoustoelectrically enhanced Brillouin processes may be the key to significant performance improvements in chip-scale non-reciprocal technologies based on traveling-wave optomechanics. Some of the most promising broadband technologies for on-chip isolators rely on inter-modal waveguide acousto-optics, in which a moving acoustic Bragg grating produces unidirectional optical mode conversion due to a nonreciprocal phase-mismatch \cite{huang2011complete,poulton2012design}.  However, it has been challenging to achieve the scattering efficiencies necessary to produce large contrast optical nonreciprocity with low insertion loss over broad bandwidths \cite{sohn2018time,kittlaus2018non,kittlaus2021electrically}.  

With recourse to acoustoelectric phonon gain, the scattering efficiency $\eta_{\rm ef}$ (in the case of lossless optical fields) can be improved as \cite{kittlaus2018non}

\begin{equation}
\begin{aligned}
\eta_{\rm ef}= \tanh^2\bigg(\frac{G_{\rm B} P L \Gamma}{4(\Gamma-G_{\rm AE})}\bigg).
\end{aligned}
\label{eq:sol8}
\end{equation}

As a result, acoustoelectrically enhanced transduction may bring unity efficiencies within reach in the InGaAsP-LN-Si system---even with intrinsic scattering efficiencies of $\sim10^{-3}$---as plotted in Fig. \ref{fig:device}c. Moreover, the acoustoelectric effect yields additional suppression of unwanted backward propagating phonons on account of the attenuation experienced by contradirectionally propagating phonons, further enhancing the overall nonreciprocity \cite{hackett2021amp}. As a result, the AEB interactions in this exemplar system or others may play a key role in enabling the first generation of practical acousto-optic on-chip isolators.

\subsubsection{Optomechanical delay}

Control of the phonon dissipation through acoustoelectric phonon gain may also be used to enhance the performance of Brillouin-based memory and optomechanical delay \cite{merklein2017chip}.  The most apparent improvement is the ability to extend the phonon lifetime ($\tau \propto 1/\Gamma$), and hence the delay time by 

\begin{equation}
\begin{aligned}
\tau^{\rm AEB}=\tau\Big(\frac{\Gamma}{\Gamma-G_{\rm AE}} \Big).
\end{aligned}
\label{eq:delay}
\end{equation}

A less obvious, but equally useful benefit of increasing the phonon lifetime is that one can achieve unity photon-phonon conversion lengths (in the case of optomechanical memory for instance) over much shorter distances. The conversion length scales as  

\begin{equation}
\begin{aligned}
L^{\rm AEB}=L\Big(\frac{\Gamma-G_{\rm AE}}{\Gamma} \Big),
\end{aligned}
\label{eq:delay}
\end{equation}
hence simultaneously permitting substantially longer delays in a much smaller geometry, as shown in Fig. \ref{fig:device}d.  For the system under consideration, tunable chip light storage of more than 0.1 $\upmu$s may be possible in the AEB limit (corresponding to $\sim20$ m of fiber optic path length), with potential for orders of magnitude larger improvements with the acoustoelectrically induced coherent Brillouin limit outlined below.

\subsection{Acoustoelectrically induced coherent Brillouin (ACB) limit}\label{sec:parametric}

We now explore what we term the acoustoelectrically induced coherent Brillouin (ACB) limit, in which the mean-free path of the phonons is on par with that of the photons. In this limit, we can no longer treat the phonon field as local in space, in sharp contrast with the standard assumption for SBS processes \cite{boyd2020nonlinear}. Specifically, we consider an interaction region in which both the optical and acoustic waves can be considered lossless, with the latter enabled by acoustoelectric gain that compensates for any phonon propagation loss. Under these conditions, the spatial equations of motion (i.e., steady state in time) become

\begin{equation}
\begin{aligned}
v_{\rm g,s} \frac{\partial \bar{A}_{\rm s}[z,\omega]}{\partial z}&= -i g_0^* \bar{A}_{\rm p} \bar{B}^\dagger[z,\omega]\\
v_{\rm g,b} \frac{\partial \bar{B}[z,\omega]}{\partial z}&=-ig_0^*\bar{A}_{\rm p}\bar{A}^\dagger_{\rm s}[z,\omega].
\end{aligned}
\label{eq:par1}
\end{equation}

We note that, as a vector quantity, the phonon velocity $v_{\rm g,b}$ plays a singularly important role in the resulting coupled dynamics. Equation \ref{eq:par1} can be decoupled through differentiation, yielding 

\begin{equation}
\begin{aligned}
\frac{\partial^2 \bar{A}_{\rm s}[z]}{\partial z^2}&=\frac{|g_0|^2 |\bar{A}_{\rm p}|^2}{v_{\rm g,s} v_{\rm g,b}} \bar{A}_{\rm s}[z]\\
\frac{\partial^2 \bar{B}[z]}{\partial z^2}&=\frac{|g_0|^2 |\bar{A}_{\rm p}|^2}{v_{\rm g,s} v_{\rm g,b}} \bar{B}[z].
\end{aligned}
\label{eq:par2}
\end{equation}

In the case of a phonon field that co-propagates with the Stokes field (e.g., forward intra- or inter-modal SBS), the solution for $\bar{A}_{\rm s}[z]$ and $\bar{B}[z]$ take the form 

\begin{equation}
\begin{aligned}
\bar{A}_{\rm s}[z]&= A \sinh[\kappa z]+ B\cosh[\kappa z]\\
\bar{B}[z]&= C \sinh[\kappa z]+ D\cosh[\kappa z]
\end{aligned}
\label{eq:par3}
\end{equation}

\noindent
where $\kappa^2= (|g_0|^2 |\bar{A}_{\rm p}|^2/(v_{\rm g,s} v_{\rm g,b}))$, and $A$, $B$, $C$, and $D$ are determined by the initial conditions.  

For instance, an input Stokes wave of the form ($\bar{A}_{\rm s}[0]=\bar{B}[0]=0$) yields the solutions

\begin{equation}
\begin{aligned}
\bar{A}_{\rm s}[z]&= \bar{A}_{\rm s}[0] \cosh (\kappa z)\\
\bar{B}[z]&=-\frac{i g_0^* \bar{A}_{\rm p}}{|g_0||\bar{A}_{\rm p}|}\sqrt{\frac{v_{\rm g,s}}{v_{\rm g,b}}} \bar{A}^{\dagger}_{\rm s}[0] \sinh (\kappa z).
\end{aligned}
\label{eq:par4}
\end{equation}

We see that the dynamics in the ACB limit resemble those of optical parametric (e.g. $\chi^{(2)}$) processes \cite{boyd2020nonlinear}---distinct from those of traditional stimulated Brillouin scattering. In particular, we note there is a qualitatively different gain-like behavior. The nontrivial acoustoelectric gain enhancement can be characterized by the ratio of the standard gain per unit length in the coherent ($\kappa$) to traditional ($G_{\rm B} P$) limits, given by

\begin{equation}
\begin{aligned}
\frac{\kappa}{G_{\rm B} P}=\frac{\Gamma}{4 |g_0|}\sqrt{\frac{\hbar \omega_{\rm p} v_{\rm g,p} v_{\rm g,s}}{P v_{\rm g,b}}},
\end{aligned}
\label{eq:par5}
\end{equation}

\noindent
where $\Gamma$ is the phonon dissipation rate in the absence of acoustoelectric gain.  We note that the ACB regime is particularly advantageous (from a perspective of accessible optical gain) in the case of relatively low optical pump powers and optomechanical coupling rates, as shown in Fig. \ref{fig:parametric}a.

The optomechanical dynamics of this acoustoelectric induced coherent Brillouin limit diverge even more dramatically from those of traditional SBS if we consider a counter-propagating phonon field---such as is required for backward stimulated Brillouin scattering.  Switching the sign of the phonon group velocity yields an imaginary coupling $\kappa=i|\kappa|$, and the solution becomes Rabi-like as 

\begin{equation}
\begin{aligned}
\bar{A}_{\rm s}[z]&= \bar{A}_{\rm s}[0] \cos (|\kappa| z)\\
\bar{B}[z]&= \frac{ g_0^* \bar{A}_{\rm p}}{|g_0||\bar{A}_{\rm p}|}\sqrt{\frac{v_{\rm g,s}}{v_{\rm g,b}}} \bar{A}^{\dagger}_{\rm s}[0] \sin (|\kappa| z).
\end{aligned}
\label{eq:par6}
\end{equation}

Thus, in this case, acoustoelectric gain can lead to strong-coupling-like optomechanical dynamics---a dramatic departure from the parametric gain behavior of the co-propagating case (see Fig. \ref{fig:parametric}b).

\section{Discussion and outlook}

In this paper, we have proposed the concept for and explored the dynamics of acoustoelectrically enhanced Brillouin (AEB) interactions for the first time.  Through a Hamiltonian-based formalism, we have determined that the acoustoelectric modification of the Brillouin susceptibility drastically enhances the performance of useful Brillouin photonic devices and opens the door to new regimes of traveling-wave optomechanical dynamics. Moreover, we have proposed an experimentally realizable chip-based system that provides the necessary optomechanical and acoustoelectric degrees of freedom to demonstrate the full range of this new device physics. This work lays the foundation for a new class of powerful optomechanics-based classical and quantum signal processing applications, synergistically combining the unique properties of photons, phonons, and electrons.

Beyond the material system and device geometry proposed in this work, acoustoelectrically enhanced Brillouin physics may be accessible in a number of promising integrated phononic/photonic platforms.  Such physics requires high carrier mobility, appreciable acoustoelectric response, and sufficient traveling-wave optomechanical guidance and coupling. Example systems with these properties may include piezoelectric semiconductors such as GaN \cite{fu2019phononic}, GaAs \cite{miller1987efficiency,balram2014moving,balram2016coherent}, and GaP \cite{pustelny2009transverse,mitchell2014cavity}, or semiconductor-piezoelectric hybrids (like the example presented in this work), such as $\rm GaAs-LiNbO_3$ \cite{rotter1998giant,siddiqui2019large} or co-integrated AlN in SOI-based silicon photonics \cite{zhao2021scaling}. In addition, systems that utilize forms of non-piezoelectric-based acoustoelectric coupling, for instance deformation potential coupling \cite{weinreich1959acoustoelectric,hakim2019non}, may enable AEB physics in an even broader class of semiconductors.   We anticipate the emergence of a range of other material systems and device geometries that will be able to leverage the new concepts proposed in this work.  

By modifying the nonlinear Brillouin susceptibility through acoustoelectric gain and nonreciprocity, we are able to drastically transform the performance of important Brillouin signal processing applications---long valued for their ability to amplify, modulate, or filter light over narrow bandwidths \cite{eggleton2019brillouin}.  At the core of these modified interactions is the ability to enhance the Brillouin amplification process by orders of magnitude, enabling exceptionally narrow bandwidths and state-of-the-art gain in systems that would otherwise yield little or no net amplification.  Moreover, our noise analysis suggests that when thermomechanical noise is dominant (see Supplementary Section~\ref{sec:noise}), this approach may enable high levels of near quantum-limited amplification at cryogenic temperatures.  Building on this work, inducing optical gain via manipulation of electrical carriers may enable a new generation of flexible and narrow linewidth Brillouin laser systems with reconfigurable dynamics and improved noise performance.  

The impact of acoustoelectrics on optomechanics extends beyond traditional Brillouin photonic devices. Whereas we have focused here on traveling-wave optomechanical interactions, these same concepts are applicable to cavity optomechanical systems \cite{aspelmeyer2014cavity}. The ability to dynamically modify the phonon decay rate and cooperativity in cavity-optomechanical systems should lead to enhanced coherent photon-phonon interactions in devices such as optomechanical amplifiers \cite{safavi2011electromagnetically,ruesink2016nonreciprocity,shen2018reconfigurable,kharel2019high}, oscillators \cite{vahala2008back,tallur2011monolithic,beyazoglu2014multi}, filters \cite{deotare2012all}, and quantum memories \cite{wallucks2020quantum}.   Finally, applying these concepts to acousto-optic devices may allow externally applied electric fields to modify local phonon dissipation rates and populations, enabling faster amplitude modulation in systems that would otherwise be limited by intrinsic phonon lifetimes.

\section*{Funding Information}

This material is based upon work supported by the Laboratory Directed Research and Development program at Sandia National Laboratories. Sandia National Laboratories is a multi-program laboratory managed and operated by National Technology and Engineering Solutions of Sandia, LLC., a wholly owned subsidiary of Honeywell International, Inc., for the U.S. Department of Energy's National Nuclear Security Administration under contract DE-NA-0003525. This paper describes objective technical results and analysis. Any subjective views or opinions that might be expressed in the paper do not necessarily represent the views of the U.S. Department of Energy or the United States Government. R. Behunin acknowledges support through the Center for Materials Interfaces in Research and Applications (¡MIRA!) Exploratory Research Program. 

\section*{Author contributions}
*N.T.O., M.J.S., and R.O.B. contributed equally to this work.

M.E. and N.T.O. jointly developed concept and proposed physical system with input from M.J.S., L.H., and P.T.R..  M.J.S. examined system properties and performed numerical simulations and analysis of the optical, elastic, optomechanical, electromechanical, and acoustoelectric properties of the proposed system with assistance from N.T.O., M.E., and P.T.R.. N.T.O. and R.O.B. developed analysis of acoustoelectric-modified Brillouin optomechanical dynamics and noise.  R.O.B. developed comprehensive Hamiltonian formalism as the basis for the acoustoelectric-modified optomechanical interactions.  M.J.S. generated 3D illustrative diagrams. L.H. analyzed prospects of cw operation.  All authors contributed to the writing and composition of the manuscript.



\renewcommand{\appendixname}{Supplementary Information}
\appendix
\onecolumngrid
\newpage

\section{Physical System Design}\label{sec:SIdesign}

\begin{figure*}[b]
\centering
\includegraphics[width=1\linewidth]{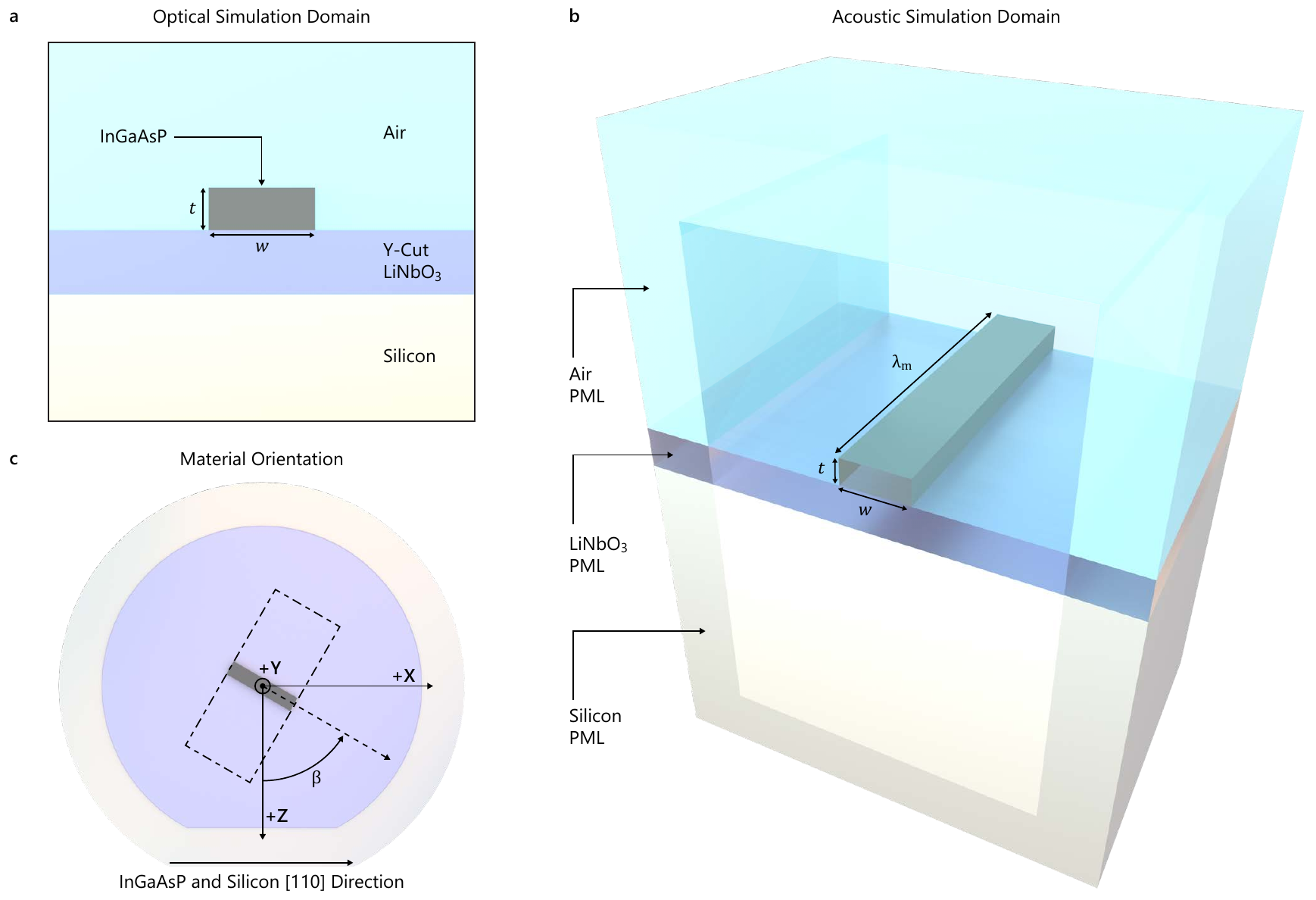}
\caption{Simulation domains (not to scale) for the epitaxial InGaAsP bonded to a Y-cut lithium niobate thin film on silicon substrate (InGaAsP-LN-Si). The materials and key design parameters are shown in the (a) optical and (b) acoustic simulation domains as well as the (c) defined material orientations.}
\label{fig:InGaAsP_SimOverview}
\end{figure*}

Phase-matched Brillouin scattering is a three-wave mixing process involving two optical modes (pump and Stokes) and one elastic mode. This supplementary section covers the design procedure for computing the optical and elastic mode triplet for acoustoelectric (AE) Brillouin interactions in a given physical system. The primary physical system focused on in this supplement is an epitaxially grown In\textsubscript{0.712}Ga\textsubscript{0.288}As\textsubscript{0.625}P\textsubscript{0.375} (referred to as InGaAsP in the manuscript for convenience) bonded to a Y-cut lithium niobate thin film on a silicon substrate (InGaAsP-LN-Si). The simulated cross-sectional 2D optical domain is shown in Fig.~\ref{fig:InGaAsP_SimOverview}a where the optical guiding layer is the patterned InGaAsP waveguide, the lower cladding is the lithium niobate, and the top and side cladding is air. Since the FEA software has a solver for time-harmonic optical field distributions, a cross-sectional 2D Mode analysis is performed to compute the propagation constant and propagating mode shapes for a given optical frequency. Due to the high index contrast of the InGaAsP waveguide, the optical fields are confined such that no perfectly matched layers (PMLs) are needed at the boundaries of the simulation domain. Additionally, the silicon is assumed to be far enough away from the optical fields as to not cause optical loss, making absorption losses in that material unnecessary to include (an assumption that is easily verified for the specific modes we find as solutions). The simulated 3D acoustic domain is shown in Fig.~\ref{fig:InGaAsP_SimOverview}b and includes the addition of PMLs to absorb any elastic energy at the edges of the simulation domain. For certain polarizations and phase-matched phonon wavelengths of the elastic modes, there is low contrast between the phonon velocities in InGaAsP and lithium niobate, which results in minimal guiding of the elastic mode in the waveguide. The amount of elastic energy absorbed by the PMLs help us characterize whether the simulation boundary has been made large enough and whether a given elastic mode is guided. For backward intra-modal and forward inter-modal SBS processes, the elastic modes in this physical system have periodic variations in their strain along the direction of propagation, which invalidates the cross-sectional 2D plain strain approximation. Therefore, analysis of these elastic modes require a 3D simulation domain with floquet periodic boundary conditions in the direction of the defined phonon wavelength. The elastic mode is guided within the InGaAsP/lithium niobate heterostructure and the silicon substrate assists in heat dissipation for continuous wave (CW) operation (see Supplementary Section~\ref{sec:CW}).

The three key geometric parameters to consider when designing an AE Brillouin device are shown in Fig.~\ref{fig:InGaAsP_SimOverview}b. The width ($w$) and thickness ($t$) of the InGaAsP waveguide not only determines the confinement and effective index of the optical modes, but it also determines the lateral and vertical confinement of the elastic mode in the InGaAsP/lithium niobate heterostructure. The phonon wavelength ($\lambda_\text{m}$) sets the operating phonon frequency and elastic mode shape, as determined by the width, thickness, and mode-dependent dispersion. Since this physical system includes anisotropic materials, Fig.~\ref{fig:InGaAsP_SimOverview}c describes all of the material orientations. The lithium niobate thin film has its +Y-axis normal to the device plane and the +X-axis is parallel to the [110] direction in the cubic InGaAsP and silicon crystals. The lithium niobate thin film has a thickness of 5~$\mu$m and the +Y-axis was chosen to be normal to the device plane because these LN-Si wafers are commercially available and have high piezoelectric coupling to both the Rayleigh and Shear Horizontal surface acoustic waves. In the plane of the device, the propagation angle ($\beta$) not only determines the optical axis of lithium niobate, but also the piezoelectric coupling strength of the elastic mode. The optomechanical coupling strength is a function of the overlap between the optical mode and elastic mode, while the acoustoelectric coupling strength is a function of the overlap between the elastic mode, piezoelectric potential, and free carriers in the InGaAsP. Therefore, careful consideration must be made when designing the optical and elastic modes to ensure good overlap between all domains, resulting in strong optomechanical and acoustoelectric coupling. 

The general outline for the design procedure is detailed below. First, the type of stimulated Brillouin scattering (SBS) is specified and the appropriate acoustic and optical phase matching conditions are applied. The two types of SBS considered here are intra-modal backward scattering (i.e. two contradirectionally propagating optical modes with the same spatial mode profile) and inter-modal forward scattering (i.e. two co-directionally propagating optical modes with different spatial mode profiles). For each case of SBS, the following design parameters must be considered to cover the full design space: optical mode pair, propagation angle, InGaAsP waveguide width, and InGaAsP waveguide thickness. The pair of optical modes used for the pump and Stokes photons will determine the phonon wavelength needed for phase matching, which together with the mode-dependent acoustic dispersion determines the elastic mode frequency. The propagation angle chosen is largely influenced by the optical and acoustic properties of the lithium niobate thin film. Longitudinal strains have high piezoelectric coupling parallel to the Z-axis, whereas shearing strains have a high piezoelectric coupling parallel to the X-axis. This corresponds to either a 0\textdegree~or 90\textdegree~propagation angle, respectively. These two propagation angles also correspond to the extraordinary and ordinary optical axes in lithium niobate. Since the optimal waveguide width strongly depends on the chosen elastic mode, the range of simulated waveguide widths must be carefully considered. Finally, the simulated InGaAsP waveguide thickness ranges from 100~nm to 300~nm to remain a single mode optical waveguide in the vertical direction and to reduce the electric field required to achieve a given phonon gain (see Eq.~\ref{eq:simple_AE}). 

After the range of design parameters is set, the simulations and calculations are carried out in the following order: optical domain simulation, acoustic domain simulation, optomechanical coupling calculation, and acoustoelectric coupling calculation. Using COMSOL Multiphysics, a commercial finite element method (FEM) software, the 2D optical domain simulation is used to identify the effective indices of the pump and Stokes photons, which are used to compute the phonon wavelength required for phase matching. Additionally, the normalized electric fields for the pump and Stokes photons are extracted from the simulation to use in the optomechanical coupling rate calculations. Then, the 3D acoustic domain is simulated for the free and grounded eigenfrequencies of the elastic modes. The normalized displacement and strain profiles for the elastic modes are extracted from the simulation to use in the optomechanical coupling rate calculations. Using all of the extracted values from simulation, the following optomechanical coupling rates are calculated: photoelastic coupling in the InGaAsP waveguide, photoelastic coupling in the surrounding lithium niobate, radiation pressure coupling in the waveguide, and total optomechanical coupling rate. Finally, the sets of elastic modes from the free and grounded eigenfrequency simulations are correlated to determine the piezoelectric coupling coefficient of each elastic mode. Using the full treatment of the normal mode theory for an AE amplifier, the acoustoelectric gain is calculated as a function of applied drift field for each elastic mode.

\subsection{InGaAsP Material}

Before any of the simulations are performed, we calculate the interpolated material parameters for the quaternary InGaAsP compound. For the InGaAsP-LN-Si physical system, the operating optical wavelength is 1.55~$\mu$m. Some of the figures and interpolation schemes consider this operating optical wavelength in terms of its photon energy $E = \hbar \omega$. The corresponding photon energy at a 1.55~$\mu$m wavelength is approximately 0.8~eV. The InGaAsP considered for this physical system will be grown on a (100) InP wafer---a typical substrate orientation for ternary and quaternary epitaxial growth---and bonded to a Y-cut lithium niobate on silicon wafer. Based on previous work in InGaAsP optical devices grown on a (100) InP wafer \cite{Chen_InGaAsP_ori_1993,Tsai_InGaAsP_ori_1997,Stulz_InGaAsP_ori_1983,Watanabe_InGaAsP_ori_2016}, the [110] direction of the InGaAsP will be oriented parallel to the [110] direction of the silicon substrate \cite{Hopcroft_Si_2010} and the +X-axis of the lithium niobate thin film. The following subsections will outline the proposed fabrication procedure, quaternary composition, and interpolation schemes used to determine the InGaAsP material parameters for use in this physical system. 

\subsubsection{Acoustoelectric Fabrication Modified for InGaAsP Waveguides}

\begin{figure}[t]
	\centering
	\includegraphics[width=1\linewidth]{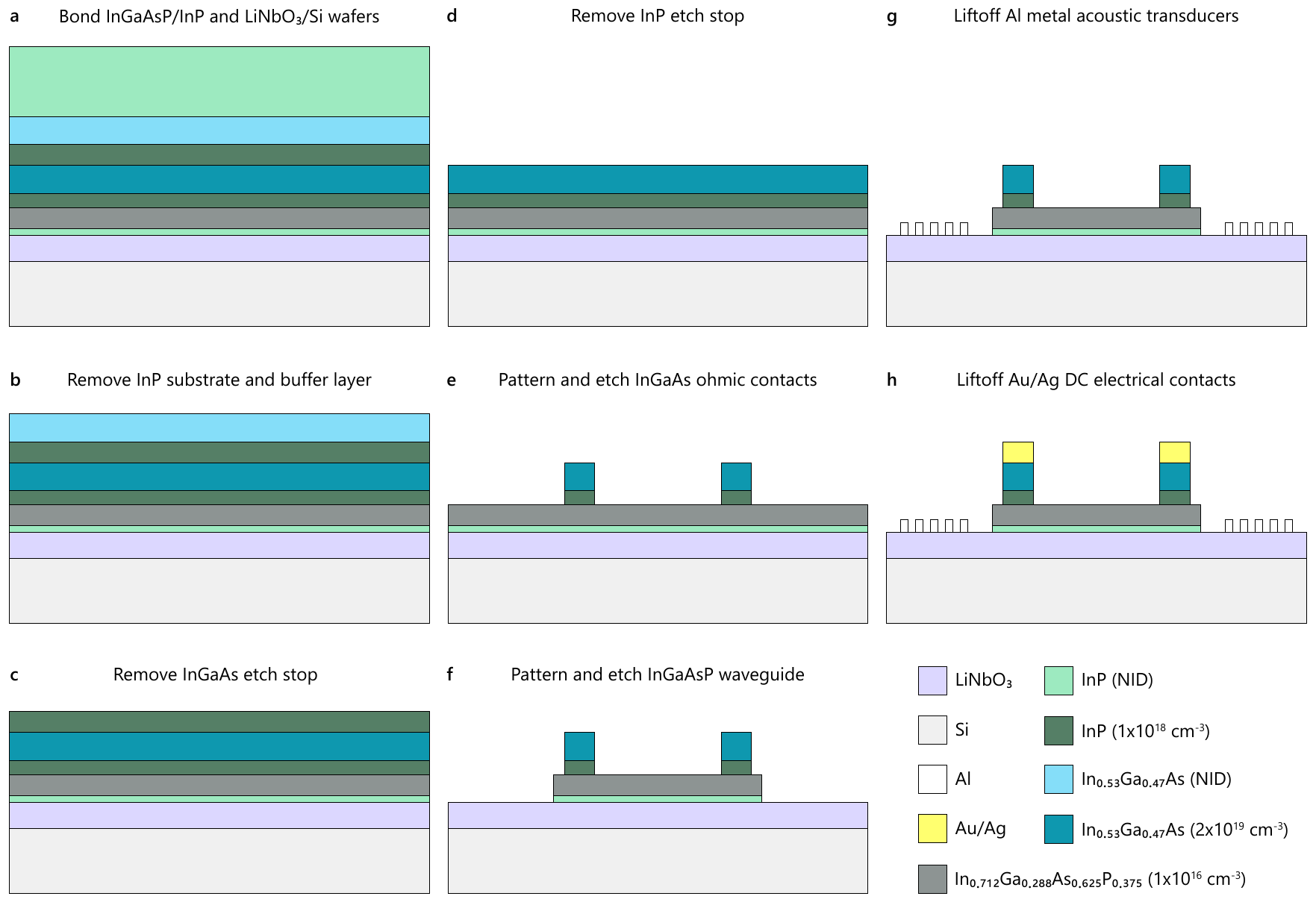}
	\caption{Fabrication process flow from our previous work in acoustoelectric devices modified to include the InGaAsP waveguide layer for acoustoelectrically enhanced Brillouin devices (not to scale).}
	\label{InGaAsP_fab}
\end{figure}

Based on our previous work in designing and fabricating high performance acoustoelectric devices \cite{hackett2019amp,hackett2021amp,siddiqui2019large,siddiqui_comparison_2020,Storey_Switch_2021}, we can leverage our experience in epitaxial growth and nanofabrication to make acoustoelectrically enhanced Brillouin devices. Based on the fabrication process flow from our previous work, the modified process is shown in Fig.~\ref{InGaAsP_fab}. The only modification in the epitaxial growth stack of the InGaAsP/InP wafer (Fig.~\ref{InGaAsP_fab}a) from our previous work is during the growth of the waveguide layer with a free carrier concentration of 1~x~10\textsuperscript{16}~cm\textsuperscript{-3}. Here, the addition of phosphorus alters the growth of a lattice-matched ternary In\textsubscript{1-x}Ga\textsubscript{x}As to a lattice-matched quaternary In\textsubscript{1-x}Ga\textsubscript{x}As\textsubscript{y}P\textsubscript{1-y}. All the other InP and InGaAs layers in the growth process (etch stop layers, ohmic contact layers, capping layer) remain unchanged from the fabrication process of previous work. The fabrication steps in Fig.~\ref{InGaAsP_fab}b-e make use of two different wet etches, one selective to InP and one selective to InGaAs \cite{Fiedler_selectiveetch_1982}, which will be the same wet etch chemistries that were used in the fabrication process of our previous acoustoelectric work. If the InGaAsP waveguide layer (Fig.~\ref{InGaAsP_fab}f) is patterned through wet etching, there are several chemistries available that etch both InGaAs and InGaAsP alloys \cite{Conway_selectiveetch_1982,Huang_selectiveetch_2005,Pasquariello_selectiveetch_2006}. Due to the isotropic nature of wet etching, the undercut makes resolving small features difficult. For the waveguide dimensions needed for Brillouin devices, anisotropic dry etching is typically required. There are several chemistries available for etching InGaAsP using reactive ion etching (RIE) including BCl\textsubscript{3} \cite{Maeda_dryetch_1999}, Cl\textsubscript{2}/H\textsubscript{2} \cite{Parker_dryetch_2011,Rommel_dryetch_2002}, and CH\textsubscript{4}/H\textsubscript{2} \cite{Muller_dryetch_1990}. The remaining fabrication steps (Fig.~\ref{InGaAsP_fab}g-h) to pattern the metal transducers and DC electrical contacts remain unchanged from our previous work. Overall, only slight modifications are needed in our established acoustoelectric device fabrication process to incorporate phosphorus into the InGaAsP waveguide layer and pattern the Brillouin devices. 

\subsubsection{Quaternary Composition for InP Lattice Matching}

The quaternary compound of indium, gallium, arsenic, and phosphorus can be expressed in terms of two composition parameters, $x$ and $y$ (In\textsubscript{1-x}Ga\textsubscript{x}As\textsubscript{y}P\textsubscript{1-y}). For lattice matching with InP, the following expression relates the two composition parameters \cite{adachi_InGaAsP_1982}

\begin{equation}
\begin{aligned}
x \approx \frac{0.1894 \, y}{0.4184 - 0.013 \, y} \, \, \, \, \left( 0 \leq y \leq 1 \right).
\end{aligned}
\label{eq:xy_relation}
\end{equation}

\noindent
Using this relationship between the two composition parameters, the interpolation schemes for all of the material parameter calculations can be expressed in terms of just one composition parameter ($y$). In order to operate at an optical wavelength of 1.55~$\mu$m, the direct band gap energy needs to be large enough to be optically transparent and the optical wavelength needs to be far enough away from the absorption edge in order to minimize loss from absorption. The direct gap energy (in eV) at 300~K is given as

\begin{equation}
\begin{aligned}
E_\text{0} = 1.35 - 0.72 \, y + 0.12 \, y^{2}.
\end{aligned}
\label{eq:Eg_300K}
\end{equation}

\noindent
For the case of the ternary compound InGaAs ($y = 1$), the band gap energy is 0.75~eV, which is too small to be optically transparent at the desired optical wavelength. The band gap energy can be increased through the introduction of phosphorus into the compound. For the physical system in this manuscript, the chosen composition parameter is $y = 0.625$, giving the full description of the quaternary composition as In\textsubscript{0.712}Ga\textsubscript{0.288}As\textsubscript{0.625}P\textsubscript{0.375}. At this composition, the band gap energy is 0.95~eV, which is now optically transparent at the operating optical wavelength and 0.15~eV away from the absorption edge. Reference~\cite{Fiedler_InPparam_1987} examines the interpolated InGaAsP absorption coefficient as a function of free carrier concentration. Given their operating optical wavelength, the absorption coefficient is proportional to the free carrier concentration when operating 0.15~eV away from the absorption edge. At our free carrier concentration of $N$~=~1~x~10\textsuperscript{16}~cm\textsuperscript{-3} (for optimal acoustoelectric interaction), they see an absorption coefficient of approximately 0.02~cm\textsuperscript{-1}. From this, we can estimate a loss of 0.1~dB/cm in the InGaAsP from optical absorption. 

\subsubsection{Refractive Index}

The treatment for calculating the index of refraction as a function of photon energy and composition parameter can be found in the appendix of Ref.~\cite{weber_index_1994}. Assuming an operating temperature of 300~K, the index of refraction can be expressed as 

\begin{equation}
\begin{aligned}
n^{2}_\text{r} = A \left( y \right) \left[ f \left( z \right) + \frac{1}{2} \left( \frac{E_\text{0}}{E_\text{0} + \Delta_{0}} \right)^{3/2} f \left( z_{0} \right) \right] + B \left( y \right),
\end{aligned}
\label{eq:index_refraction}
\end{equation}

\noindent
where the split-off valence band gap (in eV) at 300~K is 

\begin{equation}
\begin{aligned}
\Delta_{0} = 0.118 + 0.225 \, y.
\end{aligned}
\label{eq:D0_300K}
\end{equation}

\noindent
The normalized energy terms ($z$ and $z_{0}$) are a function of the operating photon energy, $E$, and the related terms can be expressed as 

\begin{equation}
\begin{aligned}
f \left( z^{\prime} \right) = \left( 2 - \sqrt{1 + z^{\prime}} - \sqrt{1 - z^{\prime}} \right) / z^{\prime \, 2},
\end{aligned}
\label{eq:function_z}
\end{equation}

\begin{equation}
\begin{aligned}
z = E / E_\text{0},
\end{aligned}
\label{eq:z_term}
\end{equation}

\begin{equation}
\begin{aligned}
z_{0} = E / \left( E_\text{0} + \Delta_{0} \right).
\end{aligned}
\label{eq:z0_term}
\end{equation}

\noindent
Finally, the fitting parameters $A$ and $B$ are a function of the composition parameter

\begin{equation}
\begin{aligned}
A \left( y \right) = 8.616 - 3.886 \, y,
\end{aligned}
\label{eq:A_term}
\end{equation}

\begin{equation}
\begin{aligned}
B \left( y \right) = 6.621 + 3.461 \, y.
\end{aligned}
\label{eq:B_term}
\end{equation}

\noindent
For the composition parameter and photon energy used in this physical system, the calculated refractive index is $n_\text{r} = 3.4013$.

\subsubsection{Dielectric Constant}

The interpolation for both the static and high frequency dielectric constants as a function of composition parameter are given in Ref.~\cite{adachi_InGaAsP_1982}. For the frequency range of elastic modes simulated, the static dielectric constant will be used, which is given as

\begin{equation}
\begin{aligned}
\epsilon_\text{s} = 12.4 + 1.5 \, y.
\end{aligned}
\label{eq:es_term}
\end{equation}

\noindent
For the composition parameter used in this physical system, the calculated dielectric constant is $\epsilon_\text{s} = 13.3375$.

\subsubsection{Density}

The density varies almost linearly between InP and InGaAs depending on the composition parameter, which is given in Ref.~\cite{adachi_InGaAsP_1982} (in units of kg/m\textsuperscript{3})

\begin{equation}
\begin{aligned}
\rho = 4787 + 712 \, y.
\end{aligned}
\label{eq:density_term}
\end{equation}

\noindent
For the composition parameter used in this physical system, the calculated density is $\rho = 5232$ kg/m\textsuperscript{3}.

\subsubsection{Elasticity and Compliance Matrix}

InGaAsP has a zincblende crystal structure, which means the elasticity matrix has three independent coefficients ($C_{11}$, $C_{12}$, $C_{44}$) and the matrix takes the form 

\begin{equation}
\begin{aligned}
\left[ C \right] = 
    \begin{bmatrix}
    C_{11} & C_{12} & C_{12} & 0 & 0 & 0\\
    C_{12} & C_{11} & C_{12} & 0 & 0 & 0\\
    C_{12} & C_{12} & C_{11} & 0 & 0 & 0\\
    0 & 0 & 0 & C_{44} & 0 & 0\\
    0 & 0 & 0 & 0 & C_{44} & 0\\
    0 & 0 & 0 & 0 & 0 & C_{44}
    \end{bmatrix}.
\end{aligned}
\label{eq:C_matrix}
\end{equation}

\noindent
The interpolation of the elasticity coefficients as a function of the composition parameter is given in Chapter~3 of Ref.~\cite{adachi1992IIIVsemi}. These coefficients (in GPa) are given as

\begin{equation}
\begin{aligned}
C_{11} = 101.1 - 1.2 \, y,
\end{aligned}
\label{eq:C11_term}
\end{equation}

\begin{equation}
\begin{aligned}
C_{12} = 56.1 - 6.8 \, y,
\end{aligned}
\label{eq:C12_term}
\end{equation}

\begin{equation}
\begin{aligned}
C_{44} = 45.6 + 3.3 \, y.
\end{aligned}
\label{eq:C44_term}
\end{equation}

\noindent
For the composition parameter used in this physical system, the calculated elasticity coefficients are $C_{11}$~=~100.35~GPa, $C_{12}$~=~51.85~GPa, and $C_{44}$~=~47.6625~GPa. Similarly, the three independent compliance matrix coefficients can be calculated using the interpolated elasticity matrix coefficients from Eq.~\ref{eq:C11_term}-\ref{eq:C44_term} with the following relations 

\begin{equation}
\begin{aligned}
S_{11} = \frac{C_{11} + C_{12}}{\left( C_{11} - C_{12} \right) \left( C_{11} + 2 \, C_{12} \right)},
\end{aligned}
\label{eq:S11_term}
\end{equation}

\begin{equation}
\begin{aligned}
S_{12} = \frac{-C_{12}}{\left( C_{11} - C_{12} \right) \left( C_{11} + 2 \, C_{12} \right)},
\end{aligned}
\label{eq:S12_term}
\end{equation}

\begin{equation}
\begin{aligned}
S_{44} = \frac{1}{C_{44}}.
\end{aligned}
\label{eq:S44_term}
\end{equation}

\noindent
Using the interpolated elasticity matrix coefficients above, the calculated compliance matrix coefficients are $S_{11}$~=~1.538~x~10\textsuperscript{-11}~Pa\textsuperscript{-1}, $S_{12}$~=~-5.239~x~10\textsuperscript{-12}~Pa\textsuperscript{-1}, and $S_{44}$~=~2.098~x~10\textsuperscript{-11}~Pa\textsuperscript{-1}.

\subsubsection{Photoelastic Coupling Constants}

Since there is no complete set of interpolated photoelastic constants for InGaAsP in the literature, the following subsection will detail the methodology used for estimating the full photoelastic coupling matrix. Using the base set of binaries \{InAs, InP, GaAs, GaP\}, Chapter~9 in Ref.~\cite{adachi1992IIIVsemi} uses interpolation methods to calculate the linear piezobirefringence (PB) coefficient for the [100] and [111] stress directions. Using these PB coefficients, the method for estimating the photoelastic coupling constants as a function of the composition parameter is outlined. Since InGaAsP has a zincblende crystal structure, there are three independent photoelastic coupling constants ($p_{11}$, $p_{12}$, $p_{44}$) and the matrix takes the form 

\begin{equation}
\begin{aligned}
\left[ p \right] = 
    \begin{bmatrix}
    p_{11} & p_{12} & p_{12} & 0 & 0 & 0\\
    p_{12} & p_{11} & p_{12} & 0 & 0 & 0\\
    p_{12} & p_{12} & p_{11} & 0 & 0 & 0\\
    0 & 0 & 0 & p_{44} & 0 & 0\\
    0 & 0 & 0 & 0 & p_{44} & 0\\
    0 & 0 & 0 & 0 & 0 & p_{44}
    \end{bmatrix}.
\end{aligned}
\label{eq:p_matrix}
\end{equation}

\noindent
Chapter~9 in Ref.~\cite{adachi1992IIIVsemi} also details the procedure to compute the relationship between photoelastic coupling constants and the interpolated linear PB coefficients ($\alpha_\text{pe}$). The interpolated results are based on fitting the experimental data of the four binary compounds \{InAs, GaAs, InP, GaP\}, which were taken using uniaxial stress measurements to extract the linear PB coefficients 

\begin{equation}
\begin{aligned}
\alpha_\text{pe} = \frac{\Delta \epsilon_{ij}}{X} = - \sum_{mn} \epsilon_{ii} \epsilon_{jj} p_{ijkl} S_{klmn}.
\end{aligned}
\label{eq:PB_general}
\end{equation}

\noindent
Here, $\epsilon_{ii,jj}$ is the component of the dielectric constant tensor in the absence of the applied stress, $p_{ijkl}$ is the photoelastic tensor component, $S_{klmn}$ is the component of elastic compliance tensor, $X$ is the directional applied stress, and $\Delta \epsilon_{ij}$ is the difference between the real part of the dielectric constant parallel and perpendicular to the applied stress direction. Since all of the higher order band gaps for the four binary compounds are far away from the first direct gap, the linear PB coefficient can be modeled with dispersive effects from only the first direct band gap and include the influence of the higher order gaps as a constant background effect. This model for the linear PB coefficient is given as 

\begin{figure}[t]
	\centering
	\includegraphics[width=0.85\linewidth]{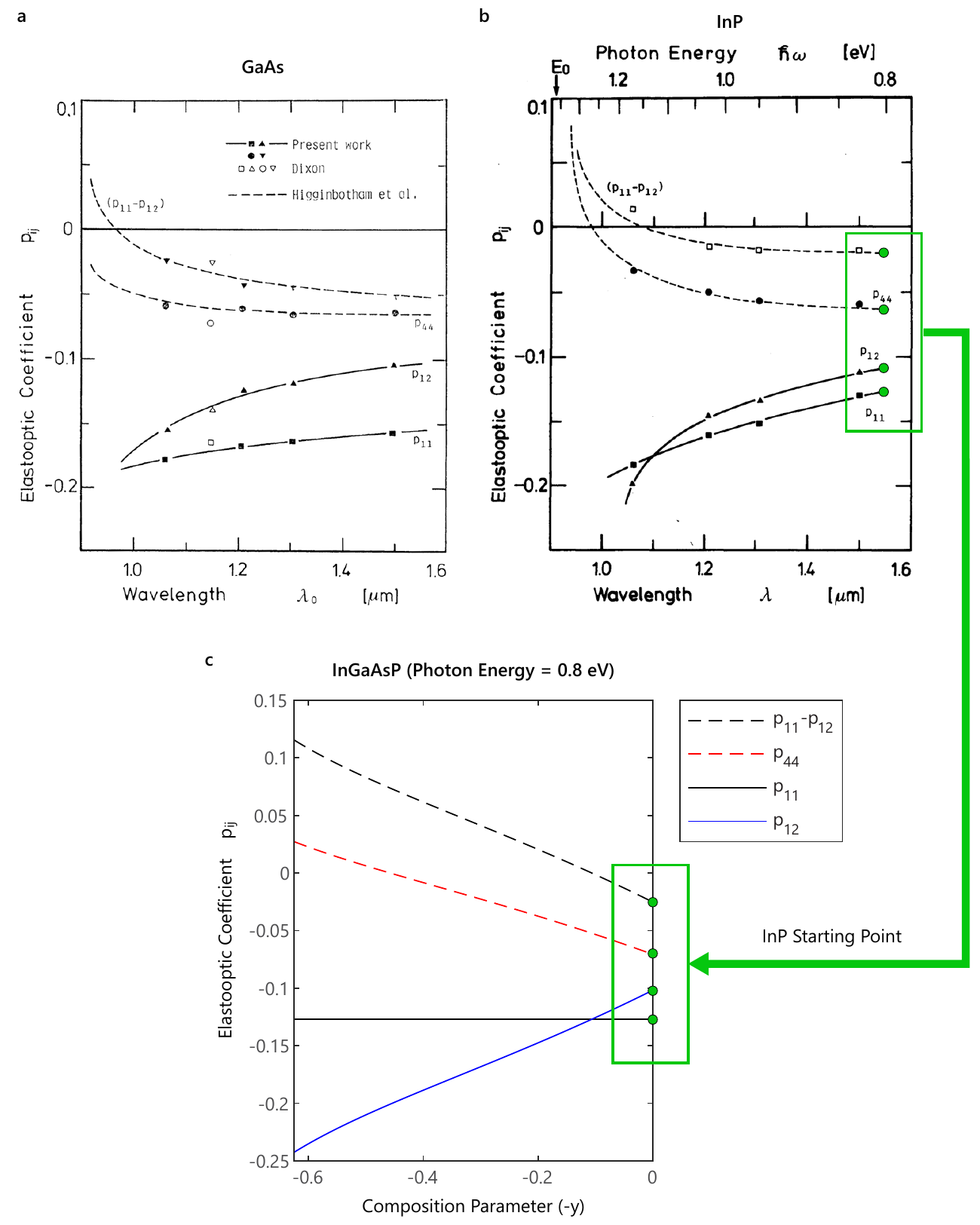}
	\caption{Dispersion of the photoelastic constants as a function of the optical wavelength for (a) GaAs \cite{Suzuki_GaAs_1984} and (b) InP \cite{Suzuki_InP_1983}. (c) A similar dispersion trend is shown for the photoelastic constants of InGaAsP as a function of composition parameter. The green points in (b) and (c) correspond to the same photoelastic constant values on the InP and InGaAsP dispersion curves.}
	\label{pe_interp_compare}
\end{figure}

\begin{equation}
\begin{aligned}
\alpha_\text{pe} = C^{*} \left\{ -g \left( z \right) + \frac{4 E_{0}}{\Delta_{0}} \left[ f \left( z \right) - \left( \frac{E_\text{0}}{E_\text{0} + \Delta_{0}} \right)^{3/2} f \left( z_{0} \right) \right] \right\} + D^{*},
\end{aligned}
\label{eq:PB_fit}
\end{equation}

\noindent
where $C^{*}$ and $D^{*}$ are parameters used to fit to the experimental data and $g \left( z \right)$ is the normalized function 

\begin{equation}
\begin{aligned}
g \left( z^{\prime} \right) = \left( 2 - \frac{1}{\sqrt{1 + z^{\prime}}} - \frac{1}{\sqrt{1 - z^{\prime}}} \right) / z^{\prime \, 2}.
\end{aligned}
\label{eq:g_term}
\end{equation}

\noindent
The fit parameters $C^{*}$ and $D^{*}$ for InGaAsP can be interpolated as a function of the fit parameters for each binary compound shown in Table~9.1 (Chapter~9) of Ref.~\cite{adachi1992IIIVsemi}. The interpolation scheme for these fit parameters is described in Ref.~\cite{adachi_journal_1983} and given as 

\begin{equation}
\begin{aligned}
\phi_\text{InGaAsP} \left( x,y \right) = \left( 1 - x \right) y \, \psi_\text{InAs} + \left( 1 - x \right) \left( 1 - y \right) \, \psi_\text{InP} + x y \, \psi_\text{GaAs} + x \left( 1 - y \right) \, \psi_\text{GaP},
\end{aligned}
\label{eq:interp_quad}
\end{equation}

\noindent
where $\phi$ is the quaternary parameter and each $\psi$ is the corresponding binary parameter. The interpolated linear PB coefficients correspond to a stress applied parallel to the [100] direction in InGaAsP ($\alpha_\text{pe}^\text{[100]}$) and the stress applied parallel to the [111] direction ($\alpha_\text{pe}^\text{[111]}$). Using Eq.~\ref{eq:PB_general}, each linear PB coefficient can be expressed in terms of the photoelastic constants, depending on the orientation of the applied stress. For a stress applied in the [100] direction, the linear PB coefficient can be expressed as 

\begin{equation}
\begin{aligned}
\alpha_\text{pe}^\text{[100]} = -n_\text{r}^{4} \left( p_{11} - p_{12} \right) \left( S_{11} - S_{12} \right),
\end{aligned}
\label{eq:PB_100}
\end{equation}

\noindent
and for a stress applied in the [111] direction, the linear PB coefficient can be expressed as 

\begin{equation}
\begin{aligned}
\alpha_\text{pe}^\text{[111]} = -n_\text{r}^{4} p_{44} S_{44}.
\end{aligned}
\label{eq:PB_111}
\end{equation}

\noindent
Using the previously calculated index of refraction and compliance matrix coefficients, the relationships between photoelastic constants can be expressed as 

\begin{equation}
\begin{aligned}
p_{11} - p_{12} = \frac{-n_\text{r}^{4} \, \alpha_\text{pe}^\text{[100]}}{ \left( S_{11} - S_{12} \right)},
\end{aligned}
\label{eq:p11mp12_term}
\end{equation}

\begin{equation}
\begin{aligned}
p_{44} = \frac{-n_\text{r}^{4} \, \alpha_\text{pe}^\text{[111]}}{S_{44}}.
\end{aligned}
\label{eq:p44_term}
\end{equation}

\noindent
For the composition parameter and photon energy used in this physical system, the interpolated relationships between the photoelastic constants are $p_{11} - p_{12} = 0.115$ and $p_{44} = 0.027$.

It is not possible with PB measurements to determine the magnitude and sign of $p_{11}$ and $p_{12}$ \cite{Suzuki_GaAs_1984,Suzuki_InP_1983}. In order to estimate the $p_{11}$ and $p_{12}$ photoelastic constants for simulating Brillouin optomechanics in InGaAsP, the measured dispersion relations of the GaAs and InP binary compounds will be used to estimate the dispersive trends of the InGaAsP interpolation scheme. As the photon energy approaches the direct band gap energy, the photoelastic constants experience large dispersion \cite{Suzuki_GaAs_1984,Suzuki_InP_1983,Tada_ZnSe_1977}. While most dispersion measurements of the binary compounds are plotted as a function of the operating optical wavelength, the interpolated InGaAsP dispersion will be plotted as a function of the composition parameter. The operating optical wavelength will be held constant at 1.55~$\mu$m and the composition parameter will be swept from $y$~=~0 (InP) to $y$~=~0.625 (InGaAsP). Since the photoelastic constants of InP at 1.55~$\mu$m are known from Ref.~\cite{Suzuki_InP_1983}, they can be used as a starting point for the InGaAsP dispersion curve. The first step is to estimate the sign of the $p_{11}$ and $p_{12}$ constants in InGaAsP. The binary compounds of InGaAsP, all with different band gap energies, have similar signs and magnitudes of their photoelastic constants at photon energies below the band gap energy. Reference~\cite{Maciejko_pe_InGaAsP_1989} has the photoelastic constants of InP, which are $p_{11} = -0.150$, $p_{12} = -0.115$, $p_{44} = -0.056$. Reference~\cite{Balram_pe_GaAs_2014} has the photoelastic constants of GaAs, which are $p_{11} = -0.165$, $p_{12} = -0.140$, $p_{44} = -0.072$. Finally, Ref.~\cite{Dixon_GaP_pe_1967} has the photoelastic constants of GaP, which are $p_{11} = -0.151$, $p_{12} = -0.082$, $p_{44} = -0.074$. From the signs of the binary compounds' photoelastic constants, the signs of $p_{11}$ and $p_{12}$ in InGaAsP are estimated to be negative for photon energies below the band gap. 

The next step is to determine the magnitude of the dispersive photoelastic constants in InGaAsP as a function of the composition parameter. At the starting point ($y$~=~0), the band gap of InP is 1.35~eV which is much larger than the photon energy of 0.8~eV. At the desired composition ($y$~=~0.625), the band gap reduces to 0.95~eV, which is much closer to the photon energy. With regard to the dispersion trends, lowering the band gap has a similar effect to increasing the photon energy (reducing the operating optical wavelength). To illustrate these similarities, the dispersion trends of GaAs and InP as a function of operating optical wavelength are plotted along side the dispersion of InGaAsP as a function of composition parameter in Fig.~\ref{pe_interp_compare}. All three dispersion curves show a similar trend as the photon energy approaches the band gap. The values of $p_{44}$ and ($p_{11} - p_{12}$) become less negative until crossing into positive values. The values of $p_{11}$ and $p_{12}$ become more negative, with the magnitude of $p_{12}$ eventually becoming larger than the magnitude of $p_{11}$. The green points in Fig.~\ref{pe_interp_compare} correspond to the starting InP photoelastic constants for the InGaAsP interpolation scheme. These starting InP values at an operating optical wavelength of 1.55~$\mu$m are calculated from the fit lines in Ref.~\cite{Suzuki_InP_1983} ($p_{11} = -0.127$, $p_{12} = -0.109$, $p_{44} = -0.064$). For the InGaAsP dispersion curve in Fig.~\ref{pe_interp_compare}c, the $p_{44}$ value was explicitly calculated from Eq.~\ref{eq:p44_term}. Conservatively, the value of $p_{11}$ is estimated to be the same as the starting InP value. Then, the value of $p_{12}$ can be estimated using the calculated difference ($p_{11} - p_{12}$) from Eq.~\ref{eq:p11mp12_term}. Using this methodology, the estimated InGaAsP photoelastic constants used in our simulations are $p_{11} = -0.127$, $p_{12} = -0.242$, and $p_{44} = 0.027$.

\subsubsection{Carrier Transport}

From Chapter~10 in Ref.~\cite{adachi1992IIIVsemi}, the mobility of InGaAs at room temperature is similar in magnitude to the mobility of InGaAsP at the composition parameter used in this physical system. The mobility, $\mu$, of InGaAs thin films made at Sandia for use in AE delayline devices \cite{hackett2021amp} is 2000~cm\textsuperscript{2}/Vs, which is lower than the calculated values for bulk InGaAs as a result of surface effects limiting the mobility. Since these same surface effects will be present in the fabricated InGaAsP thin film, its reasonable to assume a similar limit to the mobility for the physical system in this work ($\mu$~=~2000~cm\textsuperscript{2}/Vs). The optimal carrier concentration for AE interactions with MHz and GHz elastic modes typically falls within the range of 10\textsuperscript{15} to 10\textsuperscript{17}~cm\textsuperscript{-3} \cite{hackett2021amp}. The carrier concentration must also be low enough so significant optical losses are not introduced \cite{hava1993theoretical}. Therefore, the simulations and calculations for this physical system will use a carrier concentration of $N$~=~1~x~10\textsuperscript{16}~cm\textsuperscript{-3}. 

The effective mass and plasma frequency of the InGaAsP are necessary to compute the Brillouin nonlinear susceptibility. The effective mass of the electrons in the conduction band edge is given by a linear interpolation as a function of the InGaAsP composition parameter \cite{adachi_InGaAsP_1982}

\begin{equation}
\begin{aligned}
m^{*} = \left( 0.080 - 0.039 \, y \right) \, m_\text{e}.
\end{aligned}
\label{eq:mc_eff}
\end{equation}

\noindent
For the composition parameter used in this physical system, the effective mass is $m^{*} = 0.055625 \, \, m_\text{e}$, where $m_\text{e}$ is the mass of an electron. Finally, the plasma frequency in a semiconductor material is given as

\begin{equation}
\begin{aligned}
\omega_\text{e}^{2} = \frac{N e^{2}}{m^{*} \epsilon_{0}},
\end{aligned}
\label{eq:plasma_f2}
\end{equation}

\noindent
where $e$ is the charge of an electron and $\epsilon_{0}$ is the permittivity of free space. Given the carrier concentration and effective mass of this physical system, the plasma frequency is $\omega_\text{e}/2 \pi = 3.8069$~THz. 

\newpage

\begin{table*}[ht]
    \caption{Summary of the calculated InGaAsP parameters for the physical system used in this paper.}
    \centering
    \ra{1.2}
    \begin{tabularx}{6.5cm}{@{} A B @{}}
    \toprule
    {\makecell[t]{InGaAsP\\ Parameter}} & {\makecell[t]{Calculated\\ Value}} \\
    \midrule
       $\lambda_\text{optical}$ & 1.55~$\mu$m
        \\
        $E$ & 0.7999~eV
        \\
        $y$ & 0.625
        \\
        $x$ & 0.288
        \\
        $C_{11}$ & 100.35~GPa
        \\
        $C_{12}$ & 51.85~GPa
        \\
        $C_{44}$ & 47.6625~GPa
        \\
        $S_{11}$ & 1.538~x~10\textsuperscript{-11}~Pa\textsuperscript{-1}
        \\
        $S_{12}$ & -5.239~x~10\textsuperscript{-12}~Pa\textsuperscript{-1}
        \\
        $S_{44}$ & 2.098~x~10\textsuperscript{-11}~Pa\textsuperscript{-1}
        \\
        $\rho$ & 5232~kg/m\textsuperscript{3}
        \\
        $E_{0}$ & 0.9469~eV
        \\
        $\Delta_{0}$ & 0.2586~eV
        \\
        $\epsilon_\text{s}$ & 13.3375
        \\
        $A \left( y \right)$ & 6.1873
        \\
        $B \left( y \right)$ & 8.7841
        \\
        $n_\text{r}$ & 3.4013
        \\
        $C^{*}$ [100] & -1.327~x~10\textsuperscript{-11}~cm\textsuperscript{2}/dyn
        \\
        $C^{*}$ [111] & -0.799~x~10\textsuperscript{-11}~cm\textsuperscript{2}/dyn
        \\
        $D^{*}$ [100] & 1.932~x~10\textsuperscript{-11}~cm\textsuperscript{2}/dyn
        \\
        $D^{*}$ [111] & 2.315~x~10\textsuperscript{-11}~cm\textsuperscript{2}/dyn
        \\
        $\alpha_\text{pe}^\text{[100]}$ & -3.186~x~10\textsuperscript{-11}~cm\textsuperscript{2}/dyn
        \\
        $\alpha_\text{pe}^\text{[111]}$ & -0.764~x~10\textsuperscript{-11}~cm\textsuperscript{2}/dyn
        \\
        $p_{11} - p_{12}$ & 0.115
        \\
        $p_{11}$ & -0.127
        \\
        $p_{12}$ & -0.242
        \\
        $p_{44}$ & 0.027
        \\
        $\mu$ & 2000~cm\textsuperscript{2}/Vs
        \\
        $N$ & 1~x~10\textsuperscript{16}~cm\textsuperscript{-3}
        \\
        $m^{*}$ & 0.055625~$m_\text{e}$
        \\
        $\omega_\text{e}$ & $2\pi \cdot 3.8069$~THz.
        \\
    \bottomrule
    \end{tabularx}
    \label{table:AEBmodes}
\end{table*}

\newpage

\subsection{Optical Simulations}

Given a physical system and its material parameters, the first step in designing an acoustoelectric Brillouin device is to examine the phase-matching criteria between the optical mode pair and the phonon wavelength. This section of the supplement will cover both backward intra-modal and forward inter-modal Brillouin scattering processes. In the case of a backward intra-modal Stokes process, a higher energy pump photon creates a forward scattered phonon and backward scattered lower energy Stokes photon in the same optical mode. Similarly, in the case of a forward inter-modal Stokes process, a higher energy pump photon from one optical mode creates a forward scattered phonon and lower energy Stokes photon in a second optical mode. Due to momentum conservation, the phonon wave vector can be expressed as the difference between the pump and Stokes photon wave vectors 

\begin{equation}
\begin{aligned}
q_\text{m} = k_\text{p} \left( \omega_\text{p} \right) - k_\text{s} \left( \omega_\text{s} \right),
\end{aligned}
\label{eq:InterB_ap}
\end{equation}

\noindent
where $k_\text{p}$ is the pump photon wave vector, $k_\text{s}$ is the Stokes photon wave vector, $\omega_\text{p}$ is the pump photon frequency, and $\omega_\text{s}$ is the Stokes photon frequency. The required energy conservation can be expressed as 

\begin{equation}
\begin{aligned}
\Omega_\text{m} = \omega_\text{p} - \omega_\text{s},
\end{aligned}
\label{eq:InterBf_ap}
\end{equation}

\begin{figure*}[b]
\centering
\includegraphics[width=1\linewidth]{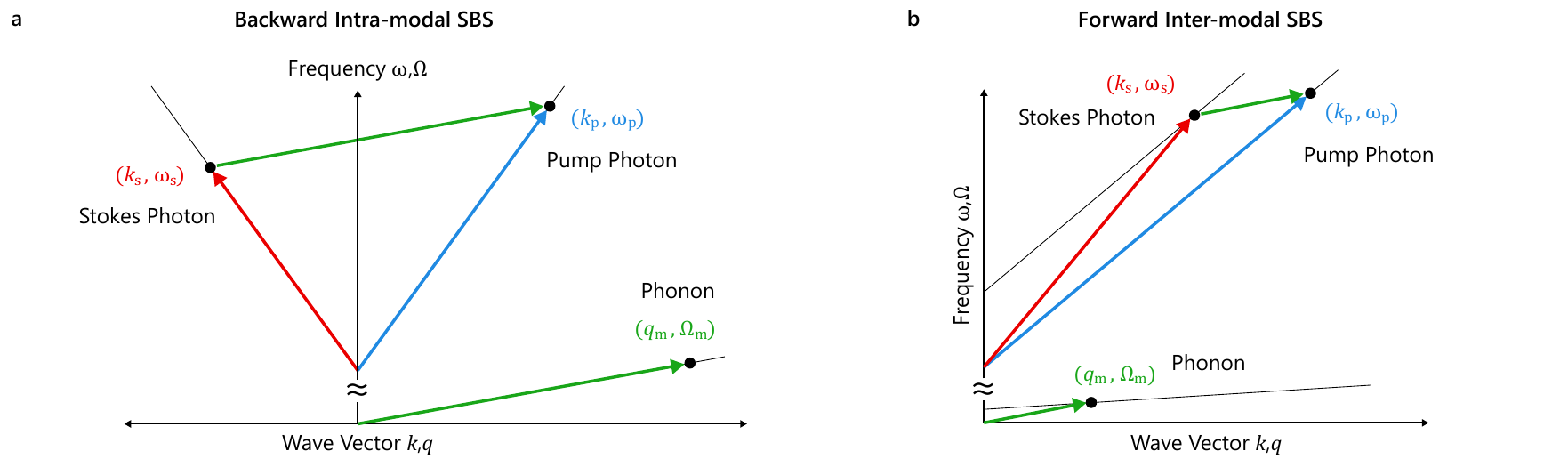}
\caption{Illustrations of (a) backward intra-modal and (b) forward inter-modal dispersion diagrams showing the phase-matched pump photon, Stokes photon, and phonon.}
\label{fig:opt_dispersion}
\end{figure*}

\noindent
where $\Omega_\text{m}$ is the phonon frequency. Illustrations of the intra-modal and inter-modal dispersion diagrams with phase-matched phonons are shown in Fig.~\ref{fig:opt_dispersion}. As the dispersion diagrams show, the phase-matched phonon wave vector is much larger for backward intra-modal scattering than forward inter-modal scattering. In general, this means backward scattering Brillouin processes require small phonon wavelengths while forward scattering requires much larger phonon wavelengths. The phase-matched phonon wavelength can be calculated using Eq.~\ref{eq:InterB_ap} along with expressions for the photon and phonon wave vectors. The optical wave vectors ($k_\text{p,s}$) can be expressed as 

\begin{equation}
\begin{aligned}
k_\text{p,s} = n_\text{eff}^\text{p,s} \, \frac{2\pi}{\lambda_\text{p,s}},
\end{aligned}
\label{eq:opticalB_ap}
\end{equation}

\noindent
where $n_\text{eff}^\text{p}$ is the pump optical mode's effective index, $n_\text{eff}^\text{s}$ is the Stokes optical mode's effective index, $\lambda_\text{p}$ is the pump photon wavelength, and $\lambda_\text{s}$ is the Stokes photon wavelength. Similarly, the wave vector of the phonon ($q_\text{m}$) is given as 

\begin{equation}
\begin{aligned}
q_\text{m} = \frac{2\pi}{\lambda_\text{m}},
\end{aligned}
\label{eq:acousticB_ap}
\end{equation}

\noindent
where $\lambda_\text{m}$ is the phonon wavelength. In the case of backward intra-modal Brillouin scattering, the Stokes photon wave vector is approximately equal to the pump photon wave vector, but opposite in sign ($k_\text{s} \approx - k_\text{p}$). Using Eq.~\ref{eq:InterB_ap}, the phonon wave vector can be expressed in terms of the pump wave vector 

\begin{equation}
\begin{aligned}
q_\text{m} = 2 \, k_\text{p} \left( \omega_\text{p} \right).
\end{aligned}
\label{eq:IntraB_ap}
\end{equation}

\noindent
The phonon wavelength can then be solved for in terms of the optical wavelength and effective mode index

\begin{equation}
\begin{aligned}
\lambda_\text{m} = \frac{\lambda_\text{p}}{2 \, n_\text{eff}^\text{p}}.
\end{aligned}
\label{eq:IntraLm_ap}
\end{equation}

\begin{figure*}[tb]
\centering
\includegraphics[width=1\linewidth]{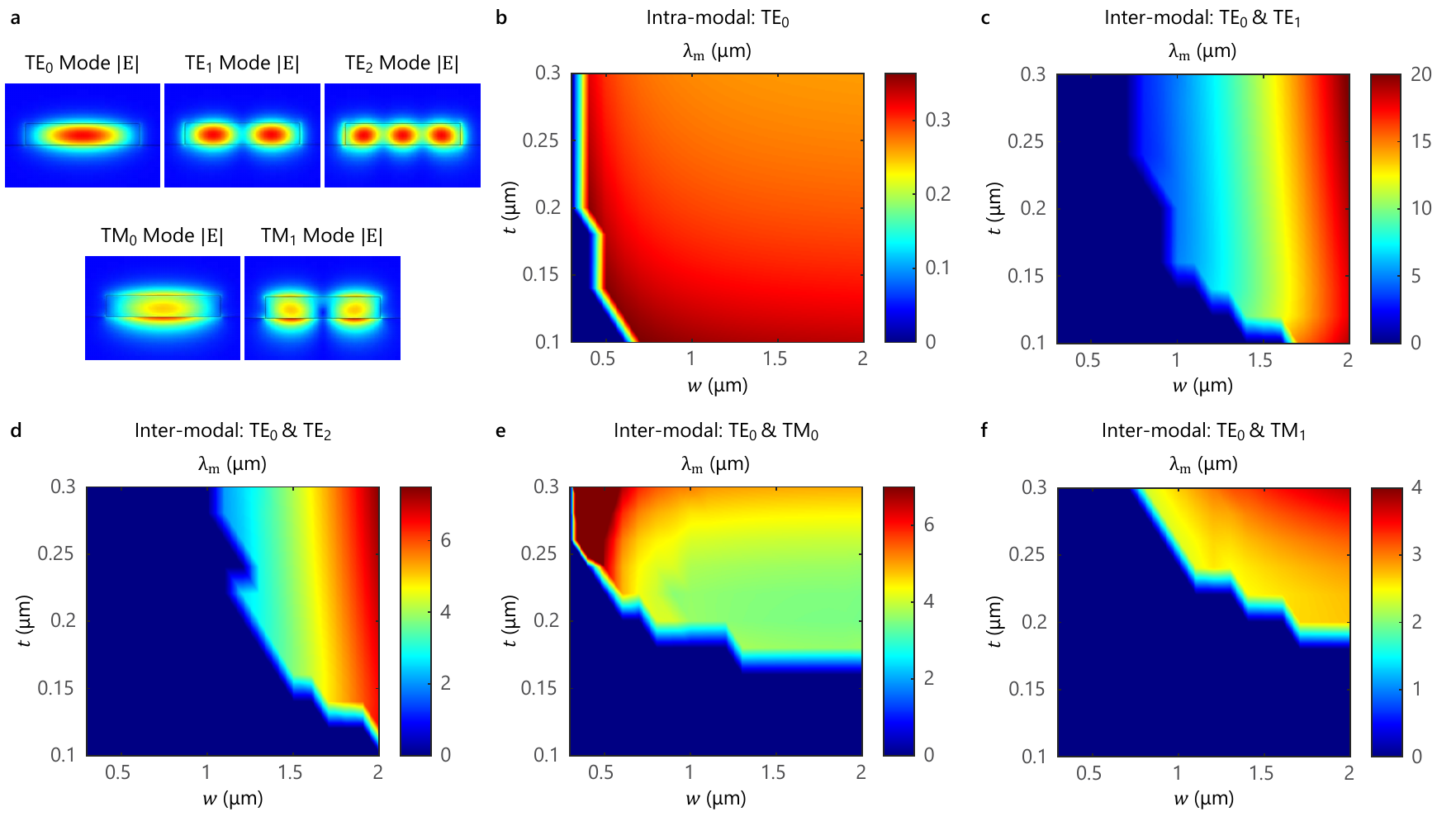}
\caption{Optical simulation results for both backward intra-modal and forward inter-modal phase matching conditions. (a) Example TE-like and TM-like optical mode shapes for the InGaAsP-LN-Si physical system. Surface plots of the phase-matched phonon wavelength for (b) backward intra-modal and (c-f) forward inter-modal Brillouin scattering as a function of the waveguide width and thickness.}
\label{fig:opt_sims_ap}
\end{figure*}

\noindent
The phase-matched phonon wavelength for backward intra-modal scattering is smaller than half of the optical wavelength and strongly depends on the optical mode's effective index. Using the TE\textsubscript{0} optical mode in the InGaAsP-LN-Si physical system, the phase-matched phonon wavelength for backward intra-modal Brillouin scattering as a function of waveguide geometry is shown in Fig.~\ref{fig:opt_sims_ap}b. The phonon wavelength, which has an average value around 300~nm, is almost independent of both the waveguide width and thickness. This phonon wavelength is the same order of magnitude as the waveguide dimensions, which leads to elastic modes with the strain almost entirely confined in the InGaAsP waveguide. These elastic waveguide modes will have strain profiles characteristic of either the traditional Rayleigh or Shear Horizontal (SH) surface acoustic wave (SAW) modes, which means these waveguide modes will have the same favorable propagation angle as the SAW mode it resembles most. 

This phase-matching analysis can then be extended to forward inter-modal Brillouin scattering. Since the phonon frequency is much smaller than the photon frequencies ($\Omega_\text{m}\,$$<<$$\,\omega_\text{p,s}$), both the pump and Stokes photons can be approximated as having the same optical wavelength when solving for the phase-matched phonon wavelength. Using this assumption, the required phonon wavelength for forward inter-modal Brillouin scattering is 

\begin{equation}
\begin{aligned}
\lambda_\text{m} = \frac{\lambda_\text{p}}{n_\text{eff}^\text{p} - n_\text{eff}^\text{s}}.
\end{aligned}
\label{eq:InterLm_ap}
\end{equation}

\noindent
Due to the relatively small difference between the pump and Stokes effective mode indices, the phonon wavelength for forward inter-modal Brillouin scattering is typically larger than the optical wavelength. For different combinations of TE-like and TM-like optical modes (mode shapes shown in Fig.~\ref{fig:opt_sims_ap}a), the phase-matched phonon wavelengths for forward inter-modal Brillouin scattering range from 2~$\mu$m to 20~$\mu$m. This wavelength range is consistent with traditional SAW devices in both bulk and thin film lithium niobate material systems. Therefore, the desired elastic modes for forward inter-modal Brillouin scattering will be traditional SAW modes with high piezoelectric coupling coefficients ($k^{2}$) for a strong AE interaction. The two SAW modes with the highest $k^{2}$ in Y-cut lithium niobate are the Rayleigh \cite{yamanouchi_propagation_1972} and Shear Horizontal \cite{siddiqui_comparison_2020} elastic modes. Additionally, these SAW modes will be horizontally guided and laterally confined by the density mismatch and sound speed difference between the InGaAsP and lithium niobate. As a result of each SAW mode's strain profile, the Rayleigh SAW will have the strongest optomechanical coupling to optical modes of the same parity, while the SH SAW will have the strongest optomechanical coupling to optical modes of different parity. Therefore, the TE\textsubscript{0}/TE\textsubscript{1} and TE\textsubscript{0}/TM\textsubscript{0} optical mode pairs (Fig.~\ref{fig:opt_sims_ap}c,e) are considered for the SH SAW elastic mode and the TE\textsubscript{0}/TE\textsubscript{2} and TE\textsubscript{0}/TM\textsubscript{1} optical mode pairs (Fig.~\ref{fig:opt_sims_ap}d,f) are considered for the Rayleigh SAW elastic mode.

\subsection{Acoustic Simulations}

The next stage in the design process is to consider the strain profiles and piezoelectric coupling coefficient, $k^{2}$, for each elastic mode given the phase-matched phonon wavelength. The overlap between the strain profiles and optical fields determines the strength of the optomechanical coupling, and a high $k^{2}$ is an indicator of a strong AE interaction. In the presence of a perfect conductor, the piezoelectric coupling coefficient of a surface acoustic wave can be calculated by the shift in acoustic velocity as the perfect conductor is brought from infinitely far away to the surface of the piezoelectric material \cite{Hutson_AEk2_1962,Ingebrigtsen_SAW_1969}. In FEA simulation, this can be achieved through two different eigenfrequency simulations. One will have an electrically free lithium niobate surface and the other will have a grounded lithium niobate surface. The piezoelectric coupling coefficient can be computed as 

\begin{equation}
\begin{aligned}
k^2 = \frac{\left|v_{\text{free}}^{\, 2} - v_{\text{ground}}^{\, 2} \right|}{v_{\text{free}}^{\, 2}} = \frac{\left|f_{\text{free}}^{\, 2} - f_{\text{ground}}^{\, 2} \right|}{f_{\text{free}}^{\, 2}},
\end{aligned}
\label{eq:k2_definition_ap}
\end{equation}

\noindent
Here, $v_\text{free}$ is the acoustic velocity with a free piezoelectric surface and $v_\text{ground}$ is the acoustic velocity with a grounded piezoelectric surface. Since the phonon wavelength is the same for both simulations, the piezoelectric coupling coefficient can be expressed in terms of the acoustic eigenfrequency for the free piezoelectric surface ($f_\text{free}$) and grounded piezoelectric surface ($f_\text{ground}$).

\begin{figure*}[b]
\centering
\includegraphics[width=1\linewidth]{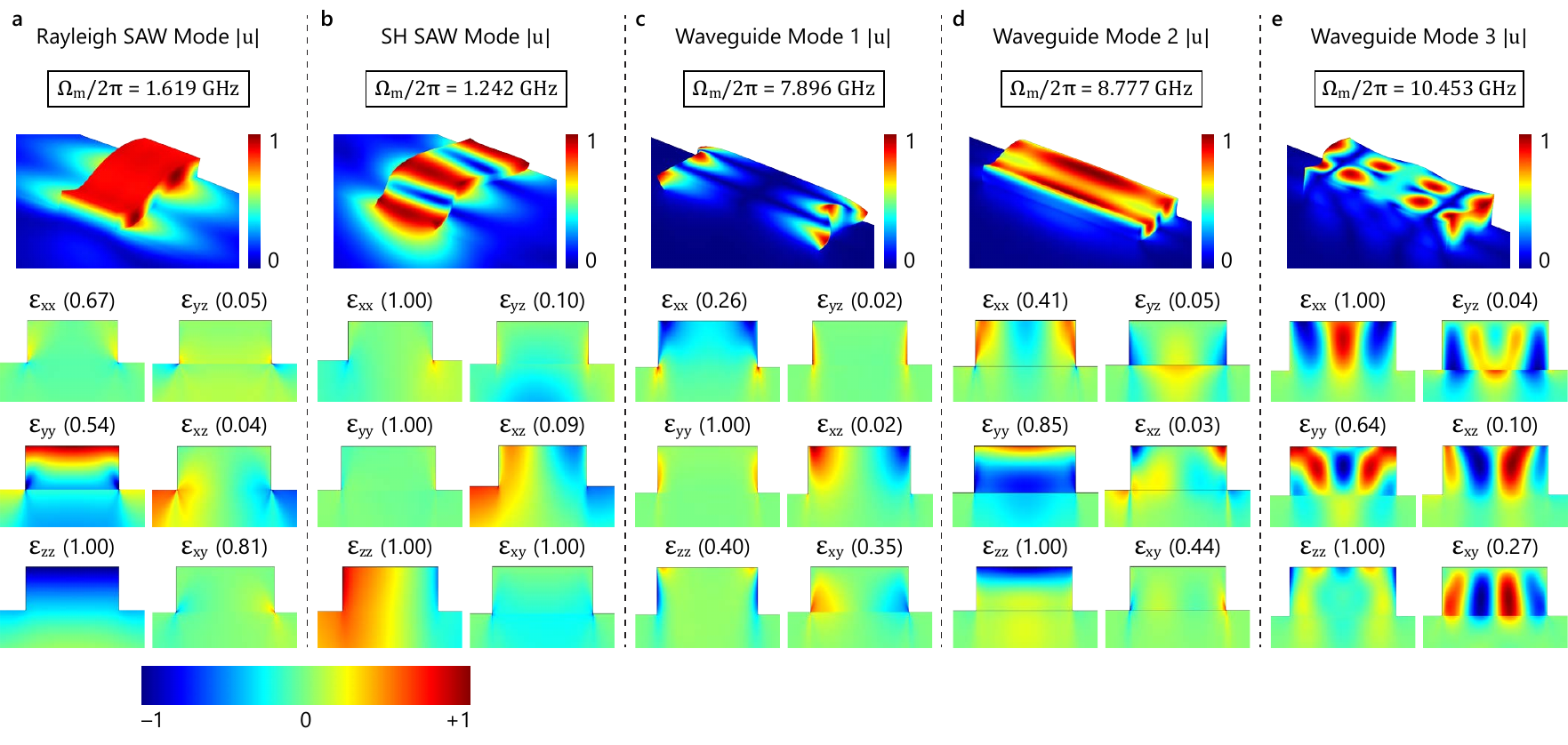}
\caption{Acoustic mode simulation results for both inter-modal and intra-modal SBS processes. Example periodic mode shape and cross-sectional strain profiles are shown for (a) Rayleigh SAW mode, (b) SH SAW mode, (c) waveguide mode 1, (d) waveguide mode 2---elastic mode highlighted in the main manuscript, and (e) waveguide mode 3. For each normalized strain profile, the relative magnitudes between the strain components are shown in parenthesis.}
\label{fig:acoustic_sims_ap}
\end{figure*}

In the case of an unguided Rayleigh SAW mode in Y-cut lithium niobate, the $k^{2}$ is highest in the 0\textdegree~propagation direction with a value around 5~\%. Similarly, the unguided SH SAW mode has its highest $k^{2}$ around 18~\% in the 90\textdegree~propagation direction. The definition of the calculated $k^{2}$, however, must be adjusted with the inclusion of the InGaAsP waveguide into the acoustic simulation. In order to consider the interaction strength between the elastic mode and free carriers in the InGaAsP, the grounded eigenfrequency is simulated with the perfect conductor placed only at the interface between the InGaAsP waveguide and lithium niobate surface. This means the Rayleigh and SH SAW modes used for forward inter-modal Brillouin scattering will only have a fraction of the $k^{2}$ found in unguided lithium niobate propagation. As the phonon wavelength approaches the waveguide width from the large wavelength side, a larger fraction of the elastic mode is guided within the waveguide, resulting in a piezoelectric coupling coefficient approaching the unguided $k^{2}$ limit. For backward intra-modal Brillouin scattering, where the phase-matched phonon wavelength is smaller than the InGaAsP waveguide width, almost all of the elastic energy is confined in the waveguide. For these elastic waveguide modes, the piezoelectric coupling strength depends on the displacement profile within the waveguide. 

The inclusion of the InGaAsP waveguide on top of the lithium niobate guides the Rayleigh and SH SAW modes within the InGaAsP, with the amount of guiding dependent on the relative size difference between the waveguide width and phonon wavelength. Example elastic mode shapes and strain profiles for guided Rayleigh and SH SAW modes are shown in Fig.~\ref{fig:acoustic_sims_ap}a-b. Additionally, as more of the elastic energy is guided within the InGaAsP waveguide, there is a larger overlap between the optical fields and elastic strains. Therefore, it is advantageous in the design process for inter-modal Brillouin devices to reduce the required phonon wavelength to as close to the waveguide dimensions as possible in order to achieve higher acoustoelectric and optomechanical coupling strength. Operating near the optical mode cutoff, however, comes at the expense of additional optical losses. This trade-off between optical losses and coupling strength will need to be carefully considered when designing acoustoelectric Brillouin devices.

For backward intra-modal Brillouin scattering, example elastic mode shapes and strain profiles for three elastic waveguide modes are shown in Fig.~\ref{fig:acoustic_sims_ap}c-e. For waveguide mode 1, the displacement is primarily at the edges of the waveguide, which has poor overlap with the TE\textsubscript{0} optical fields. Since the $\varepsilon_\text{zz}$ and $\varepsilon_\text{xz}$ strain components are similar to the SH SAW mode, the propagation angle with the highest $k^{2}$ is 90\textdegree. Waveguide mode 2---the elastic mode highlighted in the main manuscript---has displacement throughout the entire width of the waveguide and therefore a much larger overlap with the optical fields. Waveguide mode 2 and the Rayleigh SAW mode have similar strain profiles and the same dominant strain components ($\varepsilon_\text{yy}$ and $\varepsilon_\text{zz}$), which means waveguide mode 2 will have the highest $k^{2}$ at a 0\textdegree~propagation angle. Finally, waveguide mode 3 has a mix of shear and longitudinal strains with several nodes and anti-nodes along the width of the waveguide. As a result, there is no clear preference in propagation direction and waveguide mode 3 has similar $k^{2}$ values in both 0\textdegree~and 90\textdegree~propagation directions.

\subsection{Optomechanical Coupling Calculation}

After simulating the optical modes, acoustic modes, and piezoelectric coupling coefficient, the next step in the design process is to compute the optomechanical coupling coefficient ($g_{0}$). The two main contributions to optomechanical coupling are from the photoelastic ($g_\text{pe}$) and radiation pressure ($g_\text{rp}$) effects, where $g_{0} = g_\text{pe} + g_\text{rp}$. This section of the supplement will outline the theory for photoelastic and radiation pressure optomechanical coupling for Stokes scattering processes and then implement this theory for the InGaAsP-LN-Si physical system. 

\subsubsection{Photoelastic Coupling}

The photoelastic coupling coefficient (in units of (rad/s)$\sqrt{\text{m}}$) is given as \cite{kharel2016noise}
 
\begin{equation}
\label{FW_pe_Stokes_full2}
g_\text{pe} = \left| C_\text{m} \right| \left| C_\text{p} \right| \left| C_\text{s} \right| \, \xi \iint_A \left( \mathrm{D}_\text{p}^{i} \right)^{*} \mathrm{D}_\text{s}^{j} \, p^{ijkl} \varepsilon^{kl} \, dA.
\end{equation}


\noindent
where $\mathrm{D}_\text{p}$ and $\mathrm{D}_\text{s}$ are the electric displacement fields of the pump and Stokes photons extracted from simulation, $p$ is the fourth-rank photoelastic tensor, $\varepsilon$ is the second-rank strain tensor extracted from simulation, and $\xi$ is the frequency constant defined as 

\begin{equation}
\label{xi_Stokes_ap}
\xi = \frac{1}{\epsilon_{0}} \sqrt{\frac{\omega_\text{p}}{2}} \sqrt{\frac{\omega_\text{s}}{2}} \sqrt{\frac{\hbar \Omega_\text{m}}{2}}.
\end{equation}

\noindent
The tensor product is integrated over the cross-sectional area perpendicular to the wave propagation and the elastic strain and electric displacement fields need to be normalized as described in Ref.~\cite{kharel2016noise}. A scaling factor can be implemented for the phonon displacement ($\left | C_\text{m} \right|$), pump photon electric displacement field ($\left | C_\text{p} \right|$), and Stokes photon electric displacement field ($\left | C_\text{s} \right|$) 

\begin{equation}
\label{A2_ap}
\left| C_\text{m} \right|^2 = \frac{1}{\Omega_\text{m}^2 \iint_A \rho \left( \left| \mathrm{u}_\text{x} \right|^2 + \left| \mathrm{u}_\text{y} \right|^2 + \left| \mathrm{u}_\text{z} \right|^2 \right) \, dA},
\end{equation}

\begin{equation}
\label{B2_ap}
\left| C_\text{p} \right|^2 = \frac{1}{\frac{1}{\epsilon_0} \iint_A \frac{1}{\epsilon_\text{r}} \left( \left| \mathrm{D}_\text{x}^\text{p} \right|^2 + \left| \mathrm{D}_\text{y}^\text{p} \right|^2 + \left| \mathrm{D}_\text{z}^\text{p} \right|^2 \right) \, dA},
\end{equation}

\begin{equation}
\label{C2_ap}
\left| C_\text{s} \right|^2 = \frac{1}{\frac{1}{\epsilon_0} \iint_A \frac{1}{\epsilon_\text{r}} \left( \left| \mathrm{D}_\text{x}^\text{s} \right|^2 + \left| \mathrm{D}_\text{y}^\text{s} \right|^2 + \left| \mathrm{D}_\text{z}^\text{s} \right|^2 \right) \, dA},
\end{equation}



\noindent
where $\mathrm{u}$ is the elastic displacement field, $\epsilon_\text{r}$ is the relative permittivity, and $\mathrm{D}^\text{p,s}$ is the electric displacement field (pump or Stokes). Since the elastic strain is the symmetric gradient of the displacement, the $\left | C_\text{m} \right|$ scaling factor is used for the strain tensor in Eq.~\ref{FW_pe_Stokes_full2}. To compute this coupling coefficient, the simulation data is extracted and converted into matrices. The tensor product within the integration is then computed through matrix multiplication \cite{Qiu_SBSMatrix_2013} and integrated.

A compacted Voigt notation is used to reduce the second and fourth-rank tensors so that matrix multiplication can be used within the integration. The compacted notation is 1~=~(xx), 2~=~(yy), 3~=~(zz), 4~=~(yz)~=~(zy), 5~=~(xz)~=~(zx), 6~=~(xy)~=~(yx). This reduces the photoelastic coupling coefficient to 

\begin{equation}
\label{FW_pe_Stokes_full}
g_\text{pe} = \left| C_\text{m} \right| \left| C_\text{p} \right| \left| C_\text{s} \right| \, \xi \iint_A \left[ \mathrm{D} \right] \left[ p \right] \left[ \varepsilon \right] \, dA,
\end{equation}

\noindent
where the electric displacement field product ($\left[ \mathrm{D} \right]$) is a [1x6] matrix, the photoelastic constants ($\left[ p \right]$) are a [6x6] matrix, and the strain ($\left[ \varepsilon \right]$) is a [6x1] matrix. InGaAsP has a cubic crystal structure with 3 independent photoelastic constants and the coupling matrix takes the form 

\begin{equation}
\label{p_matrix}
\left[ p \right] = 
\begin{bmatrix}
p_{11} & p_{12} & p_{12} & 0 & 0 & 0\\
p_{12} & p_{11} & p_{12} & 0 & 0 & 0\\
p_{12} & p_{12} & p_{11} & 0 & 0 & 0\\
0 & 0 & 0 & p_{44} & 0 & 0\\
0 & 0 & 0 & 0 & p_{44} & 0\\
0 & 0 & 0 & 0 & 0 & p_{44}
\end{bmatrix}.
\end{equation}

\noindent 
The strain matrix is given as  

\begin{equation}
\label{S_matrix}
\left[ \varepsilon \right] = 
\begin{bmatrix}
\varepsilon_\text{xx}\\
\varepsilon_\text{yy}\\
\varepsilon_\text{zz}\\
2 \, \varepsilon_\text{yz}\\
2 \, \varepsilon_\text{xz}\\
2 \, \varepsilon_\text{xy}
\end{bmatrix},
\end{equation}

\noindent 
and the electric displacement field product matrix is given as

\begin{equation}
\label{D_matrix}
\left[ \mathrm{D} \right] = 
\begin{bmatrix}
\left\{ \left( \mathrm{D}_\text{x}^\text{p} \right)^{*} \mathrm{D}_\text{x}^\text{s} \right\} & \left\{ \left( \mathrm{D}_\text{y}^\text{p} \right)^{*} \mathrm{D}_\text{y}^\text{s} \right\} & \left\{ \left( \mathrm{D}_\text{z}^\text{p} \right)^{*} \mathrm{D}_\text{z}^\text{s} \right\} & 2 \left\{ \left( \mathrm{D}_\text{y}^\text{p} \right)^{*} \mathrm{D}_\text{z}^\text{s} \right\} & 2 \left\{ \left( \mathrm{D}_\text{x}^\text{p} \right)^{*} \mathrm{D}_\text{z}^\text{s} \right\} & 2 \left\{ \left( \mathrm{D}_\text{x}^\text{p} \right)^{*} \mathrm{D}_\text{y}^\text{s} \right\} 
\end{bmatrix}.
\end{equation}

\noindent
To include the contribution of lithium niobate in the photoelastic coupling coefficient, the treatment above is modified to use the photoelastic coupling matrix of a trigonal crystal structure when integrating over regions of lithium niobate 

\begin{equation}
\label{pLN_matrix}
\left[ p \right] = 
\begin{bmatrix}
p_{11} & p_{12} & p_{13} & p_{14} & 0 & 0\\
p_{12} & p_{11} & p_{13} & -p_{14} & 0 & 0\\
p_{31} & p_{31} & p_{33} & 0 & 0 & 0\\
p_{41} & -p_{41} & 0 & p_{44} & 0 & 0\\
0 & 0 & 0 & 0 & p_{44} & p_{41}\\
0 & 0 & 0 & 0 & p_{14} & \frac{1}{2} \left(p_{11} - p_{12} \right)
\end{bmatrix}.
\end{equation}

\noindent
Here, there are eight independent photoelastic constants ($p_{11}$, $p_{33}$, $p_{44}$, $p_{12}$, $p_{13}$, $p_{31}$, $p_{14}$, $p_{41}$) and these values can be found in Ref.~\cite{Weis_LN_1985,Andrushchak_LN_2009}. Since the optical modes are guided within the InGaAsP, there is little overlap between the elastic mode and optical fields in the lithium niobate, which means the contribution to the overall photoelastic coupling is very small.

\subsubsection{Radiation Pressure Coupling}

The radiation pressure coupling coefficient (in units of (rad/s)$\sqrt{\text{m}}$) is given as \cite{kharel2016noise}

\begin{equation}
\label{rp_Stokes_full_ap}
g_\text{rp} = \left| C_\text{m} \right| \left| C_\text{p} \right| \left| C_\text{s} \right| \, \xi \iint_A \epsilon_{0} \left[ \left( \mathbf{E}_\text{p}^{\parallel *} \cdot \mathbf{E}_\text{s}^{\parallel} \right) \nabla \epsilon - \left( \mathbf{D}_\text{p}^{\perp *} \cdot \mathbf{D}_\text{s}^{\perp} \right) \nabla \epsilon^{-1} \right] \cdot \mathbf{u} \, \, dA,
\end{equation}


\noindent
where $\mathbf{E}_\text{p,s}^{\parallel}$ is the component of the electric field parallel to the defined normal and $\mathbf{D}_\text{p,s}^{\perp}$ is the component of the electric displacement field perpendicular to the defined normal. Since the physical system presented in this paper only has discrete permittivities, there exists permittivity discontinuities at the material interfaces. Since the gradient of the permittivity is zero within the materials, this integral collapses to a line integral around the boundaries of the waveguide. At the waveguide boundaries, the permittivity is a step function between the permittivity of the waveguide ($\epsilon_{1}$) and surrounding material ($\epsilon_{2}$). Since the gradient of a unit step function is a delta function, the radiation pressure coupling coefficient can be written as \cite{Johnson_Maxwell_2002,Eichenfield_OMC_2002}

\begin{equation}
\label{FW_rp_Stokes_int_line}
g_\text{rp} = \left| C_\text{m} \right| \left| C_\text{p} \right| \left| C_\text{s} \right| \, \xi \int_L \epsilon_{0} \left[ \left( \mathbf{E}_\text{p}^{\parallel *} \cdot \mathbf{E}_\text{s}^{\parallel} \right) \Delta \epsilon - \left( \mathbf{D}_\text{p}^{\perp *} \cdot \mathbf{D}_\text{s}^{\perp} \right) \Delta \epsilon^{-1} \right] \hat{n} \cdot \mathbf{u} \, \, dL,
\end{equation}

\noindent
where $\Delta \epsilon = \epsilon_{1} - \epsilon_{2}$, $\Delta \epsilon^{-1} = \epsilon_{1}^{-1} - \epsilon_{2}^{-1}$, and $\hat{n}$ is the normal of the line integral path. 

\begin{figure*}[t!]
\centering
\includegraphics[width=1\linewidth]{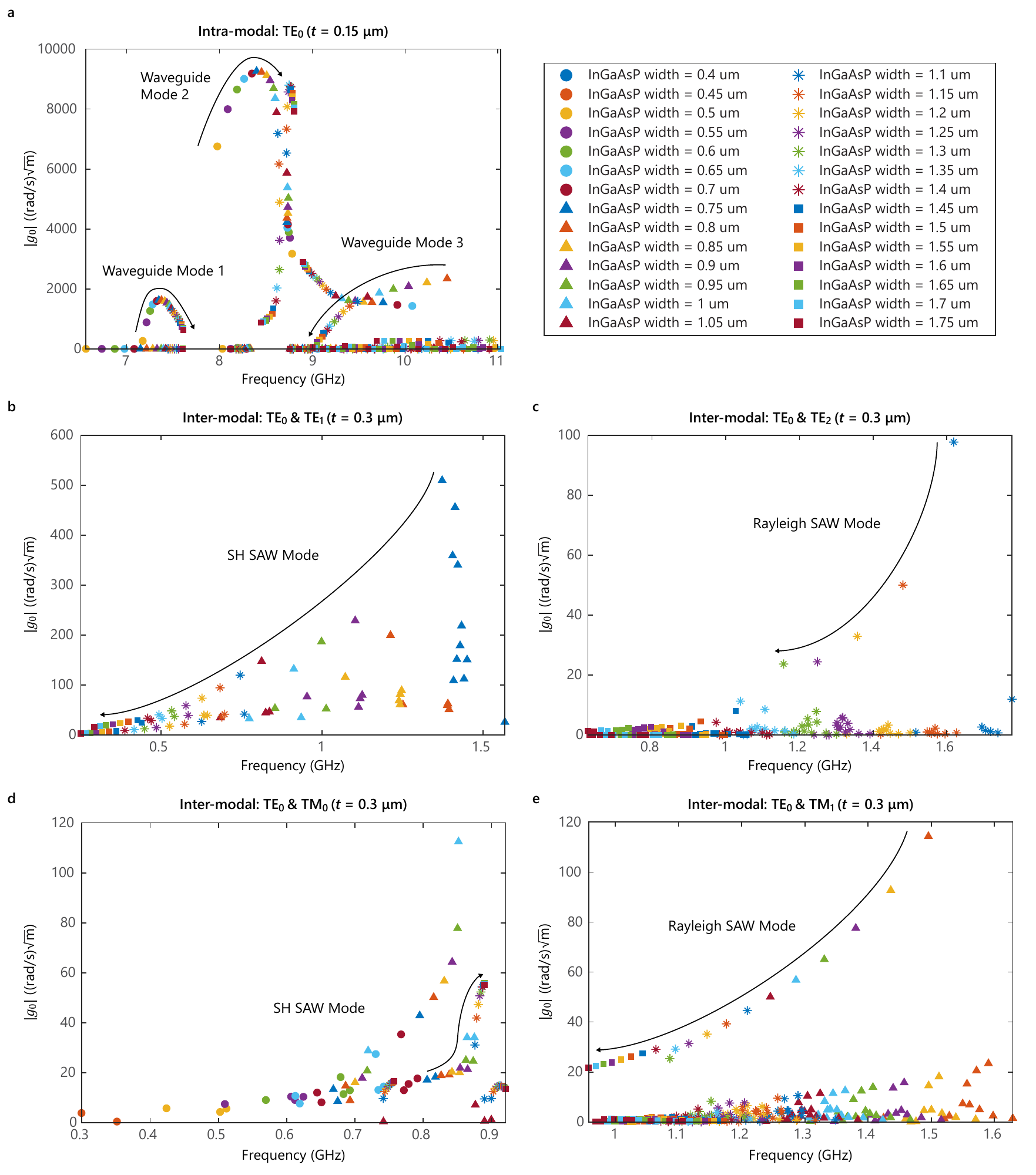}
\caption{Calculated optomechanical coupling coefficient as a function of the simulated elastic modes for various waveguide geometries. The optomechanical coupling for key elastic modes is plotted for (a) backward intra-modal and (b-e) forward inter-modal Brillouin scattering. For each elastic mode, the black arrow indicates the trend in optomechanical coupling as the waveguide width is increased.}
\label{fig:OM_coupling_full}
\end{figure*}

The optomechanical coupling for key simulated elastic modes is plotted in Fig.~\ref{fig:OM_coupling_full} at a fixed waveguide thickness as the waveguide width is swept from 0.4~$\mu$m to 1.75~$\mu$m (black arrows). For the case of backward intra-modal Brillouin scattering, the optomechanical coupling trends for three elastic waveguide modes are shown in Fig.~\ref{fig:OM_coupling_full}a. Both waveguide mode 1 and 2 have an optimal waveguide width which maximizes the optomechanical coupling coefficient. This is a result of two competing trends found in this physical system. Since all the elastic modes examined in this supplement have a slower phase velocity in InGaAsP compared to lithium niobate, a larger fraction of the strain is guided within the InGaAsP as the waveguide width is increased, increasing the overlap with the optical fields. On the other hand, increasing the waveguide width reduces the magnitude of the optical fields at the boundaries of the InGaAsP waveguide, reducing the overlap with the strain. When directly comparing the first two waveguide modes, waveguide mode 2 will have a larger optomechanical coupling coefficient compared to waveguide mode 1, which is a result of the different strain profiles between the two elastic modes. Waveguide mode 1 has a majority of its strain around the edges of the InGaAsP waveguide, which has a lower overlap with the optical fields. For waveguide mode 3, there is only a reduction in optomechanical coupling as the waveguide width is increased. Since waveguide mode 3 has several displacement nodes along the waveguide width, there exists a minimum waveguide width where this mode is allowed. Overall, waveguide mode 2 has the highest optomechanical coupling, which is why this simulated elastic mode is used throughout the main manuscript. 

For the cases of forward inter-modal Brillouin scattering shown in Fig.~\ref{fig:OM_coupling_full}b-e, the trends in optomechanical coupling as a function of waveguide width are influenced by additional factors. For each simulated optical mode, a different set of waveguide cutoff dimensions exists, creating different cutoff waveguide widths for each optical mode pair. In addition, the phase-matched phonon wavelength varies as a function of waveguide width, which will change the confinement of the elastic mode and overlap with the optical fields. For the optical mode pairs shown in Fig.~\ref{fig:OM_coupling_full}b,c,e, the dominant factor in the optomechanical coupling trends for the Rayleigh and SH SAW modes is the change in phase-matched phonon wavelength. As the waveguide width is increased, the phase-matched phonon wavelength also increases, reducing the confinement and overlap with the optical fields. In the case of the TE\textsubscript{0} \& TM\textsubscript{0} optical mode pair (Fig.~\ref{fig:OM_coupling_full}d), the phase-matched phonon wavelength does not change significantly with waveguide width and the trend in optomechanical coupling is similar to those found in backward intra-modal Brillouin scattering.

\subsection{Phonon Dissipation}

This subsection of the supplement outlines the estimated phonon dissipation of the elastic mode used in the main manuscript. Given the simulated optimal waveguide geometry for the elastic waveguide mode 2 ($w$ = 1.1~$\mu$m and $t$ = 0.1~$\mu$m), the frequency of the simulated elastic mode is $\Omega_\text{m}$/$2 \pi$ = 8.78~GHz and the phase velocity is $v_\text{m}$ = 3072~m/s. Since the simulated elastic waveguide mode 2 in the InGaAsP-LN-Si physical system has similar waveguide dimensions and frequency to elastic modes reported in silicon with mechanical quality factors ($Q_\text{m}$) ranging from 250-1800 \cite{Laer_SBS_2015,eggleton2019brillouin}, a conservative approximation for the quality factor in our physical system is $Q_\text{m}$ = 250. The phonon dissipation rate, $\Gamma_\text{m}$, can be expressed as \cite{Laer_SBS_2015}

\begin{equation}
\begin{aligned}
\Gamma_\text{m} = \frac{\Omega_\text{m}}{Q_\text{m}}.
\end{aligned}
\label{eq:phonon_Gamma}
\end{equation}

\noindent
For the simulated elastic mode used in this paper, the phonon dissipation rate is estimated to be $\Gamma_\text{m}$/$2 \pi$ = 35.1~MHz. From Chapter~8 in Ref.~\cite{boyd2020nonlinear}, the loss coefficient of the elastic mode is given as 

\begin{equation}
\begin{aligned}
\alpha_\text{m} = \frac{\Gamma_\text{m}}{v_\text{m}}.
\end{aligned}
\label{eq:phonon_alpha}
\end{equation}

\noindent
For the simulated elastic mode used in this paper, the loss coefficient is estimated to be $\alpha_\text{m}$ = 71800~m\textsuperscript{-1}. This loss coefficient can be converted into units of [dB/cm] through the following conversion 

\begin{equation}
\begin{aligned}
\alpha_\text{m} \left[ \frac{dB}{cm} \right] = \frac{1 \, m}{100 \, cm} \left( \frac{10}{\text{ln} \left( 10 \right)} \right) \alpha_\text{m} \left[ \frac{1}{m} \right].
\end{aligned}
\label{eq:alpha_conv}
\end{equation}

\noindent
This results in an estimated loss coefficient of $\alpha_\text{m}$ = 3100~dB/cm.

\subsection{Acoustoelectric Gain Calculation} \label{AEGainCalc}

The two predominant acoustoelectric theories for an amplifier consisting of a semiconductor thin film on a piezoelectric substrate are the normal mode theory \cite{kino_normal_1971} and simple theory \cite{Adler_Simple_1971}. Later, the simple theory for AE gain was expanded to include the effects of loss in the elastic mode \cite{hackett2019amp}. Each of these theories come with its own set of limitations and assumptions. This subsection of the supplement will outline each theory, and compare the results for backward intra-modal Brillouin scattering using the simulated elastic mode in the main manuscript and an example elastic mode for forward inter-modal Brillouin scattering. 

The first AE gain theory examined is based on the normal mode theory as described by Kino and Reeder \cite{kino_normal_1971}, modified to include a dielectric gap material between the amplifier layer and the piezoelectric substrate. In the presence of a semiconductor with a finite film thickness, the elastic wave's propagation constant can be expressed as 

\begin{equation}
\begin{aligned}
q = q_\text{m} + q_\text{AE} + i \frac{\alpha_\text{AE}}{2},
\end{aligned}
\label{eq:prop_const}
\end{equation}

\noindent
where $q_\text{m}$ is the elastic wave's unperturbed propagation constant, $\alpha_\text{AE}$ is the acoustoelectric gain, and $q_\text{AE}$ is the acoustoelectric phase delay. The acoustoelectric contributions to the gain and phase delay are given as

\begin{equation}
\begin{aligned}
\alpha_\text{AE} = \frac{\left(v_\text{d}/v_\text{m}-1\right) \omega_\text{c} \epsilon_\text{s} Z^\prime_\text{m} \left( q_\text{m} h \right) q_\text{m} \text{tanh} \left(q_\text{m} t\right)}{\left(v_\text{d}/v_\text{m}-1\right)^2 + \left(R \omega_\text{c} / \Omega_\text{m} + D\right)^2},
\end{aligned} 
\label{eq:AE_gain_exp}
\end{equation}

\begin{equation}
\begin{aligned}
q_\text{AE} = \frac{1}{2} \frac{\left(R \omega_\text{c} / \Omega_\text{m} + D\right) \omega_\text{c} \epsilon_\text{s} Z^\prime_\text{m} \left( q_\text{m} h \right) q_\text{m} \text{tanh} \left(q_\text{m} t\right)}{\left(v_\text{d}/v_\text{m}-1\right)^2 + \left(R \omega_\text{c} / \Omega_\text{m} + D\right)^2}.
\end{aligned} 
\label{eq:AE_phase_exp}
\end{equation}

\noindent
Here, $\Omega_\text{m}$ is the phonon frequency, $h$ is the gap height between the lithium niobate and InGaAsP waveguide, $t$ is the InGaAsP waveguide thickness, $\epsilon_\text{s}$ is the InGaAsP permittivity, $v_\text{m}$ is the elastic wave phase velocity, and $v_\text{d}$ is the free carrier drift velocity. The dielectric relaxation frequency ($\omega_\text{c}$) and diffusion frequency ($\omega_\text{D}$) can be expressed as 

\begin{equation}
\begin{aligned}
\omega_\text{c} = \frac{e \mu N}{\epsilon_\text{s}},
\end{aligned} 
\label{eq:wc_ap}
\end{equation}

\begin{equation}
\begin{aligned}
\omega_\text{D} = \frac{v_\text{m}^{2} e}{\mu \, k_\text{B} T},
\end{aligned} 
\label{eq:wD_ap}
\end{equation}

\noindent
where $e$ is the charge of an electron, $\mu$ is the free carrier mobility, $N$ is the free carrier concentration, $k_\text{B}$ is the Boltzmann constant, and $T$ is the temperature. The space charge potential factor ($M$) above the piezoelectric surface is given as 

\begin{equation}
\begin{aligned}
M \left( q_\text{m} h \right) = \frac{\epsilon_\text{g} + \epsilon_\text{p} \text{tanh} \left(q_\text{m} h\right)}{\left( \epsilon_\text{g} + \epsilon_{p} \right) \left( 1 + \text{tanh} \left( q_\text{m} h \right) \right)},
\end{aligned} 
\label{eq:M_ap}
\end{equation}

\noindent
where $\epsilon_{p}$ and $\epsilon_{g}$ are the permittivities of the piezoelectric (lithium niobate) and gap dielectric (InP), respectively. The interaction impedance ($Z_\text{m}$) of the elastic wave above the piezoelectric surface is a measure of the overlap between the electric potential and free carriers and is given as 

\begin{equation}
\begin{aligned}
Z_\text{m} \left( q_\text{m} h \right) = Z_\text{m} \left( 0 \right) \mathrm{e}^{-2 q_\text{m} h}.
\end{aligned} 
\label{eq:Zmh_ap}
\end{equation}

\noindent
The strength of the interaction impedance decays exponentially as the height of the dielectric gap is increased, with a maximum occurring at the piezoelectric surface 

\begin{equation}
\begin{aligned}
Z_\text{m} \left( 0 \right) = \frac{k^{2}}{\Omega_\text{m} \left( \epsilon_\text{g} + \epsilon_\text{p} \right)},
\end{aligned} 
\label{eq:Zm0_ap}
\end{equation}

\noindent
where $k^{2}$ is the piezoelectric coupling coefficient of the elastic mode. The introduction of the dielectric above the semiconductor thin film will perturb the fields as a result of the top dielectric permittivity ($\epsilon_{d}$). The perturbed values for the space charge potential factor and interaction impedance are given as 

\begin{equation}
\begin{aligned}
M^\prime \left( q_\text{m} h \right) = \frac{M \left( q_\text{m} h \right)}{1 + \left( \epsilon_{d} / \epsilon_{g} - 1 \right) M \left( q_\text{m} h \right)},
\end{aligned} 
\label{eq:M_prime_ap}
\end{equation}

\begin{equation}
\begin{aligned}
Z_\text{m}^\prime \left( q_\text{m} h \right) = \frac{Z_\text{m} \left( q_\text{m} h \right)}{\left[ 1 + \left( \epsilon_{d} / \epsilon_{g} - 1 \right) M \left( q_\text{m} h \right) \right]^{2}}.
\end{aligned} 
\label{eq:Zmh_prime_ap}
\end{equation}

\noindent
Considering semiconductor films of a finite thickness, the space-charge reduction factor can be expressed as 

\begin{equation}
\begin{aligned}
R = \left(\epsilon_{s} / \epsilon_{g} \right) M^\prime \left( q_\text{m} h \right) \text{tanh} \left( q_\text{m} t \right).
\end{aligned} 
\label{eq:R_ap}
\end{equation}

\noindent
The diffusion term, which comes from carrier diffusion in the semiconductor, can be expressed as 

\begin{equation}
\begin{aligned}
D = \sqrt{\frac{\omega_\text{c}}{\omega_\text{D}}} \, \frac{\text{tanh} \left( q_\text{m} t \right)}{\text{tanh} \left( \gamma_\text{d} t \right)},
\end{aligned} 
\label{eq:H_ap}
\end{equation}

\noindent
where $\gamma_\text{d}$ is related to the Debye length ($\lambda_\text{d}$) in the semiconductor 

\begin{equation}
\begin{aligned}
\gamma_\text{d} \cong \frac{\sqrt{\omega_\text{c} \omega_\text{D}}}{v_\text{m}} = \frac{1}{\lambda_\text{d}}.
\end{aligned} 
\label{eq:Ld_ap}
\end{equation}

\noindent
Finally, the conversion of this AE gain coefficient to units of (dB/cm) is given as 

\begin{equation}
\begin{aligned}
\alpha_\text{AE} \left[ \frac{dB}{cm} \right] = \frac{1 \, m}{100 \, cm} \left( \frac{10}{\text{ln} \left( 10 \right)} \right) \alpha_\text{AE} \left[ \frac{1}{m} \right].
\end{aligned}
\label{eq:alpha_conv}
\end{equation}

The expression for the AE gain from the simple theory is given as \cite{Adler_Simple_1971}  

\begin{equation}
\begin{aligned}
\alpha_\text{AE} = q_\text{m} \left[ k^{2} \frac{\gamma_{v} \Omega_\text{m} \tau}{1 + \left( \gamma_{v} \Omega_\text{m} \tau \right)^{2}} \right],
\end{aligned} 
\label{eq:simple_AE}
\end{equation}

\noindent
where the time constant ($\tau$) and non-dimensionalized velocity ($\gamma_{v}$) are given as 

\begin{equation}
\begin{aligned}
\tau = \frac{\epsilon_\text{p} + \epsilon_\text{d}}{e \mu N q_\text{m} t},
\end{aligned} 
\label{eq:simple_AE_T}
\end{equation}

\begin{equation}
\begin{aligned}
\gamma_{v}  = 1 - \frac{v_\text{d}}{v_\text{m}}.
\end{aligned} 
\label{eq:simple_AE_G}
\end{equation}

\noindent
The simple theory has a couple of limitations when compared to the normal mode theory. First, the simple theory does not include any carrier diffusion effects. Second, there is no gap height defined in the simple theory, which means the semiconductor thin film is assumed to be in direct contact with the piezoelectric substrate. The normal mode theory can be made to match the assumptions found in the simple theory by setting the temperature to zero ($T$~=~0) for the diffusion frequency in Eq.~\ref{eq:wD_ap} and by setting the gap height to zero ($h$~=~0) in all of the expressions. 

The inclusion of loss from the elastic mode into the expression for AE gain is done by representing the free ($v_\text{f} = v_\text{fr} + i \, v_\text{fi}$) and metalized ($v_\text{l} = v_\text{lr} + i \, v_\text{li}$) phase velocities as complex values. The real component of the free velocity is simply the phase velocity of the given elastic mode ($v_\text{fr} = v_\text{m}$). The real component of the metalized velocity can be determined from the definition of the piezoelectric coupling coefficient. The imaginary part of the free and metalized velocities can be computed from the complex propagation constant of the elastic mode ($q_\text{f,l} = q_\text{m} - i \, \alpha_\text{f,l}/2$) and is given as 

\begin{equation}
\begin{aligned}
v_\text{fi,li} = \frac{\Omega_\text{m} \alpha_\text{f,l}}{2 \, q_\text{m}^{2}}.
\end{aligned} 
\label{eq:complex_kv}
\end{equation}

\noindent
The full expression for the AE gain from the simple theory with the inclusion of losses from the elastic mode is given as \cite{hackett2019amp}

\begin{equation}
\begin{aligned}
\alpha_\text{AE} = q_\text{m} \left[ k^{2} \frac{\gamma_{v} \Omega_\text{m} \tau}{1 + \left( \gamma_{v} \Omega_\text{m} \tau \right)^{2}} + \frac{2 \, v_\text{li}}{v_\text{fr}} \frac{\left(\gamma_{v} \Omega_\text{m} \tau\right)^{2}}{1 + \left( \gamma_{v} \Omega_\text{m} \tau \right)^{2}} + \frac{2 \, v_\text{fi}}{v_\text{fr}} \frac{1}{1 + \left( \gamma_{v} \Omega_\text{m} \tau \right)^{2}} \right].
\end{aligned} 
\label{eq:simple_AE_wloss}
\end{equation}

\noindent
This AE gain can again be converted into units of (dB/cm) using Eq.~\ref{eq:alpha_conv}.

\begin{figure*}[b]
\centering
\includegraphics[width=1\linewidth]{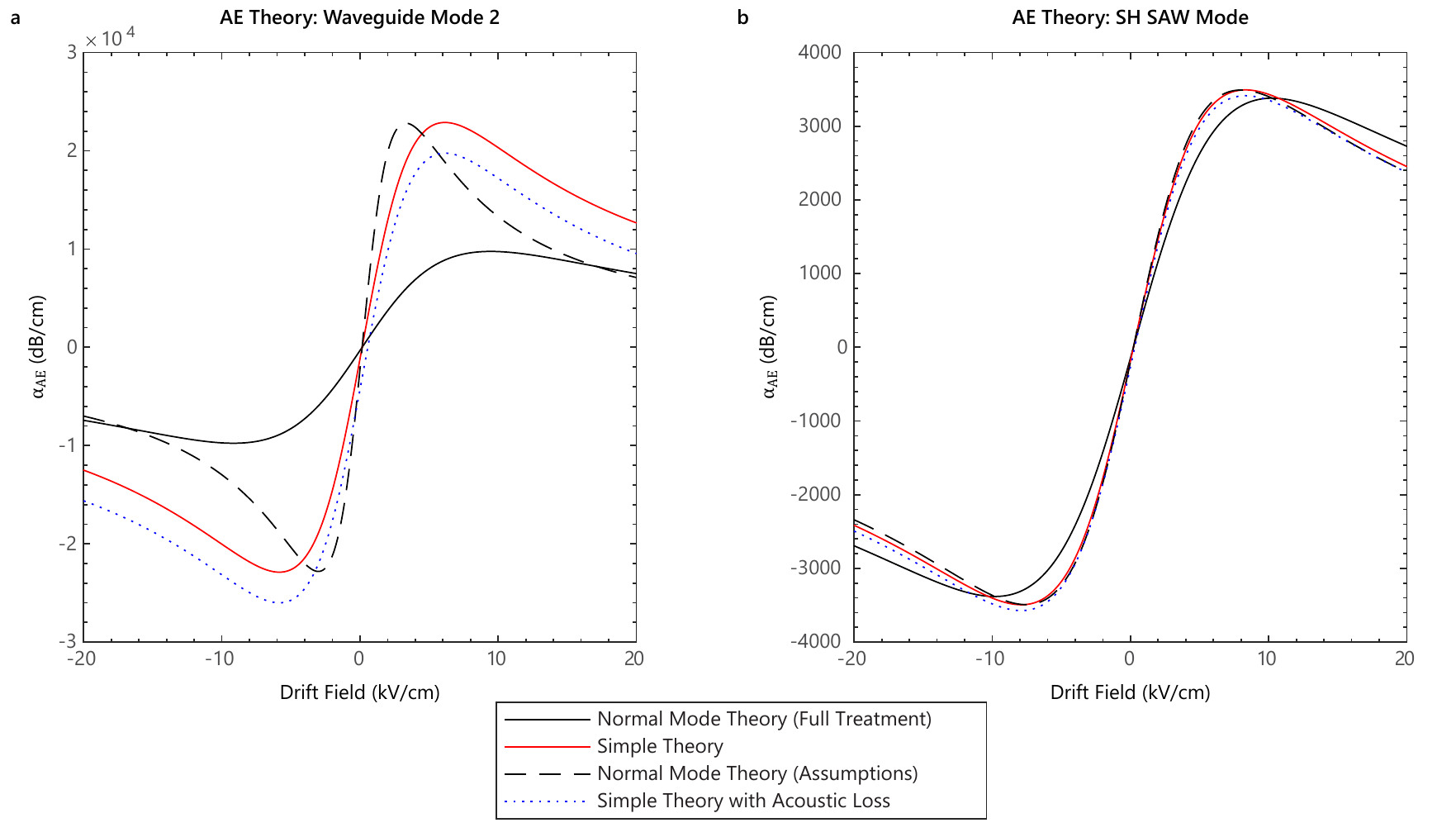}
\caption{Calculated AE gain curves as a function of applied drift field. The normal mode theory and simple theory are plotted for (a) elastic waveguide mode 2 (backward intra-modal)---elastic mode used in the main manuscript---and (b) SH SAW mode (forward inter-modal).}
\label{fig:AEcurvecomp}
\end{figure*}

A comparison of these AE theories for an example backward intra-modal and forward inter-modal Brillouin scattering process is shown in Fig.~\ref{fig:AEcurvecomp}. The four different AE theories include: the full treatment of the normal mode theory with no assumptions, the simple theory, the normal mode theory with the same assumptions as the simple theory, and the simple theory including the losses from the elastic mode. The elastic mode from the main manuscript (waveguide mode 2) is chosen as the example for backward intra-modal Brillouin scattering (Fig.~\ref{fig:AEcurvecomp}a). The full set of parameters for the AE calculations are $\Omega_\text{m}$/2$\pi$~=~8.78~GHz, $\lambda_\text{m}$~=~0.35~$\mu$m, $v_\text{m}$~=~3072~m/s, $\alpha_\text{f,l}$~=~71800~m\textsuperscript{-1}, $k^{2}$~=~5.87~\%, $t$~=~0.1~$\mu$m, $h$~=~5~nm, $\epsilon_\text{p}$~=~29.2$\epsilon_{0}$, $\epsilon_\text{s}$~=~13.3$\epsilon_{0}$, $\epsilon_\text{g}$~=~12.5$\epsilon_{0}$, $\epsilon_\text{d}$~=~$\epsilon_{0}$, $T$~=~293.15~K, $\mu$~=~2000~cm\textsuperscript{2}/Vs, $N$~=~1~x~10\textsuperscript{16}~cm\textsuperscript{-3}, and $Q_\text{m}$~=~250. For waveguide mode 2, all four AE theories show significant differences. The large difference between the full treatment of the normal mode theory and simple theory can be attributed to significant carrier diffusion effects when operating at this small phonon wavelength. The two conditions for minimal diffusion effects are $\Omega_\text{m}^{2} << \omega_\text{c} \omega_\text{D}$ and $R \omega_\text{c} / \Omega_\text{m} >> D$. Both conditions relate the phonon frequency to constants that are a function of the semiconductor properties and elastic mode's phase velocity. Assuming the elastic mode has minimal dispersion, then the amount of carrier diffusion is proportional to the phonon frequency. Most reported AE amplifiers operate at phonon frequencies in the range of MHz up to 1~GHz and the carrier diffusion is minimal in this frequency range. Backward intra-modal Brillouin scattering processes, however, have phase-matched elastic modes at frequencies at least an order of magnitude larger, which no longer satisfy the two conditions for negligible diffusion effects. Additionally, the phase-matched phonon wavelength is now the same order of magnitude as the semiconductor film thickness. As a result, the phonon wave vector-thickness product ($q_\text{m} t$) is too large for the small angle approximation in Eq.~\ref{eq:AE_gain_exp}. The full evaluation of the hyperbolic tangent in the normal mode theory (with assumptions) accounts for the discrepancy with the simple theory. Both theories have similar magnitudes in AE gain, but the difference in shape with respect to the applied drift field is a result of the invalid small angle approximation for this $q_\text{m} t$ product. Finally, the addition of losses from the elastic mode to the simple theory shifts the AE gain curve, but does not significantly change the shape of the curve. With the conservative estimate of 250 for the mechanical quality factor, the additional loss terms added to Eq.~\ref{eq:simple_AE_wloss} have a relative magnitude of 6.8~\%. Since the carrier diffusion effects are significant for backward intra-modal Brillouin scattering in the InGaAsP-LN-Si physical system, the full treatment of the normal mode theory was chosen as the AE theory to use in the main manuscript and supplement to illustrate the different regimes of AE Brillouin dynamics. If the elastic mode of an acoustoelectrically enhanced Brillouin device exhibits these losses in experiment, the full treatment of the normal mode theory would need to be expanded to include the effects of losses in the elastic mode for the most accurate AE theoretical model. Additionally, the acoustoelectric coupling defined using the Hamiltonian formulation for the acoustoelectric dynamics (see Supplementary Section~\ref{sec:AEdynamics}) is most similar to the full treatment of the normal mode theory minus the diffusion effects. Both the normal mode theory and Hamiltonian formulation take into consideration the cross-sectional overlap between the elastic wave and free carriers including the gap dielectric between the piezoelectric surface and semiconductor thin film. 

For forward inter-modal Brillouin scattering, the phase-matched phonon wavelength is similar to those found in traditional AE delayline amplifiers using the Rayleigh and SH SAW modes. As an example, the AE gain calculations for the best simulated SH SAW mode are plotted in Fig.~\ref{fig:AEcurvecomp}b. The full set of parameters for the AE calculations are $\Omega_\text{m}$/2$\pi$~=~1.24~GHz, $\lambda_\text{m}$~=~3.45~$\mu$m, $v_\text{m}$~=~4289~m/s, $\alpha_\text{f,l}$~=~1800~m\textsuperscript{-1}, $k^{2}$~=~8.84~\%, $t$~=~0.2~$\mu$m, $h$~=~5~nm, $\epsilon_\text{p}$~=~43.6$\epsilon_{0}$, $\epsilon_\text{s}$~=~13.3$\epsilon_{0}$, $\epsilon_\text{g}$~=~12.5$\epsilon_{0}$, $\epsilon_\text{d}$~=~$\epsilon_{0}$, $T$~=~293.15~K, $\mu$~=~2000~cm\textsuperscript{2}/Vs, $N$~=~1~x~10\textsuperscript{16}~cm\textsuperscript{-3}, and $Q_\text{m}$~=~1000. With a lower phonon frequency and longer phonon wavelength, the two conditions for minimal carrier diffusion effects in the normal mode theory are valid. The only significant difference between the full treatment of the normal mode theory and the simple theory is the introduction of the 5~nm InP gap dielectric between the InGaAsP and lithium niobate. The longer phonon wavelength also makes the small angle approximation for the $q_\text{m} t$ product valid. In this case, the normal mode theory with the additional assumptions is in good agreement with the simple theory. With a estimated higher mechanical quality factor at these lower phonon frequencies, the additional loss terms added to the simple theory only have a relative magnitude of 1.1~\%. This amount of loss has a minimal effect on the AE gain curve, which means all four AE theories have good agreement with each other.

\subsection{Acoustoelectric Noise Figure Calculation}

We next estimate the levels of intrinsic acoustic noise relative to thermal fluctuations in the acoustoelectric gain medium for our analysis of the acoustoelectric enhanced optical Brillouin noise figure in Supplementary Section~\ref{sec:noise}. To do this, we analyze the predicted noise figure in the absence of acoustic losses.  For the InGaAsP-LN-Si physical system, the internal noise figure resulting from the noise source in the semiconductor coupling to the elastic wave is given as \cite{kino1973noise}

\begin{equation}
\begin{aligned}
F_\text{n} = 1 + \frac{t \, e^{2} D_\text{c} \, N \left| q_\text{d} \right|^{2} Z_\text{m} \left( q_\text{m} h \right) \mathrm{e}^{-2 q_\text{m} h}}{k_\text{B} T \alpha_\text{AE}} \left( \frac{\mathrm{e}^{\alpha_\text{AE} l} - 1}{\mathrm{e}^{\alpha_\text{AE} l}} \right).
\end{aligned}
\label{eq:F_noise_eq}
\end{equation}

\noindent
Here, $D_\text{c}$ is a diffusion coefficient, $q_\text{d}$ is the perturbed carrier wave propagation constant, and the interaction impedance, $Z_\text{m}$, is the same as in Eq.~\ref{eq:Zmh_ap}-\ref{eq:Zm0_ap}. The diffusion coefficient can be expressed as $D_\text{c} = D_\text{TH} + D_\text{TR}$, where the thermal diffusion constant ($D_\text{TH}$) and trapping effects ($D_\text{TR}$) are given as

\begin{figure*}[b]
\centering
\includegraphics[width=1\linewidth]{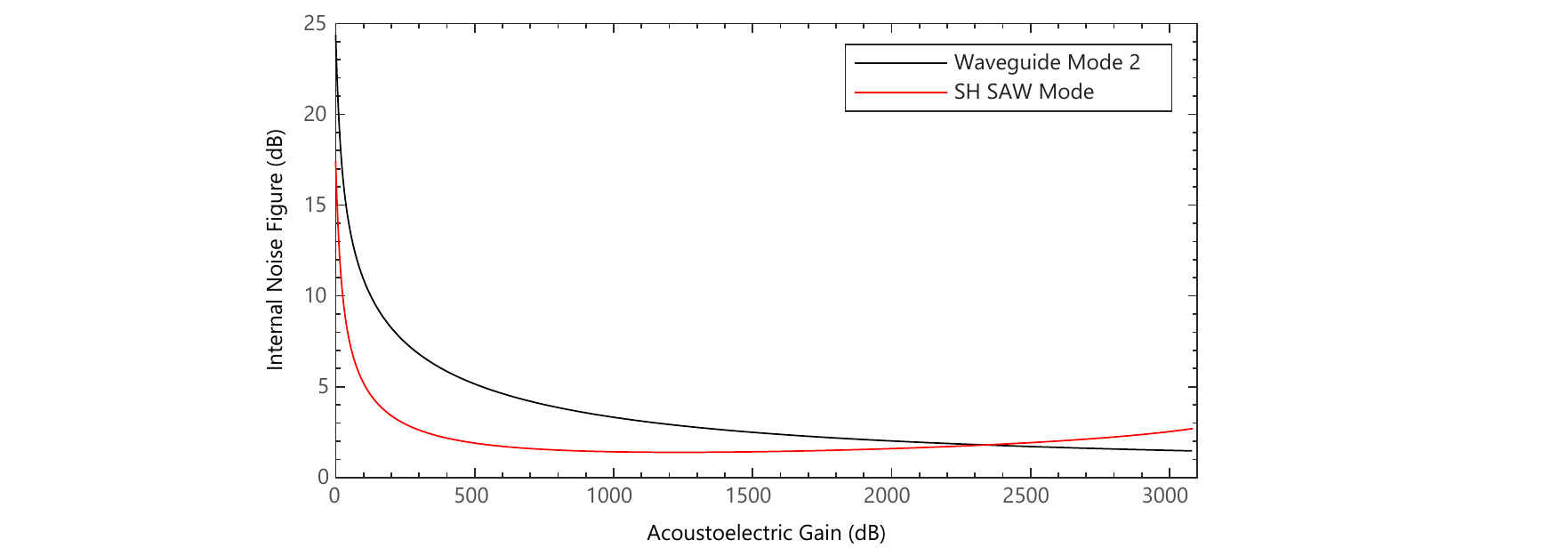}
\caption{Calculated internal noise figure as a function of acoustoelectric gain for a 1~cm long device, in the absence of acoustic loss. The acoustoelectric gain values for the elastic waveguide mode 2 (backward intra-modal)---elastic mode used in the main manuscript---and the SH SAW mode (forward inter-modal) are taken from Fig.~\ref{fig:AEcurvecomp} using the full treatment of the normal mode theory. The range of plotted acoustoelectric gain values corresponds to the enhanced Brillouin regime illustrated in Fig.~\ref{fig:AEregions}.}
\label{fig:nf}
\end{figure*}

\begin{equation}
\begin{aligned}
D_\text{TH} = \frac{\mu \, k_\text{B} T}{e},
\end{aligned}
\label{eq:DTH}
\end{equation}

\begin{equation}
\begin{aligned}
D_\text{TR} = \frac{f_\text{TR} \left( 1 - f_\text{TR} \right) v_\text{d}^{2} \, \tau_\text{TR}}{1 + \left( \Omega_\text{m} \tau_\text{TR} \right)}.
\end{aligned}
\label{eq:DTR}
\end{equation}

\noindent
The factor $f_\text{TR}$ represents the fraction of untrapped carriers and $\tau_\text{TR}$ is the carrier trapping relaxation time, which can be estimated as $\tau_\text{TR}$~=~1/$\Omega_\text{m}$. For the case where $\omega_\text{c} d / v_\text{d} >> 1$ and $q_\text{d} h << 1$, the carrier wave propagation constant can be approximated as \cite{kino1973noise}

\begin{table*}[t]
    \caption{Table summarizing the mode triplets for the acoustoelectrically enhanced backward intra-modal Brillouin dynamics in the InGaAsP-LN-Si physical system. The AE gain value is calculated using the normal mode theory and evaluated at a drift field of 2~kV/cm. For each elastic mode, the rows with the best simulated results are highlighted in gray.}
    \centering
    \ra{1.2}
    \begin{tabularx}{\textwidth}{@{} C @{} E @{} A @{} F @{} F @{} F @{} G @{} G @{} H @{} H @{} H @{} G @{} H @{}}
    \toprule
    {\makecell[t]{Optical\\ Pair}} & {\makecell[t]{Acoustic\\ Mode}} & {\makecell{$\beta$}} & {\makecell[t]{$t$\\ ($\mu$m)}} & {\makecell[t]{$w$\\ ($\mu$m)}} & {\makecell[t]{$\lambda_\text{m}$\\ ($\mu$m)}} & {\makecell[t]{$\Omega_\text{m}/2\pi$\\ (GHz)}} & {\makecell[t]{$v_\text{m}$\\ (m/s)}} & {\makecell[t]{$\left|g_\text{pe}\right|$\\ ((rad/s)$\sqrt{\text{m}}$)}} & {\makecell[t]{$\left|g_\text{rp}\right|$\\ ((rad/s)$\sqrt{\text{m}}$)}} & {\makecell[t]{$\left|g_{0}\right|$\\ ((rad/s)$\sqrt{\text{m}}$)}} & {\makecell[t]{$k^2$\\ (\%)}} & {\makecell[t]{$\alpha_\text{AE}$\\ (dB/cm)}} \\
    \midrule
        TE\textsubscript{0} / TE\textsubscript{0} & WG 1 & 0\textdegree & 0.1 & 0.9 & 0.356 & 7.597 & 2704 & 2228 & -358 & 1843 & 0.65 & 413.7
        \\
        TE\textsubscript{0} / TE\textsubscript{0} & WG 1 & 0\textdegree & 0.15 & 0.75 & 0.328 & 7.346 & 2410 & 2352 & -707 & 1626 & 0.23 & 150.0
        \\
        TE\textsubscript{0} / TE\textsubscript{0} & WG 1 & 0\textdegree & 0.2 & 0.75 & 0.302 & 7.558 & 2283 & 2545 & -643 & 1876 & 0.04 & 27.2
        \\
        TE\textsubscript{0} / TE\textsubscript{0} & WG 1 & 0\textdegree & 0.3 & 0.65 & 0.281 & 7.915 & 2224 & 2927 & -611 & 2304 & 0.02 & 14.1
        \\
    \midrule
    \rowcolor{mgray}
        TE\textsubscript{0} / TE\textsubscript{0} & WG 1 & 90\textdegree & 0.1 & 0.95 & 0.347 & 7.896 & 2740 & 1586 & -236 & 1350 & 4.22 & 2093.7
        \\
        TE\textsubscript{0} / TE\textsubscript{0} & WG 1 & 90\textdegree & 0.15 & 0.75 & 0.324 & 7.463 & 2418 & 2122 & -626 & 1496 & 1.29 & 649.5
        \\
        TE\textsubscript{0} / TE\textsubscript{0} & WG 1 & 90\textdegree & 0.2 & 0.75 & 0.301 & 7.604 & 2289 & 2362 & -601 & 1761 & 0.69 & 357.1
        \\
        TE\textsubscript{0} / TE\textsubscript{0} & WG 1 & 90\textdegree & 0.3 & 0.65 & 0.280 & 7.947 & 2225 & 2835 & -592 & 2244 & 0.19 & 100.9
        \\
    \midrule
    \rowcolor{mgray}
        TE\textsubscript{0} / TE\textsubscript{0} & WG 2 & 0\textdegree & 0.1 & 1.1 & 0.350 & 8.777 & 3072 & 7294 & 848 & 7943 & 5.87 & 3725.9
        \\
        TE\textsubscript{0} / TE\textsubscript{0} & WG 2 & 0\textdegree & 0.15 & 0.75 & 0.328 & 8.401 & 2755 & 7861 & 1461 & 9281 & 1.63 & 1054.2
        \\
        TE\textsubscript{0} / TE\textsubscript{0} & WG 2 & 0\textdegree & 0.2 & 0.7 & 0.305 & 8.579 & 2617 & 5783 & 1158 & 6906 & 0.21 & 141.0
        \\
        TE\textsubscript{0} / TE\textsubscript{0} & WG 2 & 0\textdegree & 0.3 & 0.6 & 0.285 & 9.079 & 2588 & 3200 & 307 & 3502 & 0.01 & 6.9
        \\
    \midrule
        TE\textsubscript{0} / TE\textsubscript{0} & WG 2 & 90\textdegree & 0.1 & 1.15 & 0.343 & 9.067 & 3110 & 6605 & 967 & 7572 & 0.62 & 305.7
        \\
        TE\textsubscript{0} / TE\textsubscript{0} & WG 2 & 90\textdegree & 0.15 & 0.8 & 0.321 & 8.545 & 2743 & 7518 & 1540 & 9059 & 0.03 & 15.0
        \\
        TE\textsubscript{0} / TE\textsubscript{0} & WG 2 & 90\textdegree & 0.2 & 0.75 & 0.301 & 8.672 & 2610 & 5504 & 1173 & 6676 & 0.01 & 5.1
        \\
        TE\textsubscript{0} / TE\textsubscript{0} & WG 2 & 90\textdegree & 0.3 & 0.6 & 0.285 & 9.082 & 2588 & 3043 & 383 & 3426 & 0.06 & 31.4
        \\
    \midrule
    \rowcolor{mgray}
        TE\textsubscript{0} / TE\textsubscript{0} & WG 3 & 0\textdegree & 0.15 & 0.8 & 0.325 & 10.453 & 3397 & 2186 & 161 & 2345 & 0.82 & 524.2
        \\
        TE\textsubscript{0} / TE\textsubscript{0} & WG 3 & 0\textdegree & 0.2 & 0.75 & 0.302 & 10.870 & 3283 & 2345 & 182 & 2527 & 0.66 & 437.8
        \\
    \midrule
        TE\textsubscript{0} / TE\textsubscript{0} & WG 3 & 90\textdegree & 0.15 & 0.8 & 0.321 & 10.611 & 3406 & 1978 & 113 & 2091 & 0.41 & 202.1
        \\
        TE\textsubscript{0} / TE\textsubscript{0} & WG 3 & 90\textdegree & 0.2 & 0.7 & 0.304 & 11.205 & 3406 & 2086 & 250 & 2332 & 0.00 & 0.0
        \\
    \bottomrule
    \end{tabularx}
    \label{table:AEBmodes_Intra}
\end{table*}

\begin{equation}
\begin{aligned}
q_\text{d} \approx -i \, \frac{\Omega_\text{m} \left( \epsilon_\text{g} + \epsilon_\text{p} \right)}{e \mu N t}.
\end{aligned}
\label{eq:qm_approx}
\end{equation}

\noindent
Finally, the internal noise figure can be expressed in decibels through the conversion $F_\text{n} \left[ \mathrm{dB} \right] = 10 \cdot \mathrm{log}_{10} \left( F_\text{n} \right)$.

Using the previous two example elastic modes (Fig.~\ref{fig:AEcurvecomp}), the internal noise figure is plotted in Fig.~\ref{fig:nf} as a function of the acoustoelectric gain using the full treatment of the normal mode theory for a 1~cm long device in the absence of acoustic loss. For both elastic modes, the fraction of untrapped carriers is estimated to be $f_\text{TR} = 0.95$. For amorphous InSb, the fraction of untrapped carriers is approximately 0.8, so for our nearly single crystalline epitaxial films we expect $f_\text{TR}$ to be substantially closer to 1, with 0.95 being a conservative estimate. This range of acoustoelectric gain corresponds to operating in the enhanced Brillouin regime (shown in Fig.~\ref{fig:AEregions}). The noise figure is initially large when entering the enhanced Brillouin regime from the lossy Brillouin regime ($\alpha_\text{AE} > 0$). As the acoustoelectric gain is increased, the noise figure decreases to a minimum value before increasing again as the diffusion coefficient begins to dominate at high carrier drift velocities. In the case of the elastic waveguide mode 2, the elastic mode used in the main manuscript, the noise figure does not reach its minimum value in the enhanced Brillouin regime. This means that the internal noise figure will continue to decrease for larger acoustoelectric gains up until the coherent Brillouin limit is reached.

\subsection{Mode Triplet Simulation Results}

For both backward intra-modal and forward inter-modal Brillouin scattering, an iterative design process is implemented using all the previously described metrics in this section of the supplement. First, the waveguide geometry is set and simulated to determine the optical modes, which dictates the phase-matched phonon wavelength. The elastic modes are then simulated with electrically free and grounded boundary conditions at the InGaAsP and lithium niobate interface. The piezoelectric coupling coefficient is then calculated using the difference in eigenfrequencies for each elastic mode. Finally, the optomechanical coupling coefficient is computed using the optical and elastic field profiles and the piezoelectric coupling coefficient is used to compute the AE gain using the normal mode theory. This design process is repeated for different waveguide geometries until a set of modes are identified with both high AE gain and optomechanical coupling. Tables~\ref{table:AEBmodes_Intra}~and~\ref{table:AEBmodes_Inter} show the simulation results with the rows containing the best performance for each elastic mode highlighted in gray. 

\begin{table*}[t]
    \caption{Table summarizing the mode triplets for the acoustoelectrically enhanced forward inter-modal Brillouin dynamics in the InGaAsP-LN-Si physical system. The AE gain value is calculated using the normal mode theory and evaluated at a drift field of 2~kV/cm. For each elastic mode, the rows with the best simulated results are highlighted in gray.}
    \centering
    \ra{1.2}
    \begin{tabularx}{\textwidth}{@{} C @{} E @{} A @{} F @{} F @{} F @{} G @{} G @{} H @{} H @{} H @{} G @{} H @{}}
    \toprule
    {\makecell[t]{Optical\\ Pair}} & {\makecell[t]{Acoustic\\ Mode}} & {\makecell{$\beta$}} & {\makecell[t]{$t$\\ ($\mu$m)}} & {\makecell[t]{$w$\\ ($\mu$m)}} & {\makecell[t]{$\lambda_\text{m}$\\ ($\mu$m)}} & {\makecell[t]{$\Omega_\text{m}/2\pi$\\ (GHz)}} & {\makecell[t]{$v_\text{m}$\\ (m/s)}} & {\makecell[t]{$\left|g_\text{pe}\right|$\\ ((rad/s)$\sqrt{\text{m}}$)}} & {\makecell[t]{$\left|g_\text{rp}\right|$\\ ((rad/s)$\sqrt{\text{m}}$)}} & {\makecell[t]{$\left|g_{0}\right|$\\ ((rad/s)$\sqrt{\text{m}}$)}} & {\makecell[t]{$k^2$\\ (\%)}} & {\makecell[t]{$\alpha_\text{AE}$\\ (dB/cm)}} \\
    \midrule
        TE\textsubscript{0} / TE\textsubscript{2} & Rayleigh & 0\textdegree & 0.15 & 1.55 & 4.066 & 0.865 & 3516 & 15.7 & 15.9 & 31.5 & 2.21 & 299.7
        \\
        TE\textsubscript{0} / TE\textsubscript{2} & Rayleigh & 0\textdegree & 0.2 & 1.3 & 2.902 & 1.195 & 3469 & 21.3 & 12.4 & 33.6 & 3.08 & 415.0
        \\
    \rowcolor{mgray}
        TE\textsubscript{0} / TE\textsubscript{2} & Rayleigh & 0\textdegree & 0.3 & 1.1 & 2.102 & 1.619 & 3402 & 17.5 & 81.1 & 97.7 & 4.91 & 630.3
        \\
    \midrule
        TE\textsubscript{0} / TE\textsubscript{2} & Rayleigh & 90\textdegree & 0.15 & 1.65 & 4.696 & 0.802 & 3766 & 13.5 & 13.8 & 27.3 & 0.16 & 22.4
        \\
        TE\textsubscript{0} / TE\textsubscript{2} & Rayleigh & 90\textdegree & 0.2 & 1.35 & 3.192 & 1.166 & 3723 & 27.8 & 17.4 & 45.3 & 0.22 & 30.9
        \\
        TE\textsubscript{0} / TE\textsubscript{2} & Rayleigh & 90\textdegree & 0.3 & 1.15 & 2.332 & 1.570 & 3661 & 23.1 & 63.3 & 86.4 & 2.05 & 259.3
        \\
    \midrule
        TE\textsubscript{0} / TM\textsubscript{1} & Rayleigh & 0\textdegree & 0.3 & 0.8 & 2.303 & 1.495 & 3444 & -19.2 & 133.3 & 114.3 & 3.67 & 434.4
        \\
    \midrule
        TE\textsubscript{0} / TM\textsubscript{1} & Rayleigh & 90\textdegree & 0.3 & 0.8 & 2.412 & 1.536 & 3706 & -15.7 & 129.3 & 113.6 & 1.40 & 172.6
        \\
    \midrule
        TE\textsubscript{0} / TE\textsubscript{1} & SH SAW & 0\textdegree & 0.15 & 1.05 & 5.180 & 0.702 & 3636 & 42.4 & 49.4 & 91.6 & 0.00 & 0.0
        \\
        TE\textsubscript{0} / TE\textsubscript{1} & SH SAW & 0\textdegree & 0.2 & 0.85 & 3.170 & 1.129 & 3577 & 49.9 & 75.6 & 123.7 & 0.01 & 1.3
        \\
    \rowcolor{mgray}
        TE\textsubscript{0} / TE\textsubscript{1} & SH/Ray & 0\textdegree & 0.3 & 0.7 & 2.314 & 1.418 & 3281 & 374.6 & 1049.3 & 1419.8 & 1.43 & 169.3
        \\
    \midrule
        TE\textsubscript{0} / TE\textsubscript{1} & SH SAW & 90\textdegree & 0.15 & 1.1 & 5.848 & 0.773 & 4519 & 36.2 & 44.7 & 80.9 & 2.91 & 342.6
        \\
        TE\textsubscript{0} / TE\textsubscript{1} & SH SAW & 90\textdegree & 0.2 & 1.0 & 4.759 & 0.935 & 4449 & 57.7 & 68.9 & 126.6 & 2.81 & 298.0
        \\
        TE\textsubscript{0} / TE\textsubscript{1} & SH/Ray & 90\textdegree & 0.3 & 0.75 & 2.689 & 1.373 & 3691 & 113.2 & 396.2 & 509.4 & 1.31 & 149.5
        \\
    \midrule
        TE\textsubscript{0} / TM\textsubscript{0} & SH SAW & 0\textdegree & 0.2 & 1.75 & 3.123 & 1.123 & 3507 & 34.8 & 20.7 & 58.9 & 0.12 & 15.3
        \\
        TE\textsubscript{0} / TM\textsubscript{0} & SH SAW & 0\textdegree & 0.3 & 1.5 & 4.673 & 0.772 & 3607 & 16.3 & 4.5 & 22.5 & 0.02 & 1.3
        \\
    \midrule
    \rowcolor{mgray}
        TE\textsubscript{0} / TM\textsubscript{0} & SH SAW & 90\textdegree & 0.2 & 1.75 & 3.453 & 1.242 & 4289 & 46.0 & 7.7 & 50.9 & 8.84 & 1166.1
        \\
        TE\textsubscript{0} / TM\textsubscript{0} & SH SAW & 90\textdegree & 0.3 & 1.5 & 4.972 & 0.889 & 4418 & 47.4 & 9.3 & 55.6 & 7.56 & 541.6
        \\
    \bottomrule
    \end{tabularx}
    \label{table:AEBmodes_Inter}
\end{table*}


\section{Feasibility of Continuous Operation for Acoustoelectric Brillouin Devices}\label{sec:CW}
    
Acoustoelectric devices for radiofrequency signal processing applications have historically had difficulty operating continuously with large gain, primarily due to effects related to thermal dissipation of the current required to drive the acoustoelectric current. \cite{lakin1969100, coldren1971amp, hackett2019amp, hackett2021amp}. Joule heating arises in acoustoelectric devices when a  voltage is applied to the semiconductor. If the thermal conductivity out of the device is not sufficient, this resistive heating can cause variations in the device response as temperature increases, and ultimately can lead to thermal runaway and even irreversible damage. The thermal behavior depends on the semiconductor resistance and geometry, the bias field required to achieve the necessary acoustic gain, and the material thermal properties of the semiconductor and all the other layers, including the substrate. However, in contrast to radiofrequency acoustic wave amplifiers that have been previously experimentally demonstrated, acoustoelectric-Brillouin devices have different metrics and geometries that make continuous operation significantly less challenging. 

An acoustic wave amplifier for radiofrequency signal processing applications must have significant levels of terminal gain, meaning that the acoustic gain must be large enough to overcome all acoustic losses and then provide a significant level of additional acoustic gain. This is accomplished by increasing the applied drift field, which leads to larger dissipated power and concomitant heating. Ultimately, the ability to operate continuously and to reach peak acoustoelectric gain is limited by thermal dissipation under these conditions. However, with respect to the acoustoelectric-Brillouin devices described in this work, improving the performance and observing new dynamical regimes only require that the applied drift field provides enough acoustic gain to overcome the losses inherent to the acoustoelectric effect and the intrinsic phonon losses. This requires substantially less gain---and thus substantially less heat---than achieving net terminal gain in radiofrequency acoustic amplifier systems.

To illustrate the feasibility of continuous operation, we analyze the heat generated and temperature increase for an example device, using the lithium niobate-on-silicon and In$_{0.712}$Ga$_{0.288}$As$_{0.625}$P$_{0.375}$ platform introduced in this work. We assume an interaction length of 1 cm, a semiconductor width, $w$, of 0.5-1.5 \textmu m, a semiconductor thickness, $t$, of 100-300 nm, an electromechanical coupling coefficient, $k^2$, of 1-9\%, and an acoustic wavelength of approximately 350 nm (corresponding to a phonon frequency of $\Omega_m$/2$\pi$ = 7-11 GHz). A phononic quality factor of 250 (assumed for the prior calculations) corresponds to a phonon propagation loss of 0.11 dB/$\Lambda$, where $\Lambda$ is the acoustic wavelength. Under these conditions, it is necessary to achieve an acoustic gain of 0.11 dB/$\Lambda$ or higher, which corresponds to 3100 dB/cm. The bulk semiconductor mobility for In$_{0.712}$Ga$_{0.288}$As$_{0.625}$P$_{0.375}$ is approximately 5000 cm$^2$/V$\cdot$s. Based on our previous work \cite{hackett2021amp}, we expect the as-processed thin film mobility in these type of acoustoelectric-Brillouin devices to be approximately 2000 cm$^2$/V$\cdot$s and the expected carrier concentration is $1\times10^{16}$ cm$^{-3}$ (see Supplementary Section \ref{sec:SIdesign} for more details). This carrier concentration and mobility lead to a conductivity-thickness product, $\sigma t$, that is 32-96 \textmu S, which is one of the key factors that determine the strength of the external electric field required to reach a given level of gain (see, for example, Supplementary Information section \ref{AEGainCalc}) as well as the amount of Joule heating that electric field produces to achieve that gain.  

\begin{figure*}[t]
\centering
\includegraphics[width=1\linewidth]{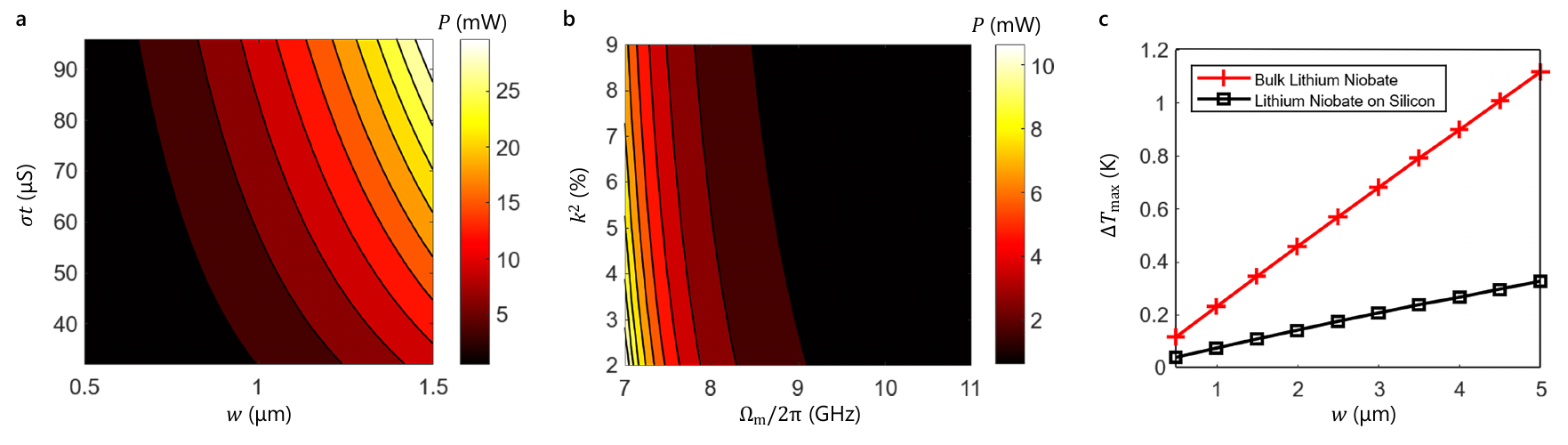}
\caption{Contour plots of the dissipated DC power as a function of (a) $\sigma t$ and $w$ and (b) $k^2$ and $\Omega$/2$\pi$. (c) $\Delta T_{\text{max}}$ as a function of $w$ for a 100 nm thick and 1 cm long In$_{0.712}$Ga$_{0.288}$As$_{0.625}$P$_{0.375}$ semiconductor layer on a bulk lithium niobate substrate or thin film lithium niobate on silicon substrate. The semiconductor acts as the heat source for the finite element method model according to the dissipated DC power.}
\label{fig:thermal1}
\end{figure*}

From these values, we can estimate the dissipated power. Figure \ref{fig:thermal1}a shows the dissipated power required to achieve an acoustic gain of 3100 dB/cm as a function of $\sigma t$ and $w$ for $\Omega$/2$\pi$ = 8.78 GHz and $k^2$ = 5.87\%. Figure \ref{fig:thermal1}b shows the dissipated power as a function of $k^2$ and $\Omega_m$/2$\pi$ for a semiconductor width of 1.1 \textmu m and $\sigma t$ = 32 \textmu S. The dissipated power varies depending on the device configuration, but can be 1 mW or less, which is significantly less than previous work where large terminal gain was achieved \cite{hackett2019amp, hackett2021amp}. Low dissipated DC powers are achieved due to two factors. One factor is the device geometry and semiconductor material parameters have been optimized to make the device highly resistive. The second factor is that $\Omega_m$ and $k^2$ are both larger, resulting in larger acoustic gain being produced with smaller bias fields.

Another aspect that should contribute to the feasibility of continuous operation in acoustoelectric Brillouin devices is that the interactions take place in a waveguide with small cross-sectional width of approximately 0.5-1.5 \textmu m. This makes the device more resistive and enables more efficient lateral heat transport by increasing the surface to volume ratio compared to wider devices. 

Beyond reducing the power dissipated, the temperature increase is also decreased by using a piezoelectric compound substrate that consists of a lithium niobate film on bulk silicon as opposed to a bulk lithium niobate substrate, both of which were demonstrated in recent work \cite{hackett2021amp}. The thermal conductivity of lithium niobate is 4.6 W/m$\cdot$K \cite{wong2002LNproperties} while the thermal conductivity of silicon is 150 W/m$\cdot$K \cite{glassbrenner1964thermal}. Therefore, the 30$\times$ increase of thermal conductivity of silicon over lithium niobate leads to a significantly smaller temperature rise for a given dissipated power. Although terminal gain was not achieved, continuous operation of an acoustic wave amplifier has been experimentally demonstrated for both an acoustic waveguide \cite{coldren1971wgamp} and a thin film lithium niobate on silicon substrate \cite{hackett2019amp}.  Continuous operation and terminal gain has been achieved in a system using a similar approach, in particular, leveraging a highly thermally conductive sapphire substrate \cite{lakin1969100}. We also reiterate that only net phonon gain, as opposed to terminal gain (which must overcome all acoustic, electronic, and transducer losses), is necessary to enhance Brillouin optomechanical interactions via the acoustoelectric effect---a much less stringent requirement.

A finite element method model was used to model conductive and convective heat transfer for this exemplary device operating in air. We model the 100 nm thick, 1.1 \textmu m wide, and 1 cm long In$_{0.712}$Ga$_{0.288}$As$_{0.625}$P$_{0.375}$ semiconductor film on a thin film lithium niobate (5 \textmu m thick) on silicon substrate or a bulk lithium niobate substrate. The In$_{0.712}$Ga$_{0.288}$As$_{0.625}$P$_{0.375}$ density is 5204 kg/cm$^3$, the heat capacity at constant pressure is 335 J/kg$\cdot$K, the thermal conductivity is 4.35 W/m$\cdot$K, and the coefficient of thermal expansion is 5.26$\times$10$^{-6}$ 1/K \cite{adachi1992IIIVsemi}. Figure \ref{fig:thermal1}c shows the maximum temperature difference $\Delta T_{\text{max}}$, defined as the difference between the maximum steady-state temperature and the ambient temperature, as a function of $w$ for the case of a bulk lithium niobate substrate and a lithium niobate film on a silion substrate. For a semiconductor width of 1.1 \textmu m on the thin film lithium niobate on silicon substrate, $\Delta T_{\text{max}}$~=~0.08~K, which is approximately 200X smaller than the temperature rise associated with achieving terminal gain in our recent acoustic wave amplifiers \cite{hackett2021amp}, where the maximum gain applied was limited by needing to avoid significant temperature increases. In conventional acoustoelectric devices, predicting device behavior is complicated by Joule heating. For example, heating leads to a temperature-dependent semiconductor conductivity which modifies the acoustic gain. The results shown here suggest that these thermal effects do not need to be considered for acoustoelectric-Brillouin devices, at least in the parameter space explored in this work, as the dissipated DC power is relatively small and thermal dissipation is effective.

\section{Acoustoelectric  dynamics} \label{sec:AEdynamics}

In this section, we formulate a model of acoustoelectric dynamics within photonic and phononic waveguide structures of arbitrary, but translationally invariant, cross-sectional geometry. This model assumes; (1) the validity of the quasistatic limit, enabling the electric field to be expressed in terms of a scalar potential, (2) describes the free carriers using a hydrodynamic description, (3) neglects the dispersion of the material lattice, and (4) captures perturbations of the charge density and velocity to first order. Assumptions (1)-(4) are well satisfied for the candidate devices considered here.

We begin with Lagrangian given by 
\begin{eqnarray}
\label{L}
    L = \int_{V_{sc}} d^3 x \bigg[ -m (\dot{n} + {\bf v}_{\rm d} \cdot \nabla n) \dot{\psi} 
    - \frac{1}{2} m n_0 (\nabla \dot{\psi})^2 -  e n \varphi + \frac{1}{2} \epsilon (\nabla \varphi)^2
    \bigg]+
    \int_{V_{out}}d^3x \frac{1}{2} \epsilon (\nabla \varphi)^2, 
\end{eqnarray}
where the displacement of the free carriers from equilibrium is expressed as the gradient of the scalar potential $\psi$ (discussed in more detail below), $n$ is the perturbation of the free-carrier density from equilibrium $n_0$, $m$ and $e$ are the free carrier mass and charge, $\epsilon$ is the spatially-dependent permittivity of the waveguide structure, and $\varphi$ is the electric potential. The term containing ${\bf v}_{\rm d}$ accounts for a constant background drift velocity where the free-carrier drift velocity ${\bf v}_{\rm d}$ is directed parallel the waveguide surfaces. For later convenience, the volume integration of the Lagrangian has been explicitly divided into regions that do ($V_{sc}$, $sc$ standing for `semiconductor') and do not contain free carriers ($V_{out}$).

The use of the displacement potential $\psi$ to represent the motion of the free carriers is justified in the quasistatic limit where the electric field is well-described by $-\nabla \varphi$. In this limit the Lorentz force acting on the charged fluid is the gradient of a scalar, and therefore is curl free. Defining the displacement of the electrons from equilibrium by ${\bf \xi}$, the linearized (hydrodynamic) equations of motion for the electron fluid about a constant drift velocity ${\bf v}_{\rm d}$ are given by $m \ddot{\bf \xi} + m {\bf v}_{\rm d} \cdot \dot{\bf \xi} = e\nabla \varphi$. Taking the curl of both sides of this equation shows $\nabla \times {\bf \xi} = 0$, justitfying the representation of the electron displacement by ${\bf \xi} = - \nabla \psi$. 

To describe the coupling between the potential and the elastic field, we use the interaction Lagrangian given by
\begin{equation}
\label{interaction}
    L_{int} = \int d^3x \ \varphi \partial_k d_{ijk} \partial_i u_j = -H_{int}
\end{equation}
which reproduces the classical equations of motion, $d_{ijk}$ is the piezoelectric tensor (or more generically, any coupling electro-mechanical coupling), $u_j$ is the {\it j}th component of the elastic displacement, and $\partial_k$ represents the $k$th component of the gradient.

Neglecting the coupling to the elastic field, the least action principle yields the following equations
\begin{eqnarray}
\label{cont}
&& \dot{n} + {\bf v}_{\rm d} \cdot \nabla n = n_0 \nabla^2 \dot{\psi}
\\
\label{hydro}
&& \ddot{\psi} + {\bf v}_{\rm d} \cdot \nabla \dot{\psi} = \frac{e}{m} \varphi
\\
\label{Gauss}
&& - \nabla \cdot \epsilon \nabla \varphi = e n.
\end{eqnarray}
Noting that the perturbation to the free-carrier velocity is given by ${\bf v} = -\nabla \dot{\psi}$, Eqs. \eqref{cont} and \eqref{hydro} reproduce the linearized continuity and hydrodynamic equations in the presence of drift, and Eq. \eqref{Gauss} gives Gauss' Law. Eliminating the charge density, we find the effective equation of motion for the electric and free-carrier displacement potential given by 
\begin{equation}
\label{Plasma-Eq}
   ( \nabla \cdot \epsilon \nabla \partial_\tau^2  
    + \epsilon_{in} \omega_e^2 \nabla^2 ) 
    \begin{bmatrix}
    \varphi
    \\
    \psi
    \end{bmatrix}  = 0
\end{equation}
where $\partial_\tau = \partial_t + {\bf v}_{\rm d} \cdot \nabla$ (and $\partial_\tau^2 = (\partial_t + {\bf v}_{\rm d} \cdot \nabla)^2$), $\epsilon_{in}$ is the permittivity within the region containing free carriers, and $\omega_e = \sqrt{e^2 n_0/(m \epsilon_{in})}$ is the permittivity normalized plasma frequency. Within the region containing free carriers (i.e. $\epsilon$ = $\epsilon_{in}$ = constant), Eq. \eqref{Plasma-Eq} admits two solution classes: (1) {\it bulk modes} that satisfy $(\partial_\tau^2 + \omega_e^2)\varphi = 0$, and (2) {\it surface modes} that satisfy $\nabla^2 \varphi = 0$ everywhere \cite{barton1979some,barton1997van}. With the appropriate boundary conditions, these modes can be fully characterized. Namely, the electric potential must be continuous across all interfaces and must also satisfy the Fourier domain boundary condition 
\begin{equation}
\label{SP-BC}
    \epsilon_{out} \hat{n}\cdot \nabla \varphi_{out} = \epsilon_{in}\bigg(1-\frac{\omega_e^2}{\beta^2}\bigg) \hat{n}\cdot \nabla \varphi_{in}
\end{equation}
where $\beta = \omega - v_{\rm d} q$, $\hat{n}$ is a unit vector oriented normal to the surface containing the free carriers, and the subscript `in' and `out' respectively denote quantities evaluated on the respective inner and outer sides of the interface between the region with free carriers. For bulk modes, the right hand side of Eq. \eqref{SP-BC} vanishes, requiring the potential outside $V_{sc}$ to vanish \cite{barton1997van}. Consequently, bulk modes do not couple to charges outside of $V_{sc}$ for this model, and therefore for the candidate system proposed in this paper---with a piezoelectric domain outside the semiconducting region---bulk modes do not produce acoustoelectric gain \cite{barton1979some,barton1997van}. However, with diffusion, bulk modes are not confined within the semiconducting region \cite{barton1979some}. As an example, for the surface modes on a planar interface without drift ($v_{\rm d} =0$) Eq. \eqref{SP-BC} yields the well-known surface plasmon frequency $\omega_S^2 = {\epsilon_{in}/(\epsilon_{in}+\epsilon_{out})\omega_e^2}$. 

\subsection{Acoustoelectric Hamiltonian}
To derive the Hamiltonian for the acoustoelectric dynamics, we select $n$ as a generalized coordinate, find the conjugate momentum ($\delta L/\delta \dot{n} \equiv P = -m \dot{\psi}$), and perform a Legendre transform of $L$ to give
\begin{eqnarray}
    H_0 = \int_{V_{sc}} d^3x \bigg[ \frac{n_0}{2m} (\nabla P)^2 + P {\bf v}_{\rm d}\cdot \nabla n + en\varphi -\frac{1}{2}\epsilon (\nabla \varphi)^2 \bigg] 
    - \int_{V_{out}} d^3x \frac{1}{2}\epsilon (\nabla \varphi)^2.  
\end{eqnarray}
The equations of motion for this system can be derived from the Poisson bracket $\{n({\bf x}),P({\bf x}')\} =  \delta^3({\bf x} - {\bf x}')$, reproducing Eqs. \eqref{cont}-\eqref{Gauss}. We quantize this system by requiring the equal-time commutation relation (ETCR) $[n({\bf x}),P({\bf x}')] = i\hbar \delta^3({\bf x} - {\bf x}')$ (i.e., we promote the classical Poisson bracket to a commutator multiplied by $i\hbar$).

Free carrier dissipation is critical to the acousto-electric effect. To model these effects we use an open systems treatment tailored to reproduce the permittivity of the Drude-Lorentz model for free carriers in the frame of the drift current, captured by the ``bath" Hamiltonian given by
\begin{eqnarray}
    H_{bath} = \int_{V_{sc}} d^3x && \int_0^\infty d\nu 
    \bigg[  \frac{1}{2m} P_\nu^2 + P_\nu {\bf v}_{\rm d}\cdot \nabla X_\nu\nonumber
     + \frac{1}{2}m \nu^2 X_\nu^2 - m \hat{c}_\nu X_\nu n \bigg]. \quad \quad 
\end{eqnarray}
Here $P_\nu$ and $X_\nu$ denote momenta and position coordinates for the $\nu$th bath mode and $\hat{c}_\nu$ is a mode-specific system bath coupling, i.e., $\hat{c}_\nu$ takes on a specific form when the fields are expanded in normal modes, depending on the normal mode eigenvalues. The classical equations of motion for the bath variables can be derived from the Poisson bracket $\{ X_\nu({\bf x}),P_{\nu'}({\bf x}')\} =  \delta(\nu-\nu')\delta^3({\bf x}-{\bf x}')$ and the quantum dynamics can be recovered by replacing the Poisson bracket a commutator divide by $i\hbar$. 

After simplification, Hamilton's equations for this coupled system give  
\begin{eqnarray}
\label{EOM-OS}
  && ( \nabla \cdot \epsilon \nabla \partial_\tau^2  
    + \epsilon_{in} \omega_e^2 \nabla^2 ) \varphi = -e n_0 \int_0^\infty d\nu \ \hat{c}_\nu \nabla^2 X_\nu \quad
    \\
    \label{EOM-OS2}
&& ( {\partial_\tau}^2 + {\nu}^2 ) X_\nu = \frac{\hat{c}_\nu}{e} \nabla \cdot \epsilon \nabla \varphi \quad
\end{eqnarray}
which can be analyzed in terms of bulk modes and surface modes. Owing to the linearity of this coupled system, Eqs. \eqref{EOM-OS} and \eqref{EOM-OS2} also give the quantum dynamics of the system when these fields (i.e., $\varphi$ and $X_\nu$) are promoted to operators. 

\subsubsection{Bulk modes}
For bulk modes, orthonormal eigenfunctions of the Helmholtz equation, satisfying $\nabla^2 (\psi_{\ell}\exp\{iqz\}/\sqrt{2\pi}) = -Q_{\sigma q}^2 (\psi_{\ell}\exp\{iqz\}/\sqrt{2\pi})$ and $\int_{V_{sc}}d^3x \ \psi^*_{\ell} \psi_{\ell'}exp\{i(q-q')z\}/(2\pi) = \delta_{\ell \ell'} \delta(q-q')$ subject to Dirichlet boundary conditions on the boundary of $V_{sc}$, which denote from hereon as $\partial V_{sc}$, yield the spatial dependence of the potential and charge density. Using these eigenfunctions as a basis
\begin{eqnarray}
   \varphi = \sum_\ell \int dq \bigg( \psi_{\ell } \frac{e^{i q z}}{\sqrt{2\pi}} \varphi_{\ell q} + c.c.\bigg)
 \\ 
   X_\nu = \sum_\ell \int dq \bigg(\psi_{\ell } \frac{e^{i q z}}{\sqrt{2\pi}} X_{\nu \ell q} +c.c. \bigg),
\end{eqnarray}
we find 
\begin{eqnarray}
\label{EOM-OS3}
   (  \partial_\tau^2  +  \omega_e^2) \varphi_{\ell q} = -\frac{e n_0}{\epsilon_{in}} \int_0^\infty d\nu \ c_{\nu \ell q} X_{\nu \ell q} \quad
    \\
    \label{EOM-OS4}
( {\partial_\tau}^2 + {\nu}^2 ) X_{\nu \ell q} = -\frac{c_{\nu \ell q} \epsilon_{in}}{e} Q^2_{\ell q} \varphi_{\ell q}.
\quad
\end{eqnarray}
Here, we assume that the waveguide is translationally invariant along the z-direction so that  $\psi_{\ell q} \propto \exp\{iqz\}/\sqrt{2\pi}$ where $q$ is the wavevector of a plane and the system-bath coupling is taken as $\hat{c}_\nu \to c_{\nu \ell q}$ when expressed in this mode basis. 
Explicitly solving for $X_{\nu \ell q}$, including both homogeneous ($X^0_{\nu \ell q}$) and particular solutions, inserting the solution into the equation for $\varphi_{\ell q}$ and assuming that $\sqrt{n_0} c_{\nu \ell q} Q_{\ell q}  = (2\gamma_{\rm e} \nu^2/\pi)^{1/2} \equiv G_\nu$ (which produces Ohmic coupling to the bath), we obtain driven damped motion for the potential, reproducing the physics of the Drude-Lorentz model in the presence of drift current \cite{shapiro2010thermal}, given by
\begin{eqnarray}
 (  \partial_\tau^2 + \gamma_{\rm e} \partial_\tau  +  \omega_e^2 ) \varphi_{\ell q} = -\frac{e n_0}{\epsilon_{in}} \int_0^\infty d\nu \ c_{\nu \ell q} X^0_{\nu \ell q}. \quad
\end{eqnarray}
Using the solution for $\varphi_{\ell q}$, $X_{\nu \ell q}$ can be obtained from Eq. \eqref{EOM-OS2} \cite{barton1997van}. One can show that the following mode expansions for the potential and the bath satisfy Eqs. \eqref{EOM-OS} \& \eqref{EOM-OS2} as well as the ETCR
\begin{eqnarray}
\label{phi-ME}
 && \varphi = \omega_e \sum_\ell \int dq  \int_0^\infty d\omega \sqrt{\frac{\hbar}{4\pi\omega \epsilon_{in}}}  \frac{1}{Q_{\ell q}}  
 \bigg( \chi(\omega) \psi_{\ell } e^{i q z}  a_{\omega \ell q}  + H.c. \bigg) \quad
 \\
&& n = \sum_\ell \int dq  \int_0^\infty d\omega \sqrt{\frac{\hbar n_0}{4\pi m \omega}}  Q_{\ell q} \bigg( \chi(\omega) \psi_{\ell } e^{i q z} a_{\omega \ell q}  + H.c. \bigg) \quad  
 \\
&& P = -i \sum_\ell \int dq  \int_0^\infty d\omega \sqrt{\frac{\hbar m \omega}{4\pi n_0}}  \frac{1}{Q_{\ell q}} \bigg( \chi(\omega) \psi_{\ell } e^{i q z}  a_{\omega \ell q}  - H.c. \bigg) \quad  
\\
&& X_\nu = \sum_\ell \int dq  \int_0^\infty d\omega \sqrt{\frac{\hbar}{4\pi m  \omega}}
  \bigg(\chi_\nu(\omega) \psi_{\ell } e^{i q z}  a_{\omega \ell q}  + H.c. \bigg) \quad 
 \\
 \label{pnu-ME}
 && P_\nu = -i\sum_\ell \int dq  \int_0^\infty d\omega \sqrt{\frac{\hbar m \omega}{4\pi }} \bigg(\chi_\nu(\omega)\psi_{\ell } e^{i q z} a_{\omega \ell q}  - H.c. \bigg) \quad 
\end{eqnarray}
when 
\begin{eqnarray}
\label{amp-CR1}
&& [a_{\omega \ell q},a^\dag_{\omega' \ell' q'}] = \delta_{\ell \ell'} \delta(\omega-\omega')\delta(q-q')
 \\
 \label{amp-CR2}
&& \left[ a^\dag_{\omega \ell q},a^\dag_{\omega' \ell' q'} \right] =  [a_{\omega \ell q},a_{\omega' \ell' q'}] = 0
  \\
&&  \chi(\omega) = \frac{G_{\omega}}{-\omega(\omega+i \gamma_{\rm e}) +\omega_e^2}
  \\
&&  \chi_\nu(\omega) = \delta(\omega-\nu) -  \frac{G_{\nu}}{-\omega^2+ \nu^2}\frac{G_{\omega}}{-\omega(\omega+i \gamma_{\rm e}) +\omega_e^2}
\end{eqnarray}
and the functions $\chi(\omega)$ and $\chi_\nu(\omega)$ satisfy the Lippmann-Schwinger orthogonality conditions \cite{barton1997van}
\begin{eqnarray}
\label{LS1}
&& \int_0^\infty d\omega \chi(\omega)\chi^*(\omega) = 1
\\
\label{LS2}
&& \chi(\omega)\chi^*(\omega')+\int_0^\infty d\nu \chi_\nu(\omega)\chi^*_\nu(\omega') = \delta(\omega-\omega')
\quad
\\
\label{LS3}
&& \int_0^\infty d\omega \chi_\nu(\omega)\chi^*_{\nu'}(\omega) = \delta(\nu-\nu')
\quad
\\
\label{LS4}
&& \int_0^\infty d\omega \chi_\nu(\omega)\chi^*(\omega) = 0.
\quad
\end{eqnarray}

Inserting Eqs. \eqref{phi-ME}-\eqref{pnu-ME} in the total Hamiltonian (i.e., $H=H_0+H_{bath}$) and using the orthonormality conditions Eqs. \eqref{LS1}-\eqref{LS4} gives the diagonalized Hamiltonian for the coupled modes of potential and charge
\begin{eqnarray}
\label{H-mode}
H =  \sum_\ell \int dq  \int_0^\infty d\omega \ \hbar(\omega+v_{\rm d} q) a^\dag_{\omega \ell q} a_{\omega \ell q}.
\end{eqnarray}
Note that the background drift current has the effect of Doppler shifting the frequency of these charge-potential modes.

\subsection{Envelope formulation of acoustoelectric dynamics}
To describe slowly varying changes to the elastic amplitude, this section develops an envelope formulation of the acoustoelectric effect. The envelope ${\Phi}_{\omega \ell}(z)$ for charge density, velocity potential and electric potential is related to the normal mode amplitudes $a_{\omega \ell q}$ by
\begin{eqnarray}
\label{env-def}
a_{\omega \ell q} = \int \frac{dz}{\sqrt{2\pi}} \ e^{i(q_{\rm m}-q)z} \Phi_{\omega \ell}(z)
\end{eqnarray}
where $q_{\rm m}$ is the carrier wavevector describing the spatial changes along the waveguide. 
Equations \eqref{env-def} and \eqref{amp-CR1} show that the envelope operators satisfy the commutation relations 
\begin{eqnarray}
&& \big[\Phi_{\omega \ell}(z), \Phi^\dag_{\omega' \ell'}(z)\big] = \delta_{\ell \ell'} \delta(\omega-\omega')\delta(z-z') 
\\
&& \big[\Phi_{\omega \ell}(z), \Phi_{\omega' \ell'}(z)\big] = 0
\\
&& \big[\Phi^\dag_{\omega \ell}(z), \Phi^\dag_{\omega' \ell'}(z)\big] = 0
\end{eqnarray}
Using the envelope description for the potential and elastic field, we find
\begin{eqnarray}
 && \varphi \approx \omega_e \sum_\ell  \int_0^\infty d\omega \sqrt{\frac{\hbar}{2\omega \epsilon_{in}}}  \frac{1}{Q_{\ell q_{\rm m}}}  \bigg( \chi(\omega) \psi_{\ell} e^{i q_{\rm m} z} \Phi_{\omega \ell}(z) + H.c. \bigg) \quad  \nonumber
 \\
 &&{\bf u} \approx \sum_\lambda \sqrt{\frac{\hbar}{2\Omega_{\lambda q_{\rm m}}}} \bigg(\vec{\mathcal{U}}_{\lambda} e^{i q_{\rm m} z}  B_\lambda(z) + H.c.\bigg).
\end{eqnarray}
where $\Omega_{\lambda q_{\rm m}}$ is the eigenfrequency for the $\lambda$th mode  with wavevector $q_{\rm m}$ and $\mathcal{U}_{\lambda,j}$ are orthonormal eigenfunctions of the elastic equation satisfying the eigenvalue equation for the medium's mechanical motion
$\partial_j C_{ijkl} \partial_k [\mathcal{U}_{\lambda,l}\exp\{iqz\}] = -\rho \Omega^2_{\lambda q}[\mathcal{U}_{\lambda q,i}\exp\{iqz\}]$ and the orthonormality condition $\int d^3x \ \rho \ \vec{\mathcal{U}}_{\lambda} \cdot \vec{\mathcal{U}}^*_{\lambda'}\exp\{i(q-q')z\}/(2\pi) = \delta_{\lambda \lambda'} \delta(q-q')$. Here, $\rho$ is the spatially-dependent mass density.
When restricted to a single mode of the elastic field (i.e. we suppress the sum over $\lambda$ from hereon), the interaction Hamiltonian Eq. \eqref{interaction} can be expressed in the envelope picture as 
\begin{eqnarray}
H_{int} =  - \sum_\ell \int_0^\infty d\omega \int dz \ \hbar (\kappa_{\omega \ell} \Phi_{\omega \ell}(z) B^\dag(z)+H.c.) \quad \quad
\end{eqnarray}
where the coupling rate (with units of $1/\sqrt{s}$) is given by
\begin{eqnarray}
 \kappa_{\omega \ell} =  \sqrt{\frac{1}{4 \epsilon_{in} \omega \Omega_{\rm m}}} \frac{\omega_e \chi(\omega)}{Q_{\ell q_{\rm m}}}   
 \int d^2 x \ \psi_{\ell} \tilde{\partial_k} d_{ijk} \tilde{\partial_i} \mathcal{U}^*_{j} \quad \quad \ \
\end{eqnarray}
with the integral taken over the waveguide cross section and $\tilde{\partial_i} = \{\partial_x,\partial_y,-i q_{\rm m}\}$ for $i$ equal $x$, $y$ and $z$ respectively. Inserting Eq. \eqref{env-def} into Eq. \eqref{H-mode} and adding $H_{int}$ yields Eq. \eqref{eq:eom} listed in the main text. 

\subsection{Acoustoelectric gain and dispersion for bulk modes}
The acoustoelectric gain and dispersion can be obtained from the coupled envelope equations. Neglecting Brillouin coupling, the Heisenberg equations give
\begin{eqnarray}
 && \dot{B}(z) + i \Omega_{\rm m} B(z) +v_{\rm g,b} \partial_z B(z) = i \sum_\ell \int_0^\infty d\omega  \ \kappa_{\omega \ell} \Phi_{\omega \ell}(z) \quad
 \\
 && \dot{\Phi}_{\omega \ell}(z) +i (\omega+v_{\rm d} q_{\rm m}) \Phi_{\omega \ell}(z) +v_{\rm d} \partial_z \Phi_{\omega \ell}(z) = i \kappa^*_{\omega \ell} B(z). \quad
\end{eqnarray}
To find steady-state time harmonic solutions, we assume that $\Phi_{\omega \ell}$ and $B$ oscillate at the same frequency $\Omega$, giving the following solution for $\Phi_{\omega \ell}$
\begin{eqnarray}
 \Phi_{\omega \ell}(z) = i \frac{\kappa^*_{\omega \ell}}{v_{\rm d}} \int_{-\infty}^z dz' e^{-i(\Omega -\omega - v_{\rm d} q_{\rm m})(z-z')/v_{\rm d}}B(z').
\end{eqnarray}
Inserting the solution for $\Phi_{\omega \ell}$ into the equation for $B$ we find
\begin{eqnarray}
&&  -i(\Omega- \Omega_{\rm m}) B +v_{\rm g,b} \partial_z B =
 - \sum_\ell \int_{-\infty}^z dz' \int_0^\infty d\omega \frac{|\kappa_{\omega \ell}|^2}{v_{\rm d}} e^{-i(\Omega -\omega - v_{\rm d} q_{\rm m})(z-z')/v_{\rm d}} B(z'). 
\end{eqnarray}
The $\omega$-integral, $\int_0^\infty d\omega \frac{|\kappa_{\omega \ell}|^2}{v_{\rm d}} e^{-i(\Omega -\omega - v_{\rm d} q_{\rm m})(z-z')/v_{\rm d}}$, exponentially decays as $\sim \exp\{-\gamma_{\rm e}(z-z')/(2v_{\rm d})\}$, which far exceeds typical spatial decay rates for phonons. Under these conditions the phonon envelope $B(z')$ can be replaced with $B(z)$ and brought outside the integral so that
\begin{eqnarray}
&& - \sum_\ell \int_{-\infty}^z dz' \int_0^\infty d\omega \frac{|\kappa_{\omega \ell}|^2}{v_{\rm d}} e^{-i(\Omega -\omega - v_{\rm d} q_{\rm m})(z-z')} B(z') \nonumber
\approx   - iP.V. \sum_\ell \int_0^\infty d\omega  \frac{|\kappa_{\omega \ell}|^2 }{\Omega -\omega - v_{\rm d} q_{\rm m}} B(z) \nonumber 
 - \pi \sum_\ell |\kappa_{\Omega-v_{\rm d}q_{\rm m},\ell}|^2 B(z)
\end{eqnarray}
where we have used $\int_0^\infty dz \ \exp{(i k z)} = i P.V. \ 1/k +\pi \delta(k)$ and P.V. denotes the Cauchy principal value. These assumptions lead to the effective acoustoelectric dynamics including gain and dispersion given by 
\begin{eqnarray}
v_{\rm g,b}\partial_z B -i(\Omega- \Omega_{\rm m}-\Delta \Omega_0) B - \frac{1}{2 }G_{AE}B  = 0. \quad  \quad
\end{eqnarray}
where
\begin{eqnarray}
&& \Delta \Omega_{\rm AE} = \sum_\ell \int_0^\infty d\omega |\kappa_{\omega \ell}|^2 P.V. \frac{1}{\Omega -\omega - v_{\rm d} q_{\rm m}}
\\
&& G_{AE} = -2 \pi \sum_\ell |\kappa_{\Omega-v_{\rm d}q_{\rm m},\ell}|^2.
\end{eqnarray}
While this analysis was completed for bulk modes, the same expressions apply for the case of the surface modes (described below). 

\subsection{Gain and dispersion for bulk plane-waves}
In the uniform plane-wave limit the coupling rate for compressional waves is given by 
\begin{eqnarray}
 \kappa_{\omega \ell} \approx  \frac{\omega_e  q_{\rm m} d_{zzz} \chi(\omega)}{\sqrt{4 \rho \epsilon \omega \Omega_{\rm m}}}   \delta_{\ell,0}
\end{eqnarray}
yielding gain and dispersion given by 
\begin{eqnarray}
 && G_{\rm AE} = - \frac{\pi \omega_e^2  d^2_{zzz}q_{\rm m}^2}{2 \rho \epsilon_{in} \Omega_{\rm m}} \frac{|\chi(\Omega-v_{\rm d} q_{\rm m})|^2}{\Omega-v_{\rm d} q_{\rm m}}
 \\
 && \Delta \Omega_{\rm AE} = \frac{\omega_e^2  d^2_{zzz}q_{\rm m}^2}{4 \rho \epsilon_{in} \Omega_{\rm m}} P.V. \int_0^\infty \frac{d \omega}{\omega}
 |\chi(\omega)|^2  \frac{1}{\Omega - \omega -v_{\rm d} q_{\rm m}}. \quad \quad
\end{eqnarray}

\subsection{Surface modes} 
The acoustoelectric dynamics for the surface modes follows by close analogy with the bulk mode analysis above. In contrast with the bulk modes, the surface modes have distinct spatial dependence being strongly localized to surfaces, nontrivial boundary conditions at interfaces, and include the effects of a singular surface charge density. 

\subsubsection{Surface mode Hamiltonian}
The Hamiltonian can be derived from Eq. \eqref{L} by: 
(1) taking $n \to \sigma \delta(x_{\perp} \in \partial V_{sc})$, where $\sigma$ is a surface charge density, $\delta(x_{\perp} \in \partial V_{sc})$ is a delta function with vanishing argument on the boundary of $V_{sc}$, and $x_{\perp}$ is the coordinate locally normal the boundary of $V_{sc}$, 
(2) performing an integration by parts and dropping terms proportional to $\nabla^2 \psi$ and $\nabla^2 \varphi$ that vanish for surface modes,
(3) finding the conjugate momentum $p = - m\dot{\psi}$ (with $\dot{\psi}$ restricted to $\partial V_{sc}$) to the surface charge density $\sigma$, and 
(4) performing a Legendre transform. These steps lead to the Hamiltonian for the surface modes $H_S$, including the impact of a bath (momenta $p_\nu$ and position $x_\nu$), given by
\begin{eqnarray}
\label{H_S}
 H_S = \oint_{\partial V_{sc}} da
 \bigg[
 \frac{n_0}{2m} p \frac{\partial p}{\partial x_\perp}
 - p {\bf v}_{\rm d}\cdot\nabla \sigma
 + e \sigma \varphi_S 
 - \frac{1}{2} \varphi_S \bigg(
 \epsilon_{in} \frac{\partial \varphi_{in}}{\partial x_\perp} 
 -\epsilon_{out} \frac{\partial \varphi_{out}}{\partial x_\perp} \bigg)
 +\int_0^\infty d\nu \bigg( \frac{p_\nu^2}{2m} +\frac{1}{2}m \nu^2 x_\nu^2 + \hat{\mu}_\nu x_\nu \sigma
 \bigg)
 \bigg]. \quad \quad
\end{eqnarray}
Here, $\varphi_S$ in the potential restricted to the boundary $\partial V_{sc}$, $\hat{\mu}_\nu$ is the mode-specific (i.e., depending on mode eigenvalues) system-bath coupling, and it is assumed that the drift velocity is parallel to the waveguide. The dynamics for this coupled system follow from Hamilton's equations, yielding
\begin{eqnarray}
\label{EOM-SM1}
 && \partial_\tau \sigma = \frac{n_0}{m} \frac{\partial p}{\partial x_\perp}
 \\
  \label{EOM-SM2}
 && \partial_\tau p = -e \varphi_S - \int_0^\infty d\nu \ \hat{\mu}_\nu x_\nu 
 \\
 \label{EOM-SM3}
 && e\sigma = \varphi_S \bigg(
 \epsilon_{in} \frac{\partial \varphi_{in}}{\partial x_\perp} 
 -\epsilon_{out} \frac{\partial \varphi_{out}}{\partial x_\perp} \bigg)
 \\
 && \partial_\tau x_\nu = p_\nu/m
 \\
 \label{EOM-SM5}
 && \partial_\tau p_\nu = - \hat{\mu}_\nu \sigma. 
\end{eqnarray}

\subsection{Quantization of surface modes}
By generalization of the classical Poisson bracket, i.e., $\{ \sigma({\bf x}), p({\bf x}')\} = \delta^2({\bf x}-{\bf x}')$ with ${\bf x}$ and ${\bf x}'$ on the boundary $\partial V_{sc}$ to the ETCR $[ \sigma({\bf x}), p({\bf x}')] = i\hbar \delta^2({\bf x}-{\bf x}')$ the charge-potential system can be quantized. The bath variables satisfy an analogous ETCR $[ x_\nu({\bf x}),p_{\nu'}({\bf x}')] = i\hbar \delta(\nu-\nu')\delta^2({\bf x}-{\bf x}')$. By expressing the charge, potential and bath variables in terms of the eigenfunctions of the equations of motion Eqs. \eqref{EOM-SM1}-\eqref{EOM-SM5}, the Hamiltonian can be expressed in terms of creation and annihilation operators for surface mode quanta. 

Noting that both the electric potential and $p$ satisfy the Laplace equation, we express $\varphi \propto \phi_{\ell}(x_\|) \exp\{iqz\}  f_{\ell q}(x_\perp)/\sqrt{2\pi}$ where 
$\nabla^2 (\phi_{\ell}(x_\|) \exp\{iqz\}  f_{\ell q}(x_\perp)/\sqrt{2\pi}) = 0$ and $\phi_{\ell }(x_\|)\exp\{iqz\}/\sqrt{2\pi}$ form a complete set of orthogonal eigenfunctions on the surface of $V_{sc}$. Here, we denote coordinates of the surface as $\{x_\|,z\}$ where $z$ is directed along the waveguide. Consequently, the eigenfunctions for the potential satisfy the following relations
\begin{eqnarray}
&& (\nabla_\|^2 + \partial_z^2)(\phi_{\ell}\exp\{iqz\}) = -K^2_{\ell q} \phi_{\ell }\exp\{iqz\},
\\
 && \oint_{\partial V_{sc}} da \ \phi_{\ell } \phi^*_{\ell'} \exp{iq(z-z')}/(2\pi) = \delta_{\ell \ell'} \delta(q-q')
\quad {\rm and}
\\
 && \frac{\partial^2}{\partial x_\perp^2} f_{\ell q}  -K^2_{\ell q} f_{\ell q} = 0.
\end{eqnarray}
Likewise, $p$ (the velocity potential) can be expressed as $p \propto \phi_{\ell }(x_\|) \exp\{iqz\} h_{\ell q}(x_\perp)/\sqrt{2\pi}$ where $\frac{\partial^2}{\partial x_\perp^2} h_{\ell q}  -K^2_{\ell q} h_{\ell q} = 0$. Critical to the surface mode dynamics are the boundary conditions for the functions $f_{\ell q}$ and $h_{\ell q}$ on the surface of $V_{sc}$. These boundary conditions, following directly from Eqs. \eqref{EOM-SM1}, \eqref{EOM-SM3}, and the ETCRs up to normalization, require 
\begin{eqnarray}
&&
\bigg[\frac{\partial h_{\ell q}(x_\perp )}{\partial x_\perp}
- \epsilon_{in} \frac{\partial f_{\ell q}(x_\perp )}{\partial x_\perp} \bigg]_{x_\perp  \in \partial V_{sc} - 0+}
=  -\epsilon_{out} \frac{\partial f_{\ell q}(x_\perp )}{\partial x_\perp} \bigg|_{x_\perp  \in \partial V_{sc} + 0+},
\\
&& 
h_{\ell q}(x_\perp )\frac{\partial h_{\ell q}(x_\perp )}{\partial x_\perp} \bigg|_{x_\perp  \in \partial V_{sc} - 0+} = K_{\ell q},
\end{eqnarray}
and $f_{\ell q}$ to be continuous. Additionally, the appropriate boundary conditions on the bounding surface of the total system must be satisfied. Within this eigenfunction basis, and setting $\hat{\mu}_\nu \sqrt{n_0 K_{\ell q}} \to G_\nu \equiv \sqrt{2\gamma_{\rm e}\nu^2/\pi}$, one can show that the following mode representations satisfy the equations of motion and ETCRs.


\begin{eqnarray}
 && 
 \label{phi-SME}
 \varphi = \omega_e \sum_\ell \int dq  \int_0^\infty d\omega \sqrt{\frac{\hbar \epsilon_{in}}{4\pi\omega K_{\ell q}}}  \bigg( 
 \Delta_{\ell q} (\omega) \phi_{\ell}(x_\|)e^{iqz}  f_{\ell q}(x_\perp) a_{\omega \ell q}  + H.c. \bigg) \quad  
 \\
&& 
\sigma = \sum_\ell \int dq  \int_0^\infty d\omega \sqrt{\frac{\hbar n_0}{4\pi m \omega K_{\ell q}}}   \bigg(
\Delta_{\ell q} (\omega) \phi_{\ell}(x_\|)e^{iqz} H_{\ell q} a_{\omega \ell q} + H.c. 
\bigg) \quad  
 \\
&& p = -i \sum_\ell \int dq  \int_0^\infty d\omega \sqrt{\frac{\hbar m \omega}{4\pi n_0 K_{\ell q}}}  
\bigg( \Delta_{\ell q}(\omega) \phi_{\ell}(x_\|) e^{iqz} h_{\ell q}(x_\perp)
a_{\omega \ell q}  - H.c. 
\bigg) \quad
\\
&& x_\nu = \sum_\ell \int dq  \int_0^\infty d\omega \sqrt{\frac{\hbar}{4\pi m \omega}}
  \bigg(
  \Delta_{\nu \ell q}(\omega) \phi_{\ell}(x_\|)e^{iqz} a_{\omega \ell q}  + H.c. 
  \bigg) \quad 
 \\
 &&
 \label{pnu-SME}
 p_\nu = -i\sum_\ell \int dq  \int_0^\infty d\omega \sqrt{\frac{\hbar m \omega}{4\pi}} \bigg(
 \Delta_{\nu \ell q}(\omega)\phi_{\ell }(x_\|) e^{iqz} a_{\omega \ell q}  - H.c. 
 \bigg) \quad 
\end{eqnarray}
when $a_{\omega \ell q}$ and $a^\dag_{\omega' \ell' q'}$ satisfy Eqs. \eqref{amp-CR1}-\eqref{amp-CR2} and 
\begin{equation}
    H_{\ell q} = \frac{\partial h_{\ell q}(x_\perp )}{\partial x_\perp}\bigg|_{x_\perp \in \partial V_{sc}}.
\end{equation}
Much like the susceptibility $\chi$ for the bulk modes, the surface modes satisfy an analogous set of Lippmann-Schwinger equations yielding
\begin{eqnarray}
\label{amp-CR}
&&  \Delta_{\ell q}(\omega) = \frac{G_{\omega}}{-\omega(\omega+i \gamma_{\rm e}) +\omega_{\ell q}^2}
  \\
&&  \Delta_{\nu \ell q}(\omega) = \delta(\omega-\nu) -  \frac{G_{\nu}}{-\omega^2 + \nu^2} \Delta_{\ell q}(\omega)
\end{eqnarray}
where the surface mode resonance frequency $\omega_{\ell q}$ is determined by the functions $f_{\ell q}$ and $h_{\ell q}$ 
\begin{eqnarray}
\omega_{\ell q}^2 = \frac{e^2 n_0}{m}\frac{f_{\ell q}}{h_{\ell q}}\bigg|_{x_\perp  \in \partial V_{sc} - 0+}.
\end{eqnarray}
The functions $ \Delta_{\ell q}$ and $ \Delta_{\nu \ell q}$ satisfy the orthonormality relations Eqs. \eqref{LS1}-\eqref{LS4} with substitution of $\chi \to  \Delta_{\ell q}$ and  $\chi_\nu \to \Delta_{\nu \ell q}$. 

Substitution of Eqs. \eqref{phi-SME}-\eqref{pnu-SME} into Eq. \eqref{H_S} leads to the Hamiltonian for the surface modes represented in terms of creation and annihilation operators. This Hamiltonian, and its formulation in terms of mode envelopes, takes the exact same form as Eqs. \eqref{H-mode} and \eqref{eq:eom} with the understanding that $\ell$ labels and counts surface modes. 

\subsubsection{Acousto-electric coupling with surface modes}
Using the interaction Hamiltonian defined in Eq. \eqref{interaction} and the formal expression for the electric potential for surface modes Eq. \eqref{phi-ME}, the coupling rate defined in the envelope picture is
\begin{eqnarray}
 \kappa_{\omega \ell} =  \omega_e\sqrt{\frac{1}{4\omega \Omega_{\rm m} K_{\ell q_{\rm m}}}}  \Delta_{\ell q_{\rm m}}(\omega)   
 \int d^2 x \ \phi_{\ell}(x_\|)f_{\ell q_{\rm m}}(x_\perp) \tilde{\partial}_k d_{ijk} \tilde{\partial}_i \mathcal{U}^*_{q_{\rm m},j} \quad \quad \ \
\end{eqnarray}
where the integral is taken over the waveguide cross-section. 

\subsection{Connection between normal mode picture and $k^2$}
In this section, we show how the acoustoelectric coupling $\kappa_{\omega \ell}$ relates to the standard expression for the coupling $k^2$. Generally, the coupling $k^2$ is defined as the fractional change in the square of the speed of sound for a free system and grounded system, or equivalently in terms of the ration of stored interaction energy to injected energy \cite{uchino2017development}. Given the direct connection between the resonance frequency and the sound speed we estimate $k^2$ by the shift in the resonance frequency $\Omega_{\rm m}$. Assuming that $\Omega_{\rm m} \gg |\Delta \Omega_{\rm AE}|$ we find
\begin{eqnarray}
\label{k^2-def}
 k^2 = \frac{\Omega_{\rm m}^2-(\Omega_{\rm m} +\Delta \Omega_{\rm AE})^2}{\Omega_{\rm m}^2} \approx -2 \frac{\Delta \Omega_{\rm AE}}{\Omega_{\rm m}}.
\end{eqnarray}
Solving for the frequency shift then allows $k^2$ to be calculated.  In the limit that $\omega_{\ell q_{\rm m}} \sim \omega_e \gg \Omega_{\rm m}$ and $\Omega_{\rm m} -v_{\rm d}q_{\rm m}$, the frequency integrals in the expression for $\Delta \Omega_{\rm AE}$ can be approximated by 
\begin{eqnarray}
&& P.V. \int_0^\infty d\omega \ \frac{|\chi(\omega)|^2}{\omega} \frac{1}{\Omega-v_{\rm d}q_{\rm m} -\omega} = P.V. \int_0^\infty d\omega \ \frac{2\gamma_{\rm e} \omega/\pi}{(\omega^2-\omega_e^2)^2 +\gamma_{\rm e}^2 \omega^2} \frac{1}{\Omega-v_{\rm d}q_{\rm m} -\omega} \approx -\frac{1}{\omega_{e}^2}
\\
 && P.V. \int_0^\infty d\omega \ \frac{|\Delta_{\ell q_{\rm m}}(\omega)|^2}{\omega} \frac{1}{\Omega-v_{\rm d}q_{\rm m} -\omega} = P.V. \int_0^\infty d\omega \ \frac{2\gamma_{\rm e} \omega/\pi}{(\omega^2-\omega_{\ell q_{\rm m}}^2)^2 +\gamma_{\rm e}^2 \omega^2} \frac{1}{\Omega-v_{\rm d}q_{\rm m} -\omega} \approx -\frac{1}{\omega_{\ell q_{\rm m}}^2}.
\end{eqnarray}
We find $k^2$ for bulk and surface modes given by 
\begin{eqnarray}
&& k^2 = \sum_{\ell} \frac{1}{2\epsilon_{in}\Omega_{\rm m}^2 Q_{\ell q_{\rm m}}^2} \bigg| 
 \int d^2x \ \psi_{\ell} \tilde{\partial_k} d_{ijk} \tilde{\partial_i} \mathcal{U}^*_{j} 
 \bigg|^2 \quad ({\rm bulk} )
\\
 && k^2 = \sum_{\ell} k_\ell^2 = \sum_{\ell} \frac{1}{2 \Omega_{\rm m}^2 K_{\ell q_{\rm m}}} \frac{\omega^2_e}{\omega_{\ell q_{\rm m}}^2}
 \bigg|    
 \int d^2 x \ \phi_{\ell}(x_\|)f_{\ell q_{\rm m}}(x_\perp) \tilde{\partial}_k d_{ijk} \tilde{\partial}_i \mathcal{U}^*_{q_{\rm m},j} \bigg|^2 \quad ({\rm surface} ).
\end{eqnarray}
Using $\epsilon(\Omega) = \epsilon_{in}[1-\omega_e^2(\Omega(\Omega +i\gamma_{\rm e}))^{-1}]$ and $\epsilon_{\ell q_{\rm m}}(\Omega) = \epsilon_{in}[1-\omega_{\ell q_{\rm m}}^2(\Omega(\Omega +i\gamma_{\rm e}))^{-1}]$
we find
\begin{eqnarray}
&& \frac{|\chi(\omega)|^2}{\omega} = -\frac{2\epsilon_{in}}{\pi \omega_e^2} {\rm Im}\left[\frac{1}{\epsilon(\omega)}\right]
 \\
 && 
 \frac{|\Delta_{\ell q_{\rm m}}(\omega)|^2}{\omega} = -\frac{2\epsilon_{in}}{\pi \omega_{\ell q_{\rm m}}^2} {\rm Im}\left[\frac{1}{\epsilon_{\ell q_{\rm m}}(\omega)}\right]
\end{eqnarray}
yielding the compact expressions for the gain given by 

\begin{eqnarray}
&& G_{AE} = 2 k^2 \Omega_{\rm m} \epsilon_{in} {\rm Im}\left[\frac{1}{\epsilon(\Omega -v_{\rm d} q_{\rm m})} \right]\quad ({\rm bulk} )
\\
 && 
 G_{AE} = \sum_\ell 2 k_\ell^2 \Omega_{\rm m} \epsilon_{in} {\rm Im}\left[\frac{1}{\epsilon_{\ell q_{\rm m}}(\Omega -v_{\rm d} q_{\rm m})} \right]\quad ({\rm surface} ).
\end{eqnarray}

\section{Noise dynamics} \label{sec:noise}

In this section, we examine the noise dynamics of acoustoelectric enhanced Brillouin interactions. We begin with the envelope equations of motion as detailed in Section \ref{sec:enveloptheory}, which are given by

\begin{equation}
\begin{aligned}
\frac{\partial \bar{B}}{\partial t}&=-i (\Omega_{\rm m}-\Omega) \bar{B}-\frac{\Gamma}{2}\bar{B} \pm v_{\rm g,b} \frac{\partial \bar{B}}{\partial z}-i g_{0}^* \bar{A}_{\rm s}^{\dagger} \bar{A}_{\rm p}+\eta\\
\frac{\partial \bar{A}_{\rm p}}{\partial t}&=-\frac{\alpha v_{\rm g,p}}{2} \bar{A}_{\rm p} \pm v_{\rm g,p} \frac{\partial \bar{A}_{\rm p}}{\partial z}-i g_{0} \bar{A}_{\rm s} \bar{B}+\xi_{\rm p}\\
\frac{\partial \bar{A}_{\rm s}}{\partial t}&=-\frac{\alpha v_{\rm g,s}}{2} \bar{A}_{\rm s} \pm v_{\rm g,s} \frac{\partial \bar{A}_{\rm s}}{\partial z}-i g^*_0 \bar{A}_{\rm p} \bar{B}^\dagger +\xi_{\rm s}
\end{aligned}
\label{eq:eomA}
\end{equation}

We again note that Eq. \ref{eq:eomA} includes the effects of dissipation in an open system, and as such, we include thermal and vacuum noise terms $\eta$, $\xi_{\rm p}$, and $\xi_{\rm s}$ according to the fluctuation-dissipation theorem for the phonon, probe, and pump fields, respectively. These Langevin terms have auto correlation functions given by $\langle \eta (z^\prime,t^\prime) \eta^\dagger(z^\prime, t^\prime) \rangle= (n_{\rm th}+1)\Gamma \delta(z^\prime-z)\delta(t^\prime-t)$, $\langle \xi_{\rm s} (z^\prime,t^\prime) \xi_{\rm s}^\dagger(z^\prime, t^\prime) \rangle= \gamma_{\rm s} \delta(z^\prime-z)\delta(t^\prime-t)$, and $\langle \xi_{\rm p} (z^\prime,t^\prime) \xi_{\rm p}^\dagger(z^\prime, t^\prime) \rangle= \gamma_{\rm p} \delta(z^\prime-z)\delta(t^\prime-t)$, respectively, where $\alpha$ is the optical spatial loss rate and $n_{\rm th}$ is the thermal phonon occupation given by the Bose-Einstein distribution \cite{kharel2016noise}.

For simplicity, we neglect noise processes intrinsic to acoustoelectric interactions, such as those from trapping effects \cite{kino1973noise}, which for the proposed system are expected to be smaller than contributions from thermomechanical phonon noise. From the normal mode theory (and given the parameters considered in this work), we estimate sources of intrinsic acoustoelectric noise to be approximately 40\% of the background thermal noise  \cite{kino1973noise} (see Section \ref{sec:SIdesign} and Fig. \ref{fig:nf}).

We once again treat the pump wave as undepleted and transform these equations into the Fourier domain as

\begin{equation}
\begin{aligned}
-i \omega \bar{A}_{\rm s}[z,\omega]+\frac{\alpha v_{\rm g,s}}{2} \bar{A}_{\rm s}[z,\omega]+v_{\rm g,s} \frac{\partial \bar{A}_{\rm s}[z,\omega]}{\partial z}&= -i g_0^* \bar{A}_{\rm p} \bar{B}^\dagger[z,\omega]+\xi_{\rm s}[z,\omega]  \\
i(\Omega-\Omega_{\rm m}-\Delta \Omega_{\rm AE}-\omega)\bar{B}[z,\omega]+\frac{\Gamma-G_{\rm AE}}{2}\bar{B}[z,\omega]+v_{\rm g,b} \frac{\partial \bar{B}[z,\omega]}{\partial z}&=-ig_0^*\bar{A}_{\rm p}\bar{A}^\dagger_{\rm s}[z,\omega]+\eta[z,\omega]
\end{aligned}
\end{equation}

We solve these equations in the acoustoelectric enhanced Brillouin (AEB) limit, in which, despite significant phonon amplification, the coherence length of the phonon field is still much smaller than that of the optical fields.  Given this hierarchy, we eliminate the spatial dynamics of the phonon field, such that

\begin{equation}
\begin{aligned}
\bar{B}[z,\omega]= \chi^{\rm AE}_{\rm B}[\omega] \left( -i g_0^* \bar{A}_{\rm p} \bar{A}_{\rm s}^\dagger[z,\omega]+ \eta[z,\omega] \right),
\end{aligned}
\label{eq:adA}
\end{equation}
and the decoupled Stokes dynamics are given by

\begin{equation}
\begin{aligned}
\frac{\partial \bar{A}_{\rm s}[z,\omega]}{\partial z}= -M \bar{A}_{\rm s}[z,\omega]+ \frac{N_{\rm s}[z,\omega]}{v_{\rm g,s}}
\end{aligned}
\label{eq:sol1s}
\end{equation}

\noindent
where $M=(-i \omega+\alpha v_{\rm g,s}/2-\chi_{\rm s}^{\rm AEB*})/v_{\rm g,s}$ and the dressed Langevin term $N_{\rm s}[z,\omega]=\xi_{\rm s}[z,\omega]- i g_0^* \bar{A}_{\rm p} \chi_{\rm B}^{\rm AE^*} \eta^\dagger [z,\omega]$.  As defined in the main text, $\chi^{\rm AE}_{\rm B}[\omega]=(i(\Omega-\Omega_{\rm m}-\Delta \Omega_{\rm AE}-\omega)+(\Gamma-G_{\rm AE})/2)^{-1}$ and $\chi_{\rm s}^{\rm AEB}[\omega]=|g_0|^2 |\bar{A}_{\rm p}|^2 \chi_{\rm B}^{\rm AE}[\omega]$.

The first order ordinary differential equation has the integral solution

\begin{equation}
\begin{aligned}
\bar{A}_{\rm s}[z,\omega]&=\frac{1}{v_{\rm s}}\int_0^z N_{\rm s}[z^\prime,\omega] e^{-M (z-z^\prime)}\\&+\Big(\bar{A}^{\rm c}_{\rm s}[0,\omega]+\bar{A}^{\rm N}_{\rm s}[0,\omega] \Big)e^{-M z}
\end{aligned}
\label{eq:sol2s}
\end{equation}

\noindent
where $\bar{A}^{\rm c}_{\rm s}[0,\omega]$ and $\bar{A}^{\rm N}_{\rm s}[0,\omega]$ represent the coherent and noise input of the Stokes wave.  

The number spectral density per unit length (defined by $\mathscr{A}_{\rm s}[z,\omega]= \lim_{T\rightarrow \infty}(1/T)|\bar{A}_{\rm s}[z,\omega]|^2$) is

\begin{equation}
\begin{aligned}
\mathscr{A}_{\rm s}[z,\omega] &= (\mathscr{A}_{\rm s}^{\rm c}[0,\omega]+ \mathscr{A}_{\rm s}^{\rm N}[0,\omega]) e^{-2 \Re[M]z}+\lim_{T \rightarrow \infty} \frac{1}{T} \int_0^z dz_1 \int_0^z dz_2 \frac{\langle N_{\rm s}[z_1,\omega] N^\dagger_{\rm s}[z_2,\omega] \rangle}{v_{\rm g,s}^2} e^{-M^*(z-z_1)-M(z-z_2)}\\
&=\underbrace{\mathscr{A}_{\rm s}^{\rm c}[0,\omega]e^{-2 \Re[M]z}}_{\rm Amplified\:signal}+\underbrace{\frac{1}{v_{\rm g,s}}e^{-2 \Re[M]z}+\frac{\alpha v_{\rm g,s} +|g_0|^2 |A_{\rm p}|^2 |\chi_{\rm B}^{\rm AE}[\omega]|^2 \Gamma n_{\rm th}}{2 v_{\rm g,s}^2 \Re[M]}\Big[1-e^{-2 \Re [M] z} \Big]}_{\rm Amplified\:vacuum\:and\:thermal\:fluctuations}
\end{aligned}
\label{eq:nils}
\end{equation}

\noindent
where $\Re[M]$ represents the real part of $M$ as defined above. 

The noise factor ($F$) is defined by the ratio of the input ($\textup{SNR}_1$) to output ($\textup{SNR}_2$) signal to noise ratios, such that 

\begin{equation}
\begin{aligned}
F\equiv \frac{\textup{SNR}_1}{\textup{SNR}_2}.
\end{aligned}
\label{eq:factor}
\end{equation}

For simplicity, we assume an input signal defined by $\mathscr{A}_{\rm s}^{\rm c}[0,\omega]=|A^{\rm in}_{\rm s}|^2 \delta (\omega)$.  The relevant input noise for an optical amplifier is the vacuum noise \cite{haus1998noise}, such that the input signal to noise ratio is given by 

\begin{equation}
\begin{aligned}
\textup{SNR}_1&=\frac{\int_{-\Delta \omega}^{\Delta \omega} d \omega^\prime |A_{\rm s}^{\rm in}|^2 \delta(\omega^\prime)}{\int_{-\Delta \omega}^{\Delta \omega} d \omega^\prime \frac{1}{v_{\rm g,s}}}\\
&=\frac{|A_{\rm s}^{\rm in}|^2 v_{\rm g,s}}{2 \Delta \omega}.
\end{aligned}
\label{eq:snr}
\end{equation}

The output SNR can be obtained from the amplified signal and noise in Eq. \ref{eq:nils}, which yields

\begin{equation}
\begin{aligned}
\textup{SNR}_2=\frac{|A_{\rm s}^{\rm in}|^2 e^{-2 \Re[M]z}}{2 \Delta \omega \Bigg[\frac{1}{v_{\rm g,s}}e^{-2 \Re[M]z}+\frac{|g_0|^2 |A_{\rm p}|^2 |\chi_{\rm B}^{\rm AE}[\omega]|^2 \Gamma n_{\rm th}}{2 v_{\rm g,s}^2 \Re[M]}\bigg[1-e^{-2 \Re [M] z} \bigg] \Bigg]}.
\end{aligned}
\label{eq:snr2}
\end{equation}

\noindent
where $ \Delta \omega$ is a bandwidth much smaller than the acoustoelectric modified Brillouin gain bandwidth (i.e., $\Delta \omega \ll \Gamma-G_{\rm AE}$).  

The resulting noise factor is given by

\begin{equation}
\begin{aligned}
F&\equiv \frac{\textup{SNR}_1}{\textup{SNR}_2}\\&=1+\frac{\alpha v_{\rm g,s} +|g_0|^2 |\bar{A}_{\rm p}|^2|\chi_{\rm B}^{\rm AE}[\omega] |^2 \Gamma n_{\rm th}}{2 v_{\rm g,s} \Re[M]} \bigg[e^{2 \Re[M]z}-1 \bigg].
\end{aligned}
\label{eq:facresult}
\end{equation}

In the limit of large Brillouin amplification and low optical loss, the noise factor simplifies to 

\begin{equation}
\begin{aligned}
F\approx 1+n_{\rm th} \left( \frac{\Gamma}{\Gamma-G_{\rm AE}} \right),
\end{aligned}
\label{eq:facresult}
\end{equation}
which is the central result of this section. This derivation suggests that for systems in which the thermomechanical noise is dominant, AEB amplifiers may achieve near quantum-limited performance in the limit of low temperatures (i.e., $k_{\rm B} T \ll \hbar \Omega_{\rm m}$).

\newpage

\section{Index of Notation}\label{sec:Notation}

\begin{table*}[h]
    \centering
    \ra{1.5}
    \begin{tabularx}{\textwidth}{@{} J K L J K @{}}
        Acoustoelectric frequency shift & $\Delta \Omega_{\rm AE}$ & & Phonon group velocity & $v_\text{g,b}$
        \\ 
        Acoustoelectric gain & $\alpha_\text{AE}$ & & Phonon lifetime & $\tau_\text{m}$
        \\ 
        Acoustoelectric gain rate & $G_{\rm AE}$ & & Phonon phase velocity & $v_\text{m}$
        \\ 
        Acoustoelectric phase delay & $q_\text{AE}$ & & Phonon scaling factor & $C_\text{m}$
        \\ 
        Acoustoelectric time constant & $\tau$ & & Phonon wave vector & $q_\text{m}$
        \\ 
        Boltzmann constant & $k_\text{B}$ & & Phonon wavelength & $\lambda_\text{m}$
        \\ 
        Compliance matrix constant & $S_{ij}$ & & Photoelastic coupling coefficient & $g_\text{pe}$
        \\ 
        Composition parameters & $x \, \, \, \, y$ & & Photoelastic matrix constant & $p_{ij}$
        \\ 
        Debye length & $\lambda_\text{d}$ & & Photon dissipation rate & $\gamma$
        \\ 
        Density & $\rho$ & & Photon energy & $E$
        \\ 
        Dielectric gap height & $h$ & & Piezoelectric coupling coefficient & $k^{2}$
        \\ 
        Dielectric gap permittivity & $\epsilon_\text{g}$ & & Piezoelectric permittivity & $\epsilon_\text{p}$
        \\ 
        Dielectric relaxation frequency & $\omega_\text{c}$ & & Plasma frequency & $\omega_\text{e}$
        \\ 
        Diffusion frequency & $\omega_\text{D}$ & & Propagation angle & $\beta$
        \\ 
        Diffusion term & $D$ & & Pump photon dissipation rate & $\gamma_\text{p}$
        \\ 
        Effective index & $n_\text{eff}$ & & Pump photon envelope operator & $A_\text{p}$
        \\ 
        Effective mass & $m^{*}$ & & Pump photon frequency & $\omega_\text{p}$
        \\ 
        Elastic mode loss coefficient & $\alpha_\text{m}$ & & Pump photon group velocity & $v_\text{g,p}$
        \\ 
        Elastic mode quality factor & $Q_\text{m}$ & & Pump photon scaling factor & $C_\text{p}$
        \\ 
        Elastic strain & $\varepsilon_{ij}$ & & Pump photon wave vector & $k_\text{p}$
        \\ 
        Elastic wave displacement & $\mathrm{u}_{i}$ & & Pump photon wavelength & $\lambda_\text{p}$
        \\ 
        Elasticity matrix constant & $C_{ij}$ & & Radiation pressure coupling coefficient & $g_\text{rp}$
        \\ 
        Electric displacement field & $\mathrm{D}_{i}$ & & Refractive index & $n_\text{r}$
        \\ 
        Electric field & $\mathrm{E}_{i}$ & & Scattering efficiency & $\eta_\text{ef}$
        \\ 
        Electron charge & $e$ & & Semiconductor permittivity & $\epsilon_\text{s}$
        \\ 
        Electron mass & $m_\text{e}$ & & Space charge potential factor & $M$
        \\ 
        Free carrier concentration & $N$ & & Space charge reduction factor & $R$
        \\ 
        Free carrier drift velocity & $v_\text{d}$ & & Stokes photon dissipation rate & $\gamma_\text{s}$
        \\
        Free carrier mobility & $\mu$ & & Stokes photon envelope operator & $A_\text{s}$
        \\ 
        Free electron scattering rate & $\gamma_\text{e}$ & & Stokes photon frequency & $\omega_\text{s}$
        \\ 
        Interaction Impedance & $Z_\text{m}$ & & Stokes photon group velocity & $v_\text{g,s}$
        \\ 
        Inverse Debye length & $\gamma_\text{d}$ & & Stokes photon scaling factor & $C_\text{s}$
        \\ 
        Linear piezobirefringence coefficient & $\alpha_\text{pe}$ & & Stokes photon wave vector & $k_\text{s}$
        \\ 
        Non-dimensionalized velocity & $\gamma_{v}$ & & Stokes photon wavelength & $\lambda_\text{s}$
        \\ 
        Optomechanical coupling coefficient & $g_{0}$ & & Temperature & $T$
        \\
        Permittivity of free space & $\epsilon_{0}$ & & Top dielectric permittivity & $\epsilon_\text{d}$
        \\ 
        Phonon dissipation rate & $\Gamma_\text{m}$ & & Waveguide thickness & $t$
        \\   
        Phonon envelope operator & $B$ & & Waveguide width & $w$
        \\
        Phonon frequency & $\Omega_\text{m}$
        \\
    \end{tabularx}
    \label{table:Notation}
\end{table*}

\newpage


\bibliography{cites}

\end{document}